\documentclass[runningheads]{svmult}

\usepackage{makeidx}   
\usepackage{graphicx}  
\usepackage{subeqnar}  
\usepackage{multicol}  
\usepackage{physprbb}  
\makeindex             

\begin{document}
\title*{Generation of Compressible Modes in MHD Turbulence}
\toctitle{Generation of Compressible Modes in MHD Turbulence}

\titlerunning{MHD Turbulence}

\author{Jungyeon Cho
\and A. Lazarian}

\authorrunning{Cho \& Lazarian}

\institute{Univ. of Wisconsin, Madison WI53706, USA}

\maketitle              

\begin{abstract}
Astrophysical turbulence is   magnetohydrodynamic (MHD) in its
nature. We discuss fundamental properties of MHD turbulence.
In particular, we discuss the generation of compressible MHD
waves by Alfvenic turbulence and show that this process
is inefficient. This allows us to study the evolution of
different types of MHD perturbations separately.
 We describe how to
separate MHD fluctuations into 3 distinct families - Alfven,
slow, and fast modes. We find that
the degree of suppression 
of slow and fast modes production by Alfvenic turbulence
depends on the
strength of the mean field.
We show that
Alfven modes in compressible regime exhibit scalings
and anisotropy similar to those in incompressible regime.
Slow modes passively mimic Alfven modes.
However, fast modes exhibit isotropy and a scaling similar to that of
acoustic turbulence both in high and low $\beta$ plasmas.
We show that our findings entail important
consequences for theories of star formation, cosmic ray propagation,
dynamics of dust, and gamma ray bursts. We anticipate many more
applications of the new insight to MHD turbulence and expect more
revisions of the existing paradigms of astrophysical processes
as the field matures.

\end{abstract}

\section{Introduction}

Astrophysics has been providing the major incentive for MHD studies.
High conductivity of astrophysical fluids makes magnetic fields ``frozen
in'', and they affect fluid motions. 
The coupled motion of magnetic field and conducting fluid
 is what a researcher has to
deal with while studying various astrophysical phenomena from star formation
to gamma ray bursts. 

Turbulence is ubiquitous in astrophysical fluids and it 
holds the key to many astrophysical
processes (stability of molecular clouds, heating of the interstellar medium, 
properties of accretion disks, 
cosmic ray transport etc). 
Why would we expect astrophysical fluids to be turbulent? 
A fluid of viscosity $\nu$ gets turbulent when the rate of viscous 
dissipation, which is $\sim \nu/L^2$ at the energy injection scale $L$, 
is much smaller than
the energy transfer rate $\sim V_L/L$, where $V_L$ is the velocity dispersion
at the scale $L$. The ratio of the two rates is the Reynolds number 
$Re=V_LL/\nu$. In general, when $Re$ is larger than $10-100$
the system becomes turbulent. Chaotic structures develop gradually as 
$Re$ increases,
and those with $Re\sim10^3$ are appreciably less chaotic than those
with $Re\sim10^8$. Observed features such as star forming clouds are
very chaotic with $Re>10^8$, which 
ensures that the fluids are turbulent.
The measured statistics of 
fluctuations ISM \cite{ArmRS95,StaL01,DesDG00}
and Solar wind fluctuations \cite{LeaSN98}
show signatures
of  
the Kolmogorov statistics obtained for 
incompressible unmagnetized turbulent fluid. 

Kolmogorov theory \cite{Kol41} provides a scaling law for incompressible 
{\it non}-magnetized hydrodynamic turbulence.
This law is true in the statistical sense and it provides a relation
between the relative velocity $v_l$ of fluid elements and their separation
$l$, namely, $v_l\sim l^{1/3}$.  An equivalent description is to 
express spectrum $E(k)$
as functions of wave number $k$ ($\sim 1/l$).
The two descriptions are related by $kE(k) \sim v_l^2$. The famous
Kolmogorov spectrum is  $E(k)\sim k^{-5/3}$. The applications of 
Kolmogorov theory range from engineering research to
meteorology (see \cite{MonY75})
but its astrophysical
applications are poorly justified.

Unlike laboratory turbulence astrophysical turbulence is magnetized
and highly compressible.
Then, why do astrophysical fluids show signatures of
Kolmogorov statistics?
Let us consider {\it incompressible} MHD turbulence first.
There have long been understanding that the MHD turbulence
is anisotropic (e.g.~\cite{SheMM83}).
A substantial progress has been achieved
recently by Goldreich \& Sridhar \cite{GolS95} (hereafter GS95) 
who made a
prediction regarding relative motions parallel and
perpendicular to magnetic field {\bf B} for incompressible
MHD turbulence. The GS95 model envisages a Kolmogorov spectrum of velocity 
and the scale-dependent anisotropy (see below).
These relations have been confirmed 
numerically 
(Cho \& Vishniac \cite{ChoV00b}; Maron \& Goldreich \cite{MarG01};
Cho, Lazarian \& Vishniac \cite{ChoLV02b}, hereafter CLV02b; 
see also review by Cho, Lazarian, \& Vishniac \cite{ChoLV02a}, 
hereafter CLV02a); 
they are in good agreement with observed and inferred astrophysical spectra 
(see CLV02a). A remarkable fact revealed in CLV02b is that
fluid motions perpendicular to {\bf B} are {\it identical} to hydrodynamic
motions. This provides an essential physical insight into why
in some respects MHD turbulence and hydrodynamic turbulence are
similar, while in other respects they are different.

However, in most cases compressibility of turbulence is important.
For instance, interstellar medium is highly compressible and star
formation requires considering supersonic compressible motions (see
reviews \cite{VazOP00,MacK03,ZweHF02}).
It can be shown
that assuming that only incompressible turbulence exists in ISM
results in grossly erroneous conclusions for cosmic ray transport
(see review \cite{LazCY03}).
It may be an important
question whether the physical pictures in incompressible
and compressible turbulence are similar. For instance, Cho, Lazarian
\& Vishniac \cite{ChoLV02c} (henceforth CLV02c) 
reported a new regime of turbulence
that takes place in a partially ionized gas. In this regime turbulent
energy protrudes to small scales through a magnetic cascade, while
the turbulent velocities are suppressed. How will compressibility
affect this regime?    

Compressible turbulence is an unsolved problem even in the absence
of magnetic fields. 
How feasible is it to strive for obtaining universal scaling relations
for compressible media in view of the fact that no such universality
exists for compressible hydro turbulence? 
The difficulty one encounters while studying compressible MHD
is that MHD turbulence is in general 
more complicated than its hydrodynamic counterpart.
In compressible regime, 
3 different types of motions (Alfven, slow, and fast modes) exist.
Alfven modes are incompressible and sometimes called {\it shear} Alfven modes.
The other two modes are
compressible modes (see \S\ref{sect_modes}). 
How do those modes interact?
Is it reasonable to talk about separate modes in highly non-linear
MHD turbulence? These and similar questions dealing with fundamental
properties of MHD turbulence we will attempt to answer below.

Thus we must consider
a more realistic case of compressible MHD turbulence.
Literature on the properties of compressible MHD is very rich (see CLV02a).
Back in 80s Higdon \cite{Hig84} theoretically studied density fluctuations
in the interstellar MHD turbulence.
Matthaeus \& Brown \cite{MatB88} studied nearly incompressible MHD at low Mach
number and Zank \& Matthaeus \cite{ZanM93} 
extended it. In an important paper
Matthaeus et al.~\cite{MatGO96}    
numerically
explored anisotropy of compressible MHD turbulence.
However, those
papers do not provide universal scalings of the GS95 type.
Some hints about effects of compressibility can be inferred from Goldreich
\& Sridhar (GS95)    
seminal paper. 
A more focused discussion was
presented in the Lithwick \& Goldreich \cite{LitG01}     
paper which deals with electron
density fluctuations in the pressure dominated plasma, 
i.e.  in high $\beta$ regime ($\beta\equiv P_{gas}/P_{mag}\gg 1$). 
Incompressible regime formally
corresponds to $\beta\rightarrow \infty$ and therefore it is natural
to expect that for $\beta\gg 1$ the GS95 picture would
persist. Lithwick \& Goldreich  \cite{LitG01}   
also speculated that for low $\beta$ plasmas the GS95
scaling of slow modes may be applicable.
A direct 
study of MHD modes in compressible low $\beta$ plasmas is given in
Cho \& Lazarian \cite{ChoL02a} (hereafter CL02),  
and more general results applicable
for a wide range of $\beta$ and Mach numbers are presented in Cho
\& Lazarian  \cite{ChoL03a} (hereafter CL03). 

The generation of slow and fast modes 
(i.e. MHD version of  ``{\it sound waves}'') has 
important astrophysical implications.
First, in the presence of damping, density and non-Alfvenic magnetic
fluctuations are generated only through compressible daughter waves 
(i.e. slow and fast waves) generated by Alfven turbulence.
These fluctuations are important for interstellar physics and
cosmic ray physics.
Second, if Alfvenic modes produce
a copious amount of compressible modes, the whole picture of independent
Alfvenic turbulence fails. Therefore, inefficient generation of
compressible modes from Alfven turbulence is a necessary condition
for independent Alfvenic cascade.

In what follows we review observational data on statistics
of turbulence, including the velocity data available through spectral line
studies \S\ref{sect_obs}. In \S\ref{sect_num} we describe our technique for 
decomposing MHD turbulence into Alfven, slow and fast modes. Mode coupling is
discussed in \S\ref{sect_result}, 
while simple theoretical arguments about mode scalings
are provided in \S\ref{sect_scaling}. 
We describe scalings of velocity fluctuations in \S\ref{sect_c6} and
magnetic and density fluctuations in 
\S\ref{sect_c7}. 
The
new regime of turbulence that emerges below the viscous cut-off is
briefly discussed in \S\ref{sect_c8}. 
\S\ref{sect_c9} deals with the applicability of our
results and with their significance for the theories of star formation,
cosmic ray propagation, gamma ray bursts etc. The summary is given
in \S\ref{sect_sum}.

\section{Observational Motivation}   \label{sect_obs}

Observations as well as space missions provide 
data on the statistics of astrophysical turbulence. 
This data suggests that in a wide variety of circumstances
astrophysical turbulence exhibits power-law spectra consistent
with Kolmogorov picture. It would be very naive to think that,
in the presence of dynamically important magnetic fields, the
turbulence may really be Kolmogorov, but it is suggestive
that, for a wide variety of circumstances, the turbulence should
allow pretty simple statistical description. 
This strongly motivates a quest
for simple relations to describe the apparently complex phenomenon.
 
Direct studies of turbulence\footnote{Indirect studies include
the line-velocity relationships \cite{Lar81} where the integrated
velocity profiles are interpreted as the consequence of turbulence.
Such studies do not provide the statistics of turbulence and their
interpretation is very model dependent.} 
have been done mostly for interstellar medium
and for the Solar wind. While for the Solar wind {\it in-situ} measurements
are possible, studies of interstellar turbulence require inverse techniques to
interpret the observational data. 

Attempts to study interstellar turbulence with statistical tools
date as far back as the 1950s \cite{Hor51,Kam55,Mun58,WilMF59}
and various directions
of research achieved various degree of success (see reviews by
\cite{KapP70,Dic85,ArmRS95,Laz99a,Laz99b,LazPE02}).

\subsection{Solar wind} 

Solar wind (see review \cite{GolR95}) is a magnetized flow of particles 
(mostly electrons and protons) from the Sun.
Studies of the solar wind allow point-wise statistics to be measured directly
using spacecrafts. These studies are the closest counterpart of 
laboratory measurements.

The solar wind flows nearly radially away from the Sun, at up to 
$\sim$700 km/s. This is much faster than both spacecraft motions and 
the Alfv\'en speed. Therefore, the turbulence is ``frozen'' and
the fluctuations at frequency $f$ are directly related to fluctuations
at the scale $k$ in the direction of the wind, as $k=2\pi f/v$, where $v$ is
the solar wind velocity \cite{Hor99}.

The solar wind shows $f^{-5/3}$ scaling on small scales.
The turbulence is strongly anisotropic (see \cite{KleBB93})
with the ratio of power in motions perpendicular to the magnetic field to 
those parallel to the magnetic field being around 30. The intermittency of 
the solar wind 
turbulence is very similar to the intermittency observed in
hydrodynamic flows \cite{HorB97}.

\subsection{Electron density statistics in the ISM}

Studies of turbulence statistics of ionized media in the interstellar space
(see \cite{SpaG90}) have provided information on
the statistics of plasma density at scales $10^{8}$-$10^{15}$~cm. 
This was based on
a clear understanding of processes of scintillations and scattering
achieved by theorists\footnote{In fact, the theory of scintillations was
developed first for the atmospheric applications.}
 (see \cite{NarG89,GooN85}).
A peculiar feature of the 
measured spectrum (see \cite{ArmRS95}) is the absence of
the slope change at the scale at which the viscosity by neutrals
becomes important. 

Scintillation measurements are the most reliable data in the
``big power law'' plot in Armstrong et al.~\cite{ArmRS95}. However
there are intrinsic limitations to the scintillations technique
due to the limited number of sampling directions, its relevance only to
ionized gas at extremely small scales, 
and the impossibility
of getting velocity (the most important!) statistics directly. Therefore
with the data one faces the problem of distinguishing actual turbulence
from static density structures. Moreover, the scintillation
data do not provide the index of turbulence directly, but only
show that the data are consistent with Kolmogorov turbulence.
Whether the (3D) index can be -4 instead of -11/3 is still
a subject of intense debate \cite{Hig84,NarG89}. 
In physical terms the former
corresponds to the superposition of random shocks rather than
eddies.

\subsection{Velocity and density statistics from spectral lines}

Atoms and molecules
in the interstellar space emit radiation at specific wavelengths.
A spectral line from atomic
hydrogen with $\lambda_0$=21~cm is particularly important in astronomy.
Astronomers observe intensity of radiation at different wavelengths
near $\lambda_0$ for different points on the sky, which
results in a 3D data cube that consists of two spatial (or, angular)
coordinates and one wavelength coordinate
(i.e. T=T($\theta_1$, $\theta_2$, $\lambda-\lambda_0$), where T is so-called
antenna temperature, which measures radiation energy).
Using the Doppler shift formula, we can convert
the wavelength dimension to the velocity dimension.
Such spectral line
data cubes are unique sources of information on interstellar turbulence. 
Doppler shifts due to supersonic motions contain information on the
turbulent velocity field which is otherwise difficult to obtain. Moreover,
the statistical samples are extremely
rich and not limited to discrete directions. In addition, line emission
allows us to study turbulence at large scales, comparable
to the scales of star formation and energy injection.

However, the problem of separating velocity and density fluctuations 
within HI data cubes is far from trivial 
\cite{Laz95,Laz99b,LazP00,LazPE02}.
The analytical
description of the emissivity statistics of channel maps (velocity slices)
in Lazarian \& Pogosyan \cite{LazP00} (see also \cite{Laz99b,LazPE02} for
reviews)
shows that the relative contribution of the
density and velocity fluctuations 
depends on the thickness of the velocity slice.
In particular, the power-law asymptote of the emissivity fluctuations 
changes
when the dispersion of the velocity at the scale under study 
becomes of the order of the velocity slice thickness (the integrated
width of the channel map).   
These results are the foundation of the Velocity-Channel Analysis (VCA) 
technique which provides velocity and density statistics
using spectral line data cubes.
The VCA has been successfully tested using data
cubes obtained via compressible magnetohydrodynamic simulations and 
has been applied
to Galactic and Small Magellanic Cloud atomic hydrogen (HI) data 
\cite{LazPV01,LazP00,StaL01,DesDG00}. 
{}Furthermore,
the inclusion of absorption effects \cite{LazP03} has increased 
the power of this technique. 
{}Finally, the VCA can be applied to different species (CO, H$_{\alpha}$ etc.)
which should further increase its utility in the future.

Within the present discussion a number of results obtained with the VCA
are important. First of all, the Small Magellanic Cloud (SMC) HI data
exhibit a Kolmogorov-type spectrum for velocity and HI density from
the
smallest resolvable scale of 40~pc to the scale of the SMC itself, i.e.
4~kpc. Similar conclusions can be inferred from the Galactic data
\cite{Gre93} for scales of dozens of parsecs, 
although the analysis has not been done systematically. Deshpande et al. 
\cite{DesDG00} studied absorption of HI on small scales toward Cas A and 
Cygnus A.
Within the VCA their results can be interpreted 
as implying
that on scales less than 1~pc the HI velocity is suppressed by
ambipolar drag and the spectrum of density fluctuations is shallow 
$P(k)\sim k^{-2.8}$. Such a spectrum \cite{Des00}
can account for the small scale structure of HI observed in absorption.     

\subsection{Magnetic field statistics}

Magnetic field statistics are the most poorly constrained aspect
of ISM turbulence. The polarization
of starlight and of the Far-Infrared Radiation (FIR) from aligned dust 
grains
is affected by the ambient magnetic fields. Assuming that dust grains are
always aligned with their longer axes perpendicular to magnetic field
(see review \cite{Laz00a}), one gets the 2D distribution of the 
magnetic field directions in the sky. Note that the alignment is
a highly non-linear process in terms of the magnetic field and
therefore the magnetic field strength is not available\footnote{The
exception to this may be the alignment of small grains which can
be revealed by microwave and UV polarimetry \cite{Laz00a}.}.

The statistics of starlight polarization (see \cite{FosLP02})
is rather rich for the Galactic plane and it allows to establish
the spectrum\footnote{Earlier papers
dealt with much poorer samples (see \cite{KapP70})
and they did not reveal power-law spectra.} $E(K)\sim K^{-1.5}$, where $K$ is a two dimensional
wave vector describing the fluctuations over sky patch.\footnote{
This spectrum is obtained by \cite{FosLP02} in terms of the expansion over the 
spherical harmonic basis $Y_{lm}$. 
For sufficiently small areas of the sky analyzed
the multipole analysis results coincide with the Fourier analysis.}

For uniformly sampled turbulence it follows from Lazarian \& Shutenkov
\cite{LazS90}
that $E(K)\sim K^{\alpha}$ for $K<K_0$ and $K^{-1}$ for $K>K_0$, where
$K_0^{-1}$ is the critical angular size of fluctuations which is
proportional to the ratio of the injection energy scale to the size
of the turbulent system along the line of sight. For Kolmogorov
turbulence $\alpha=-11/3$.

However, the real observations do not uniformly sample turbulence.
Many more close stars are present compared to the distant ones. 
Thus the intermediate slops are expected. Indeed,
Cho \& Lazarian \cite{ChoL02b} showed through direct simulations that the 
slope obtained in \cite{FosLP02} is compatible with the underlying
Kolmogorov turbulence.
At the moment FIR polarimetry
does not provide maps that are really suitable to study turbulence
statistics. This should change soon 
when polarimetry becomes possible using the
airborne SOFIA observatory.  A better understanding of grain
alignment (see \cite{Laz00a}) is required to interpret
the molecular cloud magnetic data where some of the dust is
known not to be aligned (see \cite{LazGM97} and references therein).  

Another way to get magnetic field statistics is to use synchrotron
emission. Both polarization and intensity data can be used. The angular
correlation of
polarization data \cite{BacBP01} shows the power-law spectrum 
$K^{-1.8}$ and we believe that the interpretation of it is similar to that
of starlight polarization.
Indeed, Faraday depolarization
limits the depth of the sampled region. 
The
intensity fluctuations were studied in \cite{LazS90}
with rather poor initial data and the results were inconclusive.
Cho \& Lazarian \cite{ChoL02b} 
interpreted the fluctuations of synchrotron emissivity
 \cite{GiaBF01,GiaBG02} in terms of turbulence with Kolmogorov spectrum.

\section{Numerical Approach}   \label{sect_num}

\subsection{Helmholtz decomposition for hydrodynamic turbulence} 
    \label{secion-helmholtz}

To get an insight of the turbulence cascade we have attempted
a decomposition of the MHD turbulent flow into Alfven, slow
and fast modes (see CL02, CL03). 
Our numerical method is similar to the technique utilizing the 
 ``{\it Helmholtz}''
decomposition in hydrodynamics.

Our method is different from 
Lighthill's theory  \cite{Lig52}
of {\it far field} acoustic wave generation
{}from homogeneous turbulence.
Literature on the application of Lighthill's approximation to 
astrophysical problems  is rich.
Astrophysical fluids are stratified (by gravity) and/or magnetized.
Therefore, Lighthill's theory 
requires modifications for astrophysical fluids.
Stein \cite{Ste67}     
extended Lighthill's theory to stratified astrophysical
fluids in gravitational field.
Subsequent papers (e.g. \cite{GolK90})   
further
discussed about generation of
acoustic waves in (Solar) convection zone.
On the other hand, 
Musielak \& Rosner \cite{MusR88}    
constructed a model for
weak magnetic field convection zone and 
Musielak, Rosner, \&  Ulmschneider  \cite{MusRU89}    
discussed about wave generation in an inactive flux tube.
Lee \cite{Lee93}    
explored wave generation in sunspots, 
which are the most strongly magnetized
on the surface of the Sun.
All these approaches are to calculate far-field acoustic flux.

Moyal \cite{Moy51}     
introduced a method that decomposes velocity field in
Fourier space, which is equivalent to Helmholtz's
decomposition of a vector field: ${\bf V} = {\bf V}_s + {\bf V}_c$,
where ${\bf V}_s$ is divergence-free ($\nabla \cdot {\bf V}_s$=0) field
and ${\bf V}_c$ is curl-free ($\nabla \times {\bf V}_c$=0) field.
Note that ${\bf V}_s$ represents incompressible or solenoidal part and
${\bf V}_c$ compressible or dilatational one.
In Fourier space, solenodal and dilatational components have
simple geometrical meanings: ${\bf V}_s$ is the component perpendicular
to the wave vector ${\bf k}$ and ${\bf V}_s$ parallel to ${\bf k}$.

The first (published) numerical simulations of compressible
hydrodynamic turbulence were performed by
Feiereisen, et al. \cite{FeiRF82}.                 
They studied subsonic (sonic Mach numbers, $M_s$, up to 0.32) 
homogeneous shear flows with
$64^3$ grid points.
Passot \& Pouquet \cite{PasP87}     
carried out two-dimensional
isotropic homogeneous compressible decaying 
turbulence with $256^2$ grid points.
They showed that properties of turbulence at low initial Mach numbers
($M_s<0.3$) is significantly different from those of higher Mach number
ones.
Passot, Pouquet, \& Woodward \cite{PasPW88}        
simulated two-dimensional isotropic decaying
turbulence with initial Mach numbers up to $4$.
They provided conjecture for the three-dimensional case and
discussed implications of their work on astrophysical fluids
in the interstellar medium.
Subsequent simulations \cite{KidO90a,KidO90b,StaYK90,SarEH91,LeeLM91}
addressed various issues of
compressible turbulence.
Recent high resolution three-dimensional simulations include
Porter, Pouquet, \& Woodward \cite{PorPW92},   
Porter, Woodward, \& Pouquet \cite{PorWP98},              
and Porter, Pouquet, \& Woodward  \cite{PorPW02}.         

The energy spectra of compressible hydrodynamic
turbulence are still uncertain.
For spectrum of solenoidal components,
a Kolmogorov-type dimensional analysis leads to
\begin{equation}
  E_{solenoidal}(k)\propto k^{-5/3} (kL)^{\alpha M_s^2}
\end{equation}
(\cite{KadP73}; see also \cite{PasPW88}). 
However, Moiseev et al.~\cite{MoiPT81}         
obtained slightly different
results
\begin{equation}
  E_{solenoidal}(k)\propto k^{-5/3} (kL)^{\frac{-2M_s^2}{3(3-M_s^2)}}.
\end{equation}
When, $M_s \rightarrow 0$, both results give Kolmogorov spectrum.

The energy spectrum of compressible components is more uncertain.
For example, Zakharov \& Sagdeev  \cite{ZakS70}     
derived scalings for compressible
modes:
\begin{equation}
  E_{rad}(k)\propto k^{-3/2},
\end{equation}
where the subscript {\it rad} denotes compressible components (i.e.
{\it radial} components in Fourier space).
On the other hand, Bataille \& Zhou \cite{BatZ99}            
and 
Bertoglio, Bataille, \& Marion  \cite{BerBM01}        
obtained
that the spectral index ({\it slope}) is a function of Mach number, $M_s$.
When, Mach number is of order unity, their results give a Kolmogorov
spectrum.
Recent numerical simulations \cite{PorWP98}     
with up to $1024^3$
grid points show Kolmogorov's $k^{-5/3}$ spectra both for
$E_{solenoidal}(k)$ and $E_{rad}(k)$.

The generation of compressible components from 
incompressible initial turbulence is also an unresolved issue.
Closure calculation by Bataille \& Zhou \cite{BatZ99}   
and
Bertoglio et al. \cite{BerBM01}  
predicts
that $\chi \equiv \langle V_{rad}^2 \rangle /\langle V_{solenoidal}^2\rangle
  \sim M_s^2$.
Numerical calculations of decaying turbulence with initial Mach number
of order unity  \cite{PorPW92,PorPW02}          
show that $\chi \sim 0.1$.



\subsection{MHD mode decomposition}   \label{sect_modes}

Three types of waves exist (Alfven, slow and fast)
in compressible magnetized plasma.
In this section, we describe how to separate
different MHD modes.

In the presence of magnetic field ${\bf B}$, the momentum equation
has an additional term, $(\nabla \times {\bf B})\times {\bf B}$
(divided by $1/4\pi$).
This is the so-called ${\bf J}\times {\bf B}$ term, which
can be re-written as the sum of 
the magnetic tension term, ${\bf B}\cdot \nabla {\bf B}$,
and magnetic pressure term, $\nabla^2{\bf B}$:
\begin{equation}
  (\nabla \times {\bf B})\times {\bf B}
    ={\bf B}\cdot \nabla {\bf B}-\nabla^2{\bf B}/2.
\end{equation}
In addition, when magnetic Reynolds number, $R_m=VL/\eta$, where
$\eta$ is magnetic diffusivity, is large, magnetic field lines 
move together with fluid elements, which is sometime called
that magnetic fields are {\it frozen-in}.

In some sense, the magnetic field lines are like elastic bands moving together
with fluid elements in that they have tension.
However, they are different from rubber bands in that
they are repulsive each other, which is the nature of magnetic pressure.

\begin{figure*}
\begin{center}
  \includegraphics[width=0.75\textwidth]{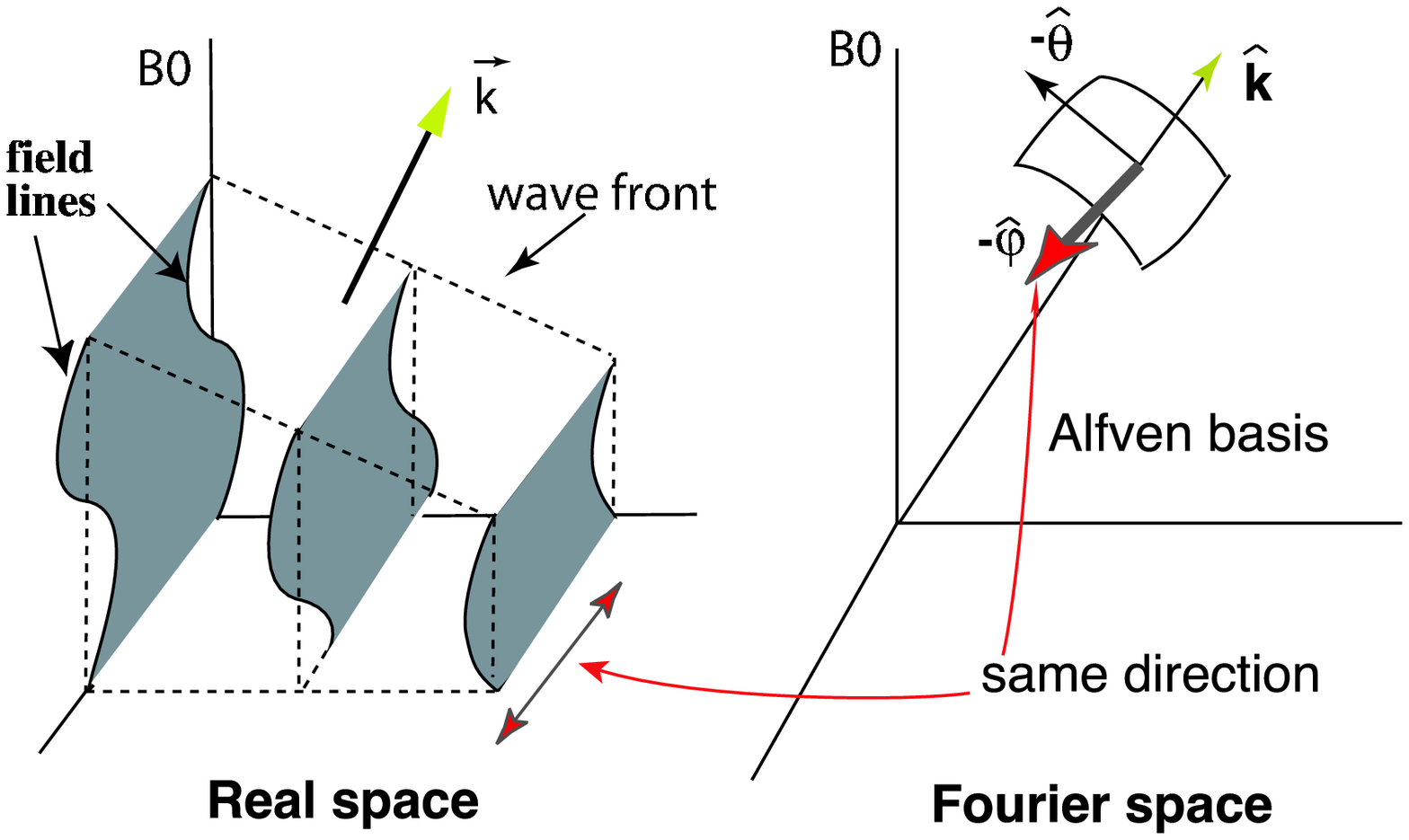} \\ 
\mbox{\bf (a)}  \\   
  \includegraphics[width=0.75\textwidth]{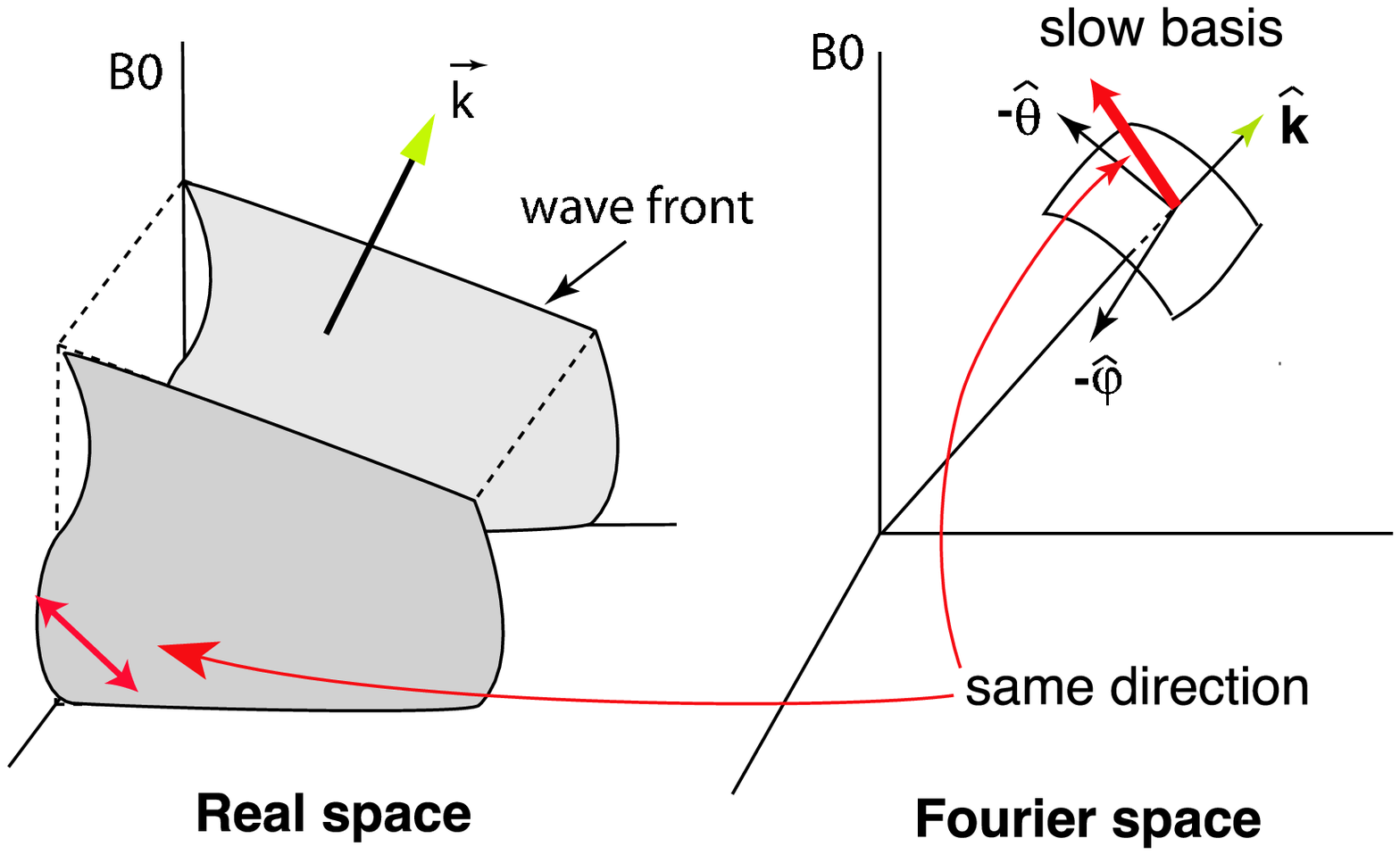} \\  
\mbox{\bf (b)} \\    
  \includegraphics[width=0.75\textwidth]{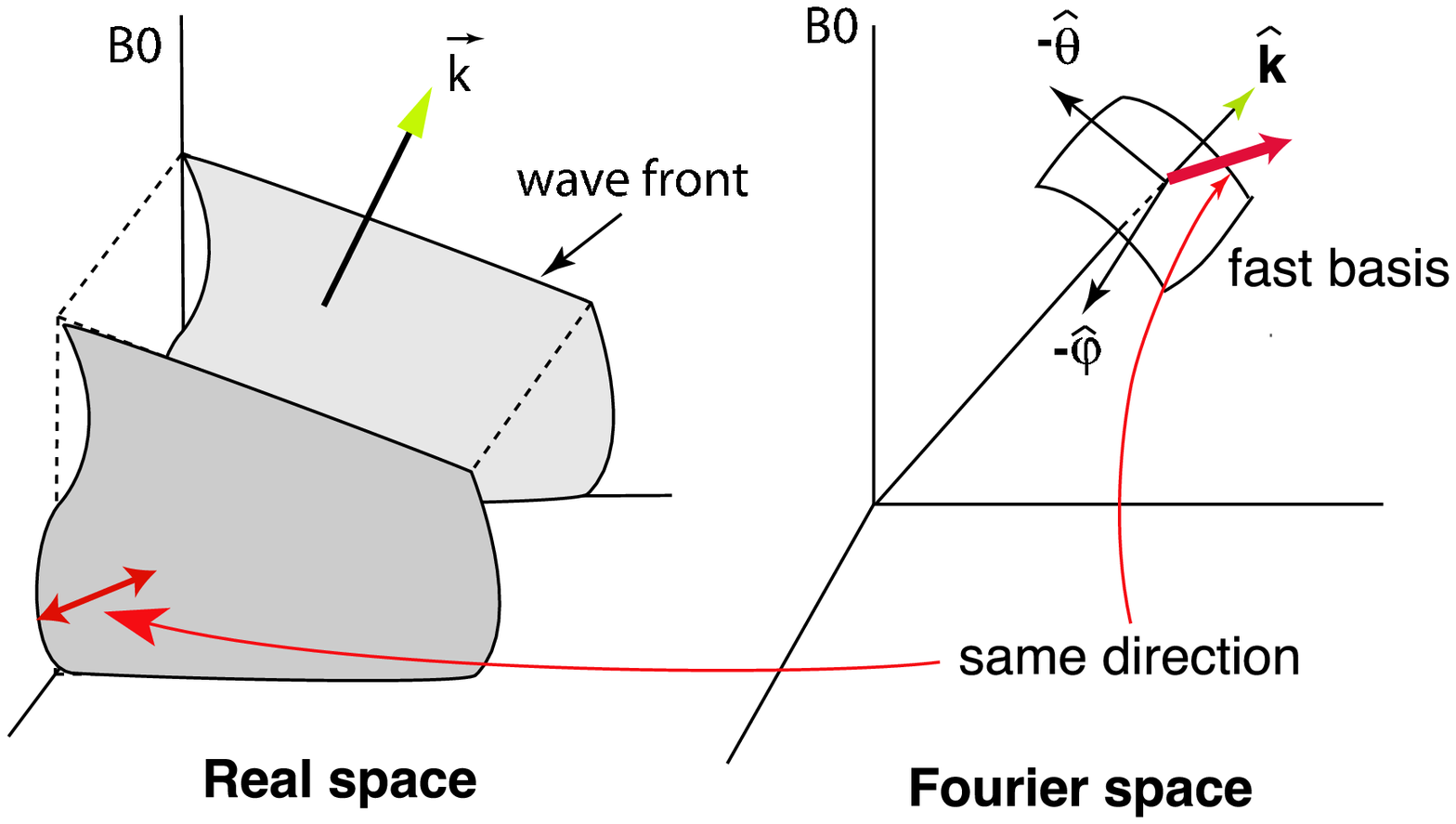} \\  
\mbox{\bf (c)}
\caption{ {\it (a)}
         Alfven wave and its direction of oscillation/displacement.
        The direction of displacement is perpendicular to
        both ${\bf B}_0$ and ${\bf k}$.
        Only magnetic tension is the restoring force.
    {\it (b)}
         Slow wave and its direction of oscillation/displacement.
        The direction of displacement is between 
        $-\hat{\bf \theta}$ and $\hat{\bf B}_0$.
    {\it (c)}
         Slow wave and its direction of oscillation/displacement.
        The direction of displacement is between 
        $\hat{\bf k}$ and $\hat{\bf k}_{\perp}$.
}
\label{fig_alf_basis}
\end{center}
\end{figure*}

\begin{figure*}
  \includegraphics[width=0.99\textwidth]{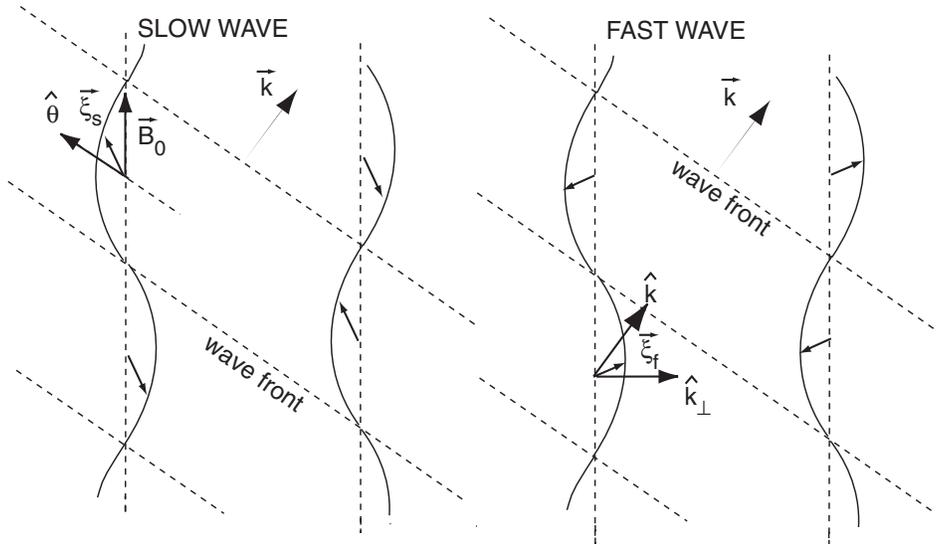}
\caption{
         Slow and fast waves in real space (${\bf B}_0 - {\bf k}$ plane).
         We show the directions of displacement vectors for
         a slow wave ({\it left panel}) and a fast wave ({\it right panel}).
         Note that $\hat{\xi}_s$ lies between
        $\hat{\theta}$ and $\hat{\bf B}_0 ~(=\hat{\bf k}_{\|})$ and 
                   $\hat{\xi}_f$ between
        $\hat{\bf k}$ and $\hat{\bf k}_{\perp}$.
         Again, $\hat{\theta}$ is perpendicular to 
             $\hat{\bf k}$ and parallel to the wave front.
         Note also that, for the fast wave, for example,
         density (inferred by the
         directions of the displacement vectors)
         becomes higher where field lines are closer, resulting
         in a strong restoring force, which is why fast waves
         are faster than slow waves.
         {}From CLV02a.
}
\label{fig_modes-real}
\end{figure*}

Because of tension and pressure, the nature of MHD waves is
much more complicated than their hydrodynamic counterpart - sound wave.
This is because we need to consider 3 different restoring forces -
magnetic tension, magnetic pressure, and gas pressure.
For Alfven waves, magnetic tension is the only the restoring force
(Fig.~\ref{fig_alf_basis}(a)).
For slow and fast waves, all 3 restoring forces are important.
For Slow waves, magnetic and gas pressure are out of phase and,
for fast modes, they are in phase (Fig.~\ref{fig_modes-real}).

The slow, fast, and Alfven bases that denote the direction of displacement
vectors for each mode are given by 
\begin{eqnarray}
   \hat{\bf \xi}_s \propto 
     ( -1 + \alpha - \sqrt{D} )
            k_{\|} \hat{\bf k}_{\|} 
     + 
     ( 1+\alpha - \sqrt{D} ) k_{\perp} \hat{\bf k}_{\perp},
  \label{eq_xis_new}
\\
   \hat{\bf \xi}_f \propto 
     ( -1 + \alpha + \sqrt{D} )
           k_{\|}  \hat{\bf k}_{\|} 
     + 
     ( 1+\alpha + \sqrt{D} ) k_{\perp} \hat{\bf k}_{\perp},  
   \label{eq_xif_new}
\\
 \hat{\bf \xi}_A \propto \hat{\bf \varphi} 
         \propto \hat{\bf k}_{\perp} \times \hat{\bf k}_{\|},
\end{eqnarray}
where $D=(1+\alpha)^2-4\alpha \cos\theta$, $\alpha=a^2/V_A^2=\beta(\gamma/2)$,
$\theta$ is the angle between ${\bf k}$ and ${\bf B}_0$, and
$\hat{\bf \varphi}$ is the azimuthal basis in the spherical polar coordinate
system (see Appendix).
These are equivalent to the expression in CL02:
\begin{eqnarray}
   \hat{\bf \xi}_s &\propto &
        k_{\|} \hat{\bf k}_{\|}+
     \frac{ 1-\sqrt{D}-{\beta}/2  }{ 1+\sqrt{D}+{\beta}/2  } 
    \left[ \frac{ k_{\|} }{ k_{\perp} }  \right]^2
     k_{\perp} \hat{\bf k}_{\perp},  \label{eq_xis}     \\
   \hat{\bf \xi}_f &\propto &
     \frac{ 1-\sqrt{D}+{\beta}/2  }{ 1+\sqrt{D}-{\beta}/2  } 
    \left[ \frac{ k_{\perp} }{ k_{\|} } \right]^2
     k_{\|} \hat{\bf k}_{\|}  +
          k_{\perp} \hat{\bf k}_{\perp}.
\end{eqnarray}
(Note that $\gamma=1$ for isothermal case.)

We can obtain slow and fast velocity by projecting velocity Fourier component 
${\bf v}_{\bf k}$ into $\hat{\bf \xi}_s$ and $\hat{\bf \xi}_f$, respectively.
In Appendix, we also discuss how to separate slow and fast magnetic modes.
We obtain energy spectra using this projection method.


\subsection{Numerical method}   \label{sect_method}

We use a hybrid
essentially non-oscillatory (ENO) scheme to solve the ideal isothermal
MHD equations.
When variables are sufficiently smooth, we use the 3rd-order
Weighted ENO scheme \cite{JiaW99}    
without
characteristic mode decomposition. 
When the opposite is true, 
we use the 3rd-order Convex ENO scheme  \cite{LiuO98}. 
Combined with a three-stage Runge-Kutta method for time integration,
our scheme gives third order accuracy in space and time.
We solve the ideal MHD equations in a periodic box:
\begin{eqnarray}
{\partial \rho    }/{\partial t} + \nabla \cdot (\rho {\bf v}) =0,  \\
{\partial {\bf v} }/{\partial t} + {\bf v}\cdot \nabla {\bf v} 
   +  \rho^{-1}  \nabla(a^2\rho)
   - (\nabla \times {\bf B})\times {\bf B}/4\pi \rho ={\bf f},  \\
{\partial {\bf B}}/{\partial t} -
     \nabla \times ({\bf v} \times{\bf B}) =0, 
\end{eqnarray}
with
    $ \nabla \cdot {\bf B}= 0$ and an isothermal equation of state.
Here $\bf{f}$ is a random large-scale driving force, 
$\rho$ is density,
${\bf v}$ is the velocity,
and ${\bf B}$ is magnetic field.
The rms velocity $\delta V$ is maintained to be unity, so that
 ${\bf v}$ can be viewed as the velocity 
measured in units of the r.m.s. velocity
of the system and ${\bf B}/\sqrt{4 \pi \rho}$ 
as the Alfv\'{e}n speed in the same units.
The time $t$ is in units of the large eddy turnover time ($\sim L/\delta V$) 
and the length in units of $L$, the scale of the energy injection.
The magnetic field consists of the uniform background field and a
fluctuating field: ${\bf B}= {\bf B}_0 + {\bf b}$.


For mode coupling studies (Fig.~\ref{fig_coupling}), 
we do {\it not} drive turbulence.
For scaling studies, 
we drive turbulence solenoidally in Fourier space and
use $216^3$ points and $\rho_0=1$. 
The average rms velocity in statistically stationary state is 
$\delta V\sim 0.7$.

For our calculations we assume that
$B_0/\sqrt{4 \pi \rho} \sim \delta B/\sqrt{4 \pi \rho} \sim \delta V$.
In this case, the sound speed is the controlling parameter and
basically two regimes can exist: supersonic and subsonic.
Note that supersonic means low-beta and subsonic means high-beta.
When supersonic, we consider mildly supersonic (or, mildly low-$\beta$)
and highly supersonic (or, very low-$\beta$).

\begin{figure*}
  \includegraphics[width=0.99\textwidth]{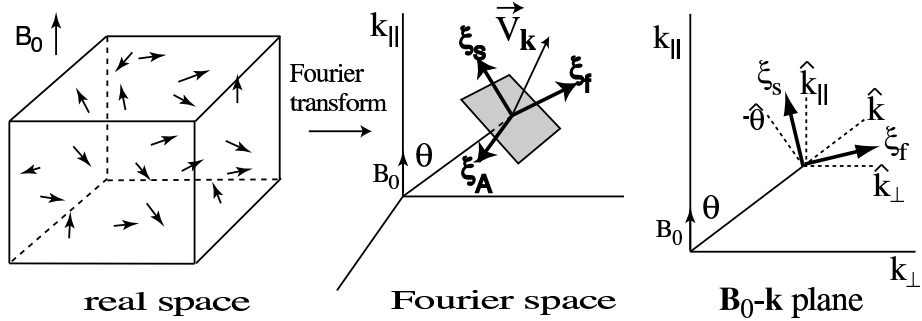}
  \caption{
      Separation method. We separate Alfven, slow, and fast modes in Fourier
      space by projecting the velocity Fourier component ${\bf v_k}$ onto
      bases ${\bf \xi}_A$, ${\bf \xi}_s$, and ${\bf \xi}_f$, respectively.
      Note that ${\bf \xi}_A = -\hat{\bf \varphi}$. 
      Slow basis ${\bf \xi}_s$ and fast basis ${\bf \xi}_f$ lie in the
      plane defined by ${\bf B}_0$ and ${\bf k}$.
      Slow basis ${\bf \xi}_s$ lies between $-\hat{\bf \theta}$ and 
      $\hat{\bf k}_{\|}$.
      Fast basis ${\bf \xi}_f$ lies between $\hat{\bf k}$ and 
      $\hat{\bf k}_{\perp}$. From CL03.
}
\label{fig_separation}
\end{figure*}

\section{Mode Coupling: Theory and Simulations}      \label{sect_result}

As mentioned above, the coupling of compressible and incompressible modes
is crucial. If Alfvenic modes produce
a copious amount of compressible modes, the whole picture of independent
Alfvenic turbulence fails. 

The generation of compressible motions 
(i.e. {\it radial} components in Fourier space) 
{}from Alfvenic turbulence
is a measure of mode coupling.
How much energy in compressible motions is drained from Alfvenic cascade?
According to the closure calculations 
(\cite{BerBM01}; see also \cite{ZanM93}),
the energy in compressible modes in {\it hydrodynamic} turbulence scales
as $\sim M_s^2$ if $M_s<1$.
We may conjecture that this relation can be extended to MHD turbulence
if, instead of $M_s^2$, we use
$\sim (\delta V)_{A}^2/(a^2+V_A^2)$. 
(Hereinafter, we define $V_A\equiv B_0/\sqrt{4\pi\rho}$.) 
However, as the Alfven modes 
are anisotropic, 
this formula may require an additional factor.
The compressible modes are generated inside so-called
Goldreich-Sridhar cone, which takes up $\sim (\delta V)_A/ V_A$ of
the wave vector space. The ratio of compressible to Alfvenic energy 
inside this cone is the ratio given above. 
If the generated fast modes become
isotropic (see below), the diffusion or, ``isotropization'' of
fast wave energy in the wave vector space increase their energy by
a factor of $\sim V_A/(\delta V)_A$. This  results in
\begin{equation}
  \frac{ (\delta V)_{rad}^2 }{ (\delta V)_A^2 }    \sim
 \left[ \frac{ V_A^2 + a^2 }{ (\delta V)^2_A } 
        \frac{ (\delta V)_A }{ V_A }   \right]^{-1},
\label{eq_high2}
\end{equation}
where $(\delta V)_{rad}^2$ and $(\delta V)_{A}^2$ are energy
of compressible\footnote{It is possible to show that 
the compressible modes inside the Goldreich-Sridhar cone
are basically fast modes.}  and Alfven modes, respectively.
Eq.~(\ref{eq_high2}) suggests that the drain of energy from
Alfvenic cascade is marginal when the amplitudes of perturbations
are weak, i.e. $(\delta V)_A\ll  V_A$. 

\begin{figure*}
\begin{tabbing}
 ~~~~~~~~ \includegraphics[width=0.40\textwidth]{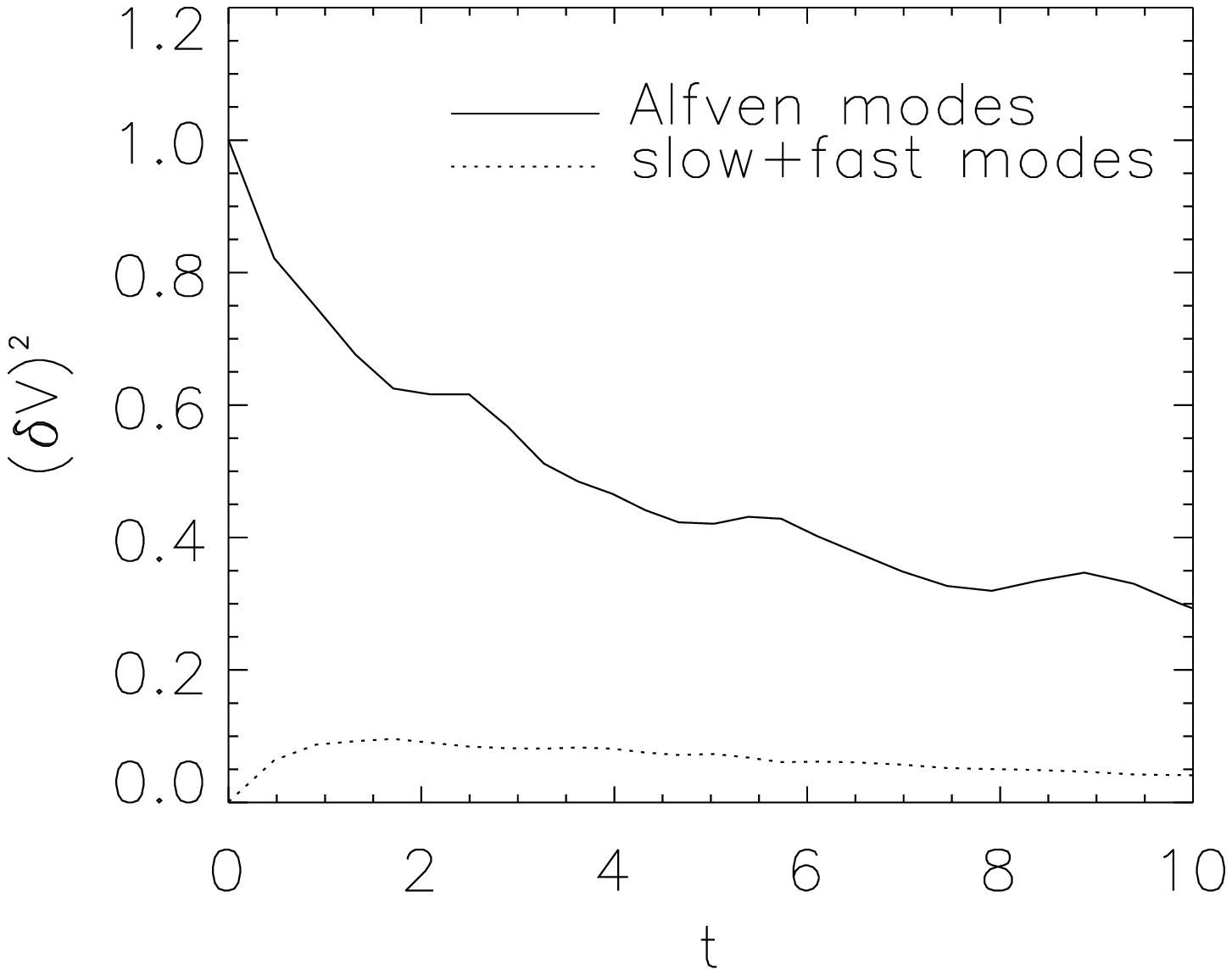}
\=
 ~~~~~~~~~ \includegraphics[width=0.40\textwidth]{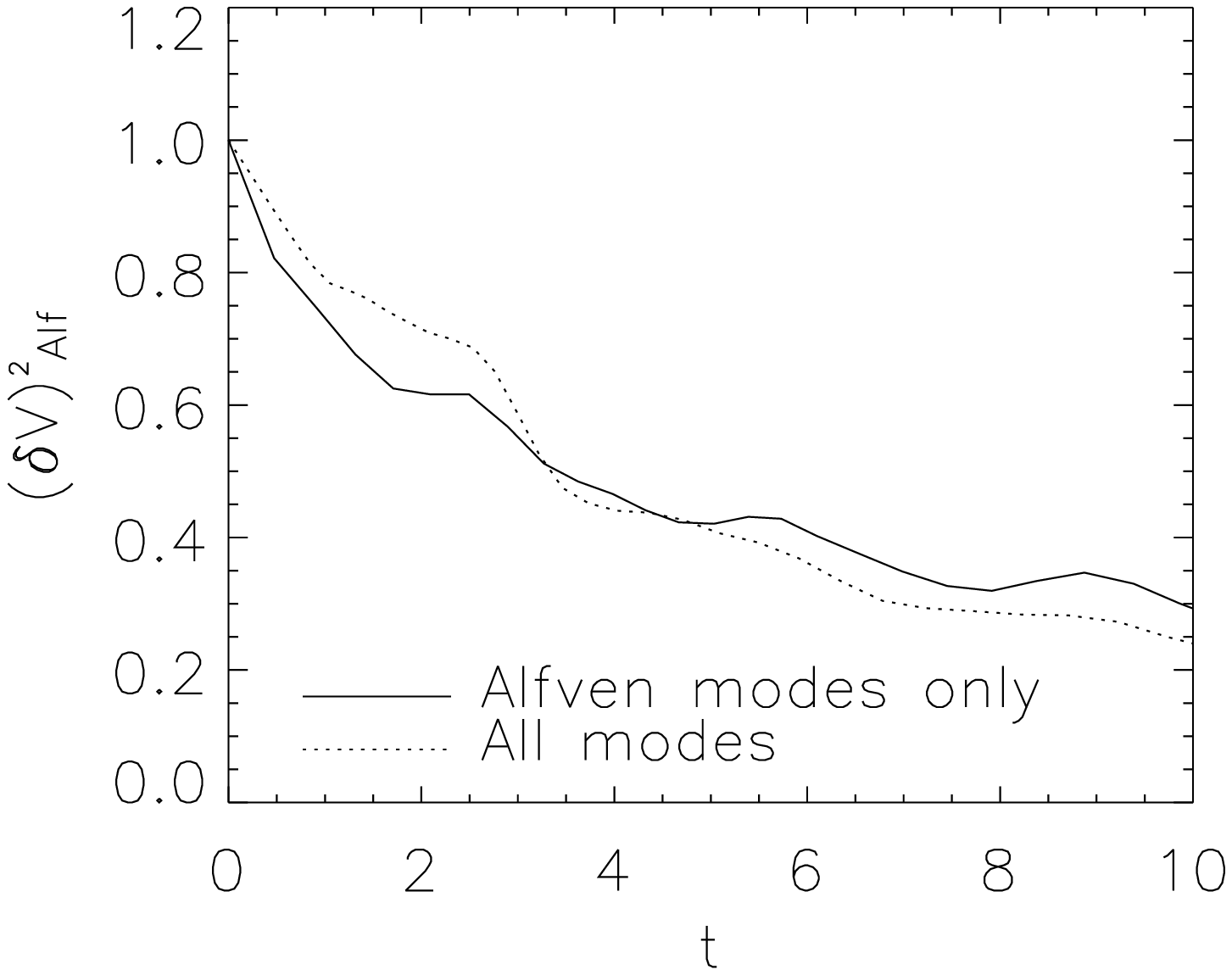}
\\
~~~~~~~~~~~~~~~~~~~~~~~~~~~~~~ (a)\>  ~~~~~~~~~~~~~~~~~~~~~~~~~~~~~~~~~~(b)
\\
  \includegraphics[width=0.49\textwidth]{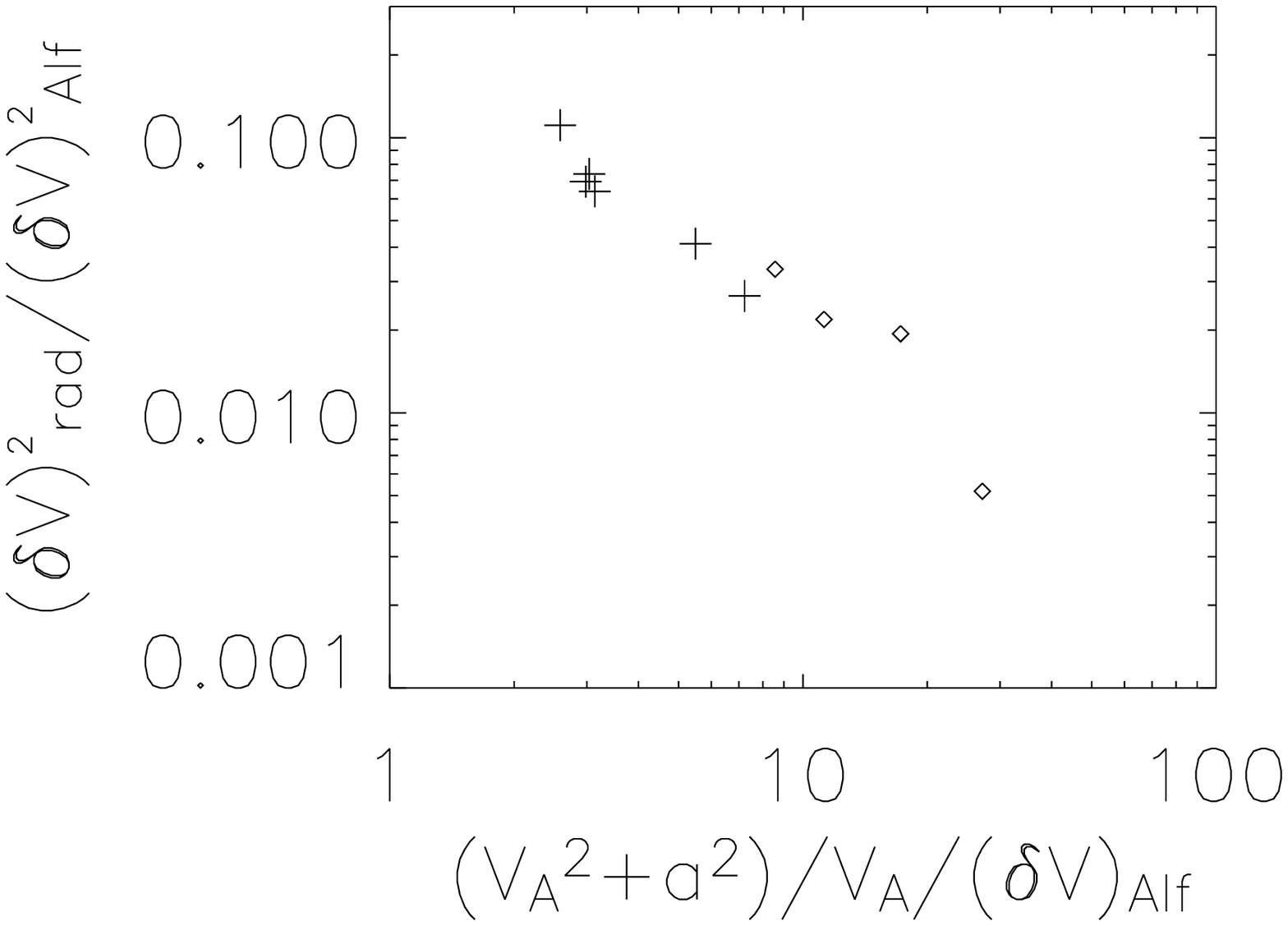}
\>
 ~ \includegraphics[width=0.49\textwidth]{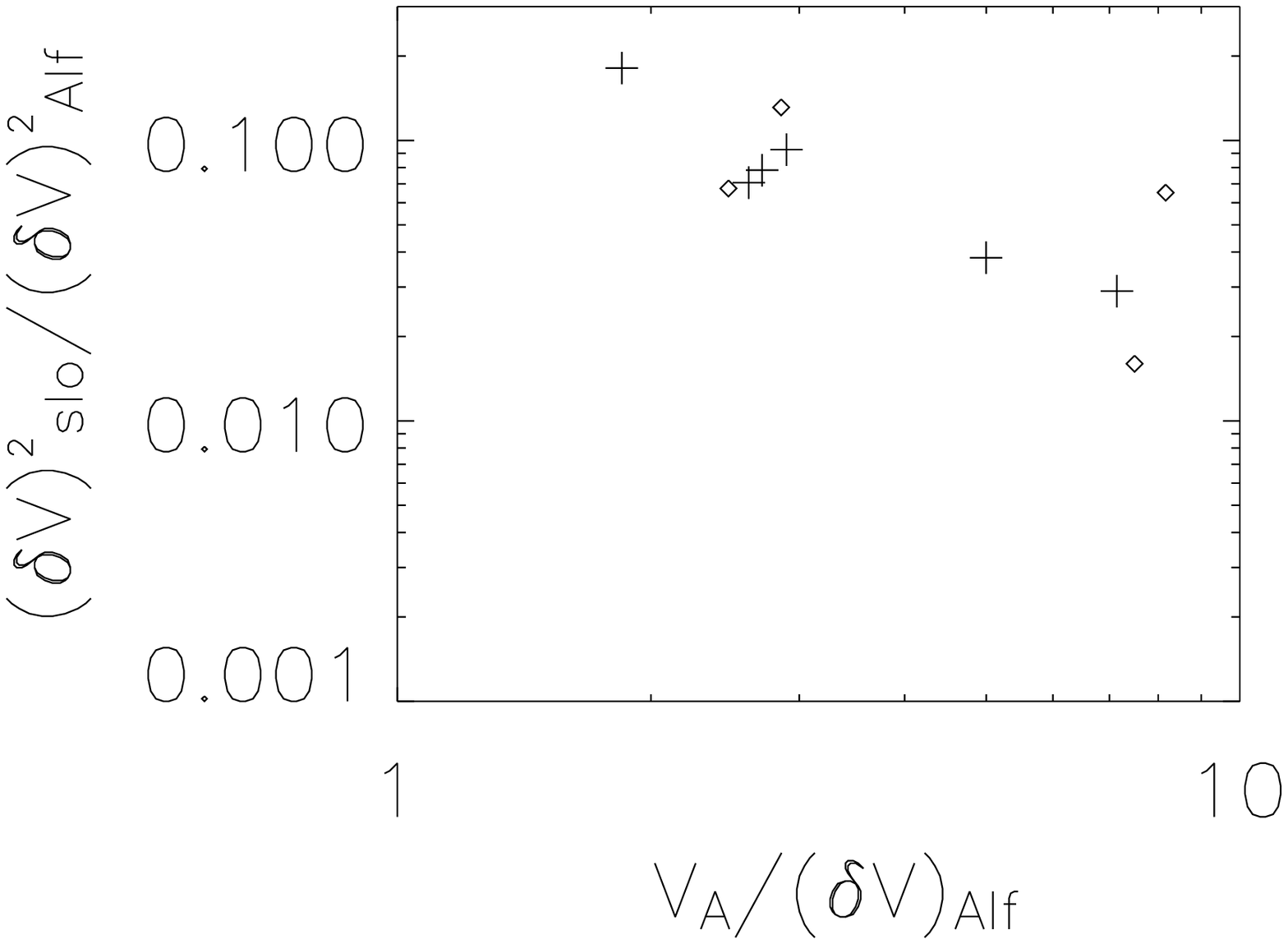}
\\
~~~~~~~~~~~~~~~~~~~~~~~~~~~~~~ (c)\>  ~~~~~~~~~~~~~~~~~~~~~~~~~~~~~~~~~~(d)
\\
 ~~~~~~~\includegraphics[width=0.40\textwidth]{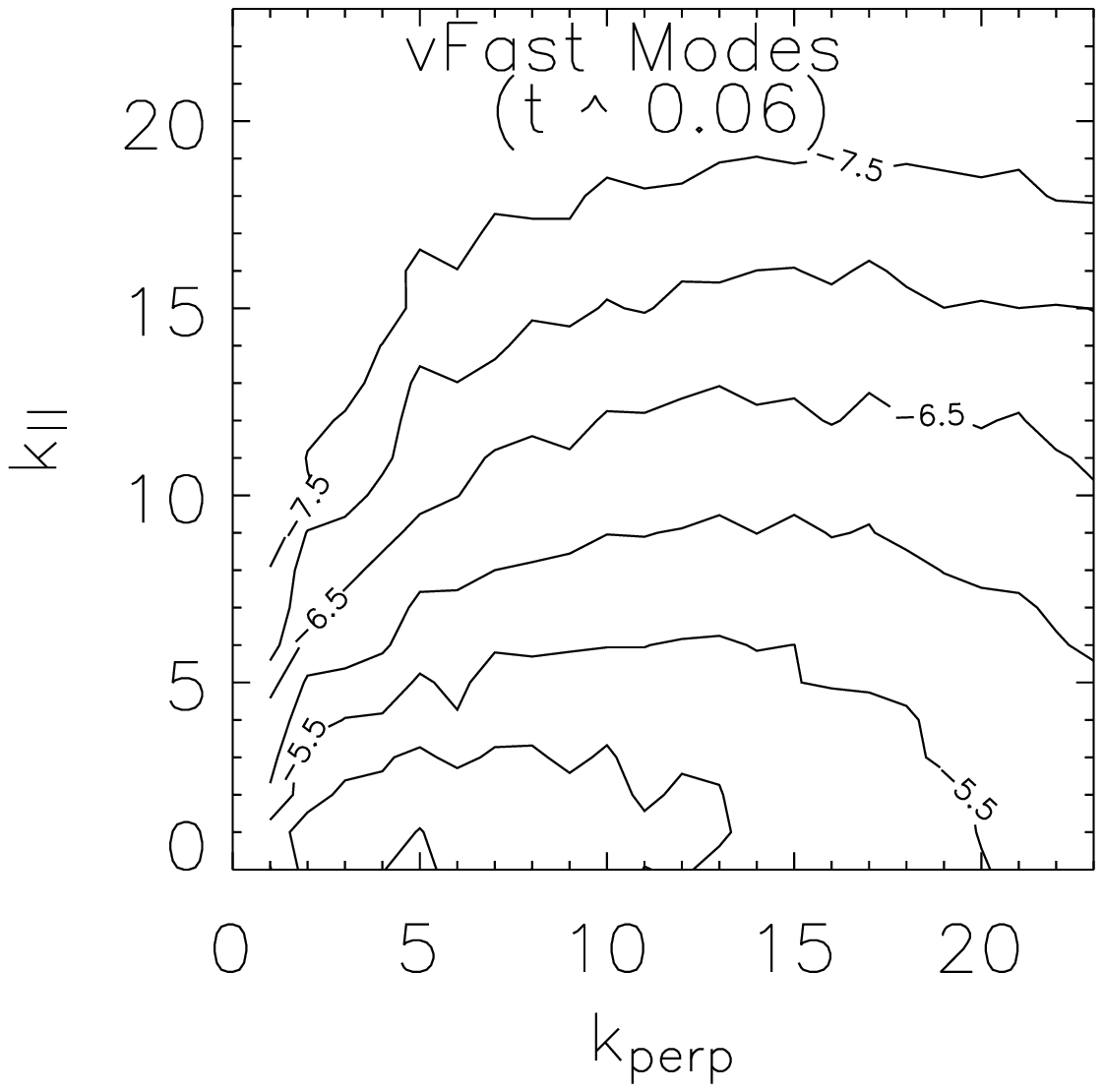}
\>
 ~~~~~~~~~\includegraphics[width=0.40\textwidth]{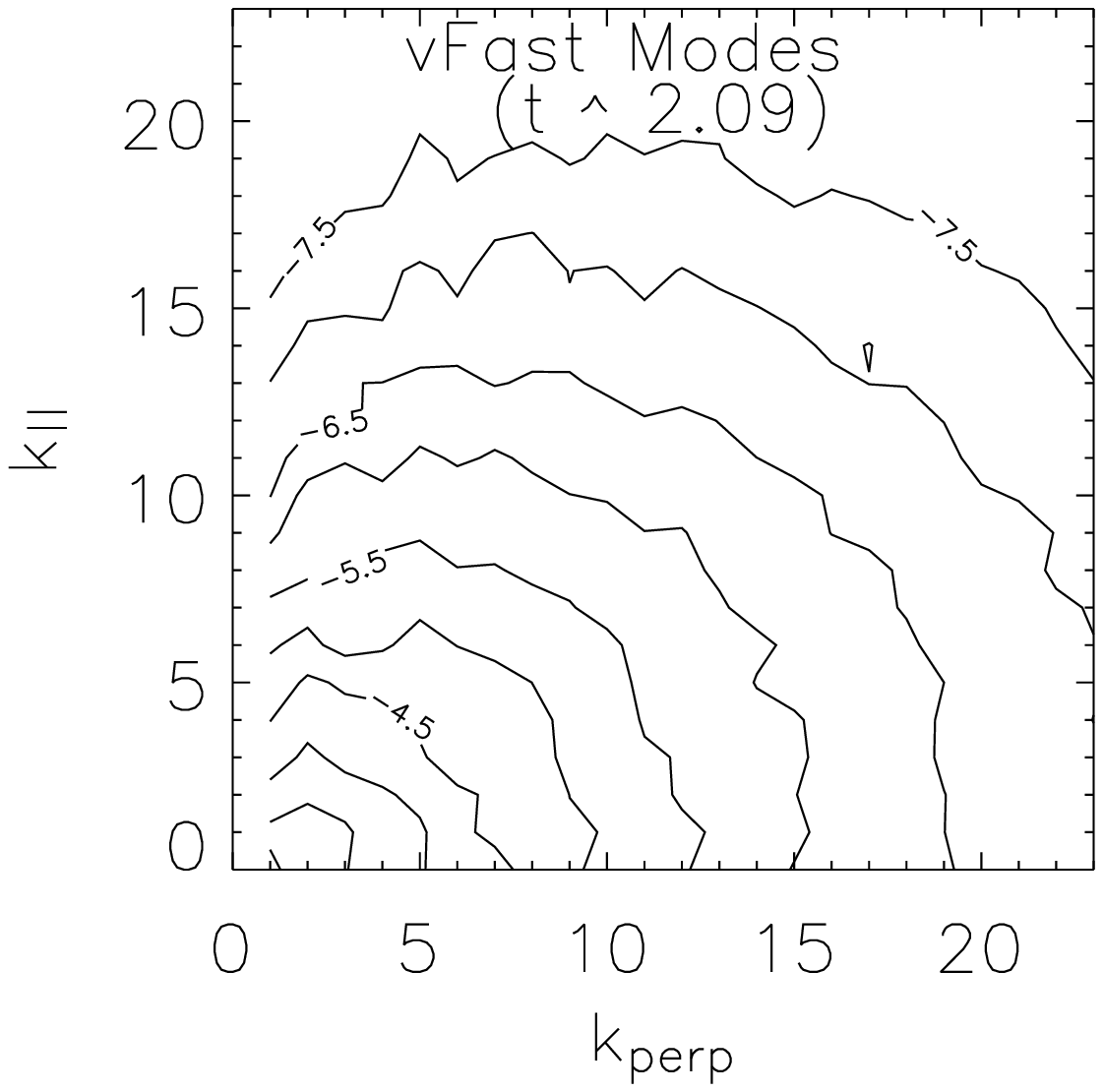}
\\
~~~~~~~~~~~~~~~~~~~~~~~~~~~~~~ (e)\>  ~~~~~~~~~~~~~~~~~~~~~~~~~~~~~~~~~~(f)
\end{tabbing}
\caption{
      Generation of compressible MHD modes.
    (a) When Alfvenic turbulence decays, generation of slow and fast
        modes is marginal. $144^3$. 
        Initially, $\beta$ (ratio of gas to magnetic pressure, $P_g/P_{mag}$) 
          $=0.2$ and 
          $M_s$ (sonic Mach number) $\sim 1.6$.
    (b) Comparison of decay rates.
        Decay of Alfven modes is not much affected by other 
       (slow and fast) modes. We use $144^3$ grid points.
        Initially, $\beta=0.2$ and 
        $M_s\sim 1.6$ for the solid line and 
        $M_s\sim 2.3$ for the dotted line. 
    (c) Square of the r.m.s. velocity of the compressible modes.
        We use $144^3$ grid points. Only Alfven modes are allowed
        as the initial condition.
        ``Pluses'' are for low $\beta$ cases ($0.02 \leq \beta \leq 0.4$).
        ``Diamonds'' are for high  $\beta$ cases ($1 \leq \beta \leq 20$).
        Fast modes follow a similar scaling.
    (d) Square of the r.m.s. velocity of the slow modes.
        See (c) for explanation.
    (e) Generation of fast modes. Snapshot is taken at t=0.06 from
        a simulation (with $144^3$ grid points) 
        that started off with Alfven modes only.
        Initially, $\beta=0.2$ 
        and 
          $M_s$ (sonic Mach number) $\sim 1.6$.
     (f) Generation of fast modes. Snapshot is taken at t$\sim$2.
         See (e) for explanation.
}
\label{fig_coupling}
\end{figure*}

Fig.~\ref{fig_coupling}(a) shows that generation of slow and fast modes 
(the dotted line) from Alfven modes (the solid line) is marginal.
The result shown in the figure is for $M_s=1.6$ at $t=0$.
We repeated similar simulation for
different Mach numbers and plasma $\beta$'s and measured
the ratios of energy in compressible modes to that in Alfven modes.
The results 
are shown in Fig.~\ref{fig_coupling}(c) and (d).
Fig.~\ref{fig_coupling}(c) suggest that the generation of
compressible motions follows equation (\ref{eq_high2}).
Fast modes also follow a similar scaling, although the scatter is a bit larger.
Fig.~\ref{fig_coupling}(e) demonstrates that fast modes are initially generated
anisotropically, which supports our theoretical consideration 
above.
Fast modes becomes isotropic later (Fig.~\ref{fig_coupling}(f)).
Fig.~\ref{fig_coupling}(d) shows that generation of slow modes follows
$(\delta V)_s^2/(\delta V)_A^2 \propto (\delta V)_A/V_A$ for low $\beta$ cases
({\it pluses} in the figure).
But, the scaling is not clear for high $\beta$ cases
({\it diamonds} in the figure).

Fig.~\ref{fig_coupling}(b) shows that dynamics of Alfven modes is 
not affected by slow modes.
The solid line in the figure is the energy in Alfven modes
when we start the decay simulation with Alfven modes only.
The dotted line is the Alfven energy when we start the simulation with
all modes. 
This result confirms that Alfven modes cascade is 
almost independent of slow and fast modes.
In this sense, coupling between Alfven and other modes is weak.

\section{Quest for Scaling Relations}    \label{sect_scaling}

\subsection{Scaling of incompressible MHD turbulence}

As mentioned in \S1, 
Goldreich \& Sridhar (GS95) made a
prediction regarding relative motions parallel and
perpendicular to magnetic field {\bf B} for incompressible
MHD turbulence. 
Here, we reconstruct GS95 model from different perspectives.

An important observation that leads to understanding of the GS95
scaling is that magnetic field cannot prevent mixing motions
of magnetic field lines if the motions
are perpendicular to magnetic field. Those motions will cause, however,
waves that will propagate along magnetic field lines.
If that is the case, 
the time scale of wave-like motion, i.e. $\sim l_{\|}/V_A$, where
$l_{\|}$ is the characteristic size of the perturbation and $V_A$ is 
the local Alfven speed, will be equal to the hydrodynamic time-scale, 
$l_{\perp}/v_l$. The mixing motions are 
hydrodynamic-like and therefore obey Kolmogorov scaling
$v_l\propto l_{\perp}^{1/3}$. Equating the two relations above, we
obtain a {\it critically balance} condition
\begin{equation}
   l_{\|}/V_A \sim l_{\perp}/v_k \mbox{ ~(or $k_{\|}V_A \sim k_{\perp}v_k$).}
\end{equation}
If conservation of energy in the turbulent cascade applies
locally in phase space then 
the energy cascade rate  
($ v_l^2/t_{cas}$) is constant):
$
    (v_l^2)/(l_{\perp}/v_l) = \mbox{constant.}
$
Combining this with the critical balance condition
we obtain 
\begin{equation}
    l_{\|}\propto l_{\perp}^{2/3}    \label{eq_ani}
\end{equation}
(or $k_{\|}\propto k_{\perp}^{2/3}$ in terms of wavevectors) and
and a Kolmogorov-like spectrum for perpendicular motions
\begin{equation}
v_l \propto l_{\perp}^{1/3}, \mbox{~~or,~} E(k) \propto k_{\perp}^{-5/3},
\end{equation}
which is not surprising since perpendicular motions are hydrodynamic.
If  we interpret $l_{\|}$ as the eddy size in the direction of the 
local\footnote{The concept of {\it local} is crucial. The GS95
scalings are obtained only in the local frame of magnetic field,
as this is the frame where magnetic field are allowed to be
mixed without being opposed by magnetic tension.} field
and $l_{\perp}$ as that in the perpendicular direction,
the relation in equation (\ref{eq_ani}) 
implies that smaller eddies are more elongated
(see Fig.~\ref{fig_ani_alf} for illustration of scale-dependent anisotropy).

Using Matthaeus et al. \cite{MatOG98}     
result, we can re-derive
GS95 model. 
Matthaeus et al.~\cite{MatOG98} showed that the anisotropy of low frequency
MHD turbulence scales linearly with the ratio of
perturbed and total magnetic field strength $b/B$ 
($=b/(b^2+B_0^2)^{1/2}$).
This scaling relation has simple geometric meaning:
perpendicular size of a large scale eddy is similar to its parallel size times
$b/B$, which is is determined by magnetic field line wandering.
Although
their analysis was based on comparing the strength of a uniform
background field and the magnetic perturbations on all scales, 
we can reinterpret this result by assuming that the strength
of random magnetic field at a scale $l$ is $b_l$, and that the
background field is the sum of all contributions from larger scales.
Then Matthaeus et al.'s result becomes a prediction that the 
anisotropy ($k_{\parallel}/k_{\perp}$) is proportional to
($b_l/B$).
We can take the total magnetic field strength $B\sim$ constant as long
as the background field is stronger than the perturbations on all scales.
Since $b_l\sim (kE(k))^{1/2} \sim k_{\perp}^{-1/3}$, we obtain
an anisotropy ($k_{\parallel}/k_{\perp}$) proportional to $k_{\perp}^{-1/3}$,
and $k_{\parallel}\propto k_{\perp}^{2/3}$.
In this interpretation, smaller eddies are more elongated because
they have a smaller $b_l/B$ ratio.

\subsection{Compressible scalings: theoretical considerations}
In \S\ref{sect_result}, we showed that Alfven modes are independent of
other modes.
Therefore, we expect GS95 scalings for Alfven modes even for
supersonic turbulence.

When Alfven cascade evolves on its own, it is natural to assume that 
slow modes passively follow the Alfven cascade and exhibit the same scaling.
Indeed, slow modes in high $\beta$ plasmas are
similar to the pseudo-Alfven modes in incompressible regime 
(see GS95; \cite{LitG01}).  
The latter modes do follow
the GS95 scaling. In low $\beta$ plasmas, 
motions of slow modes are density perturbations propagating with the
sound velocity $a$ parallel to the mean magnetic field
(see equation (\ref{xis_lowbeta})). In magnetically dominated
environments ($\beta \ll 1$), 
$a\ll V_A$ and the gaseous perturbations are essentially static.
Therefore the magnetic field mixing motions are expected to mix density
perturbations as if they were passive scalar. It is known that the 
passive scalar shows the same scaling as the velocity field of the inducing
turbulent motions. Thus the slow waves are expected to demonstrate 
GS95 scalings (see CL02).

The fast waves in low $\beta$ regime propagate at $V_A$ irrespectively
of the magnetic field direction. 
In high $\beta$ regime, the properties of fast modes are similar, 
but the propagation speed is the sound speed $a$.
Thus the mixing motions induced by Alfven waves should affect the fast wave
cascade only marginally. The latter cascade is expected to
be analogous to the acoustic wave cascade and be isotropic.

\section{Velocity scaling}             \label{sect_c6}

\subsection{Illustration of eddy structures}
\begin{figure*}
\begin{center}
\includegraphics[width=.70\textwidth]{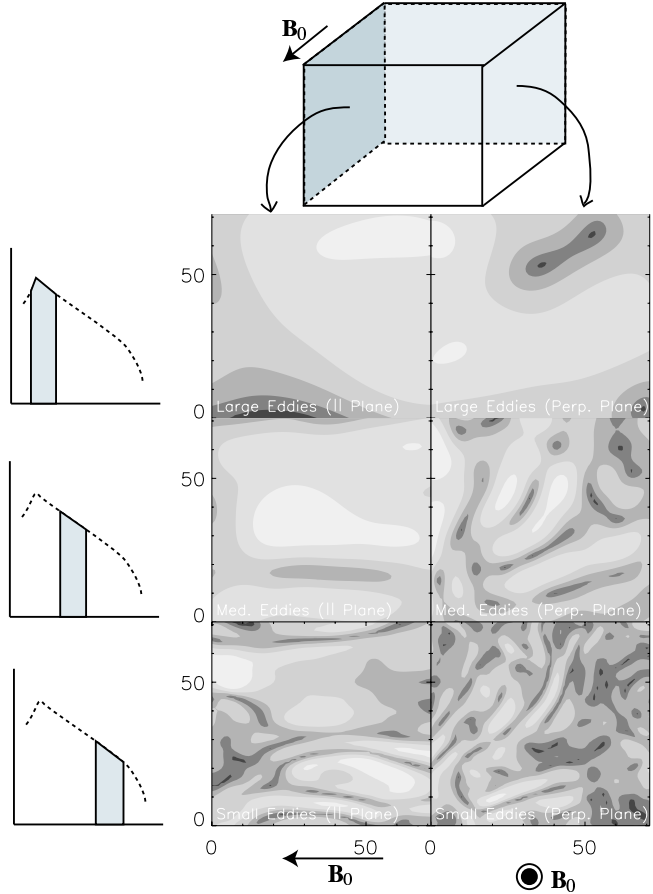}
\caption{
        Anisotropy as a function of scale.
     Alfven mode velocity show scale-dependent anisotropy.
     Lighter tones are for larger $|{\bf v}|$.
     Only part of the data cube is shown.
    Magnetic field show similar behaviors.
   {\it Large scale eddies} are obtained from the Fourier components with
      $1\leq k<3$. 
      {\it Medium scale eddies} are obtained from the Fourier components with
      $3\leq k<9$. 
      {\it Small scale eddies} are obtained from the Fourier components with
      $9\leq k<27$. 
}
\label{fig_ani_alf}
\end{center}
\end{figure*}

\begin{figure*}
\begin{center}
\includegraphics[width=.70\textwidth]{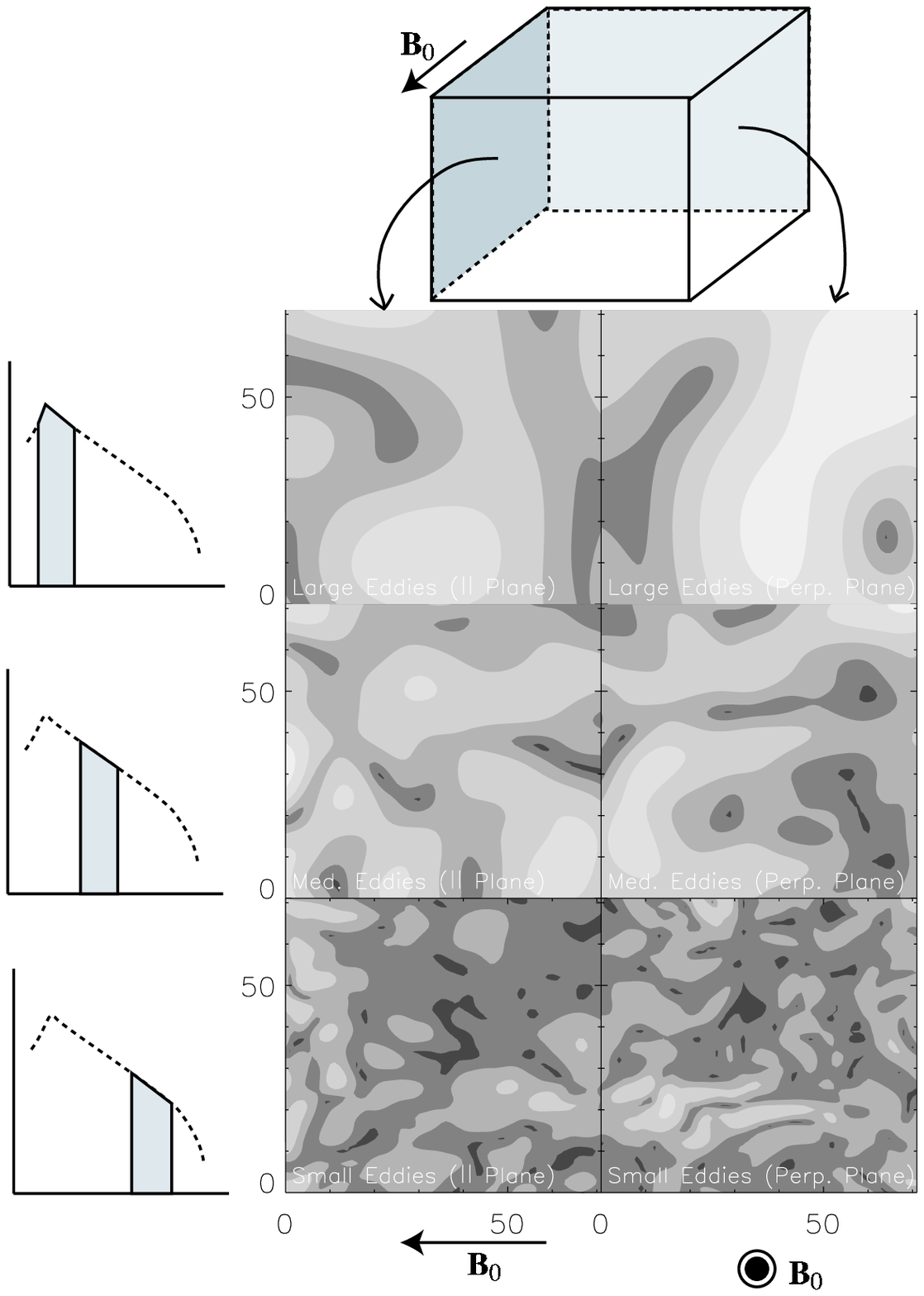}
\caption{
     Fast mode velocity show isotropy. Only part of the data cube
     is shown. Magnetic field show similar behaviors.
}
\label{fig_ani_fast}
\end{center}
\end{figure*}

{}Fig.~\ref{fig_ani_alf} and Fig.~\ref{fig_ani_fast} show 
the shapes of eddies of different sizes.
For Alfven mode eddies (Fig.~\ref{fig_ani_alf}),
left 3 panels show an increased anisotropy 
as we move from the top (large eddies) to the bottom (small eddies).
The horizontal axes of the left panels are parallel to ${\bf B_0}$.
Structures in the perpendicular plane (right panels)
do not show a systematic 
elongation. 
However, Fig.~\ref{fig_ani_fast} shows that
velocity of fast modes exhibit isotropy.
Data are from a simulation with $216^3$ grid points, $M_s=2.3$, and $\beta=0.2$.

\subsection{Alfven modes in compressible MHD}

If Alfven cascade evolves on its own, it is natural to assume that 
slow modes exhibit the GS95 scaling.
Indeed, slow modes in pressure dominated environment (high $\beta$ plasmas) are
similar to the pseudo-Alfven modes in incompressible regime 
(see GS95; \cite{LitG01}).   
The latter modes do follow
the GS95 scaling. In magnetically dominated environments (low $\beta$ plasmas), 
slow modes are density perturbations propagating with the
sound velocity $a$ parallel to the mean magnetic field
(see equation (\ref{xis_lowbeta})). Those perturbations are essentially
static for $a\ll V_A$. 
Therefore Alfvenic turbulence is expected to mix density
perturbations as if they were passive scalar. This also induces
GS95 spectrum.

\begin{figure*}[!t]
\begin{tabbing}
\includegraphics[width=.50\textwidth]{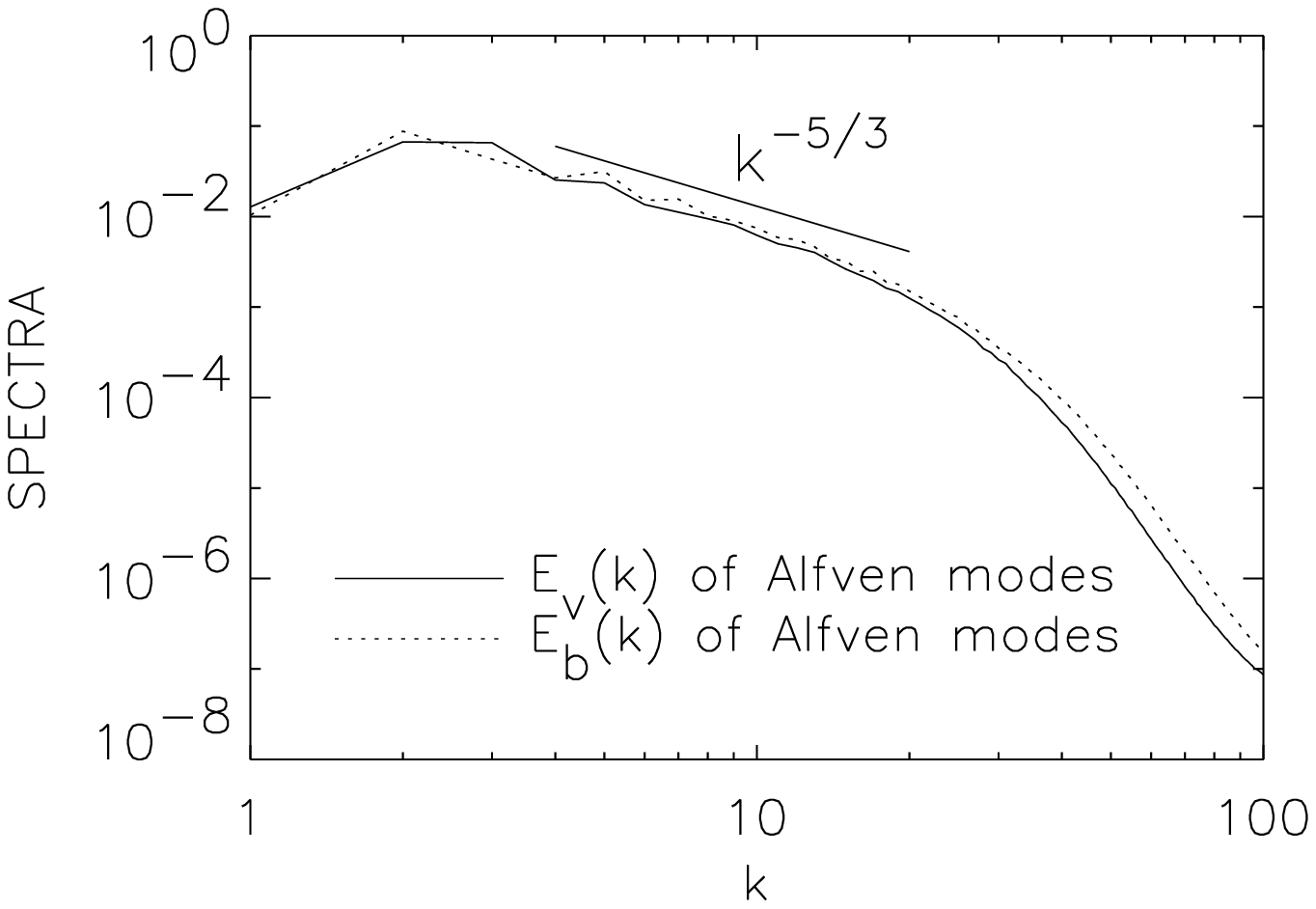}
\=
\includegraphics[width=.45\textwidth]{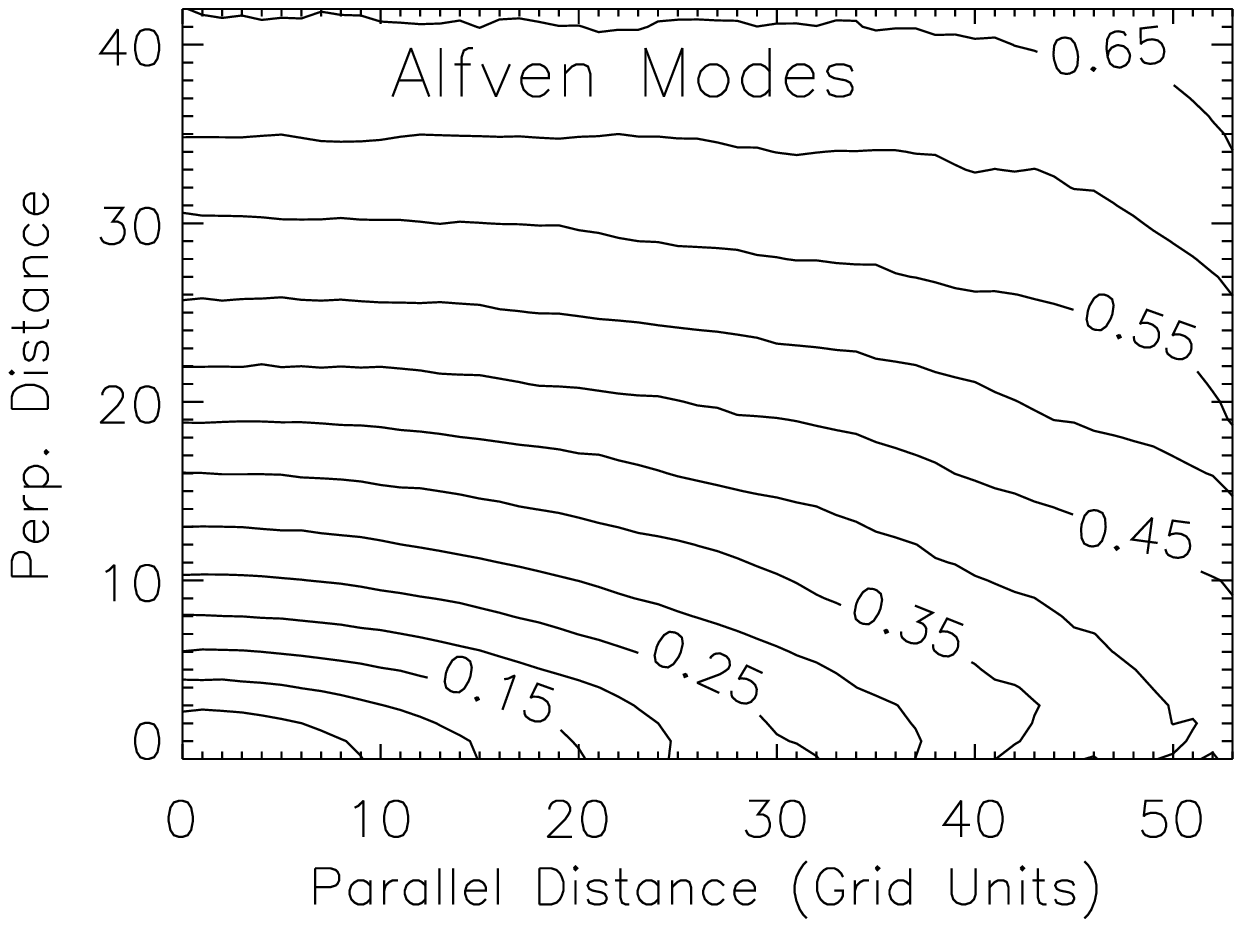}
\\
~~~~~~~~~~~~~~~~~~~~~~~~~~~~~~~(a)    \> ~~~~~~~~~~~~~~~~~~~~~~~~~(b) \\
\includegraphics[width=.50\textwidth]{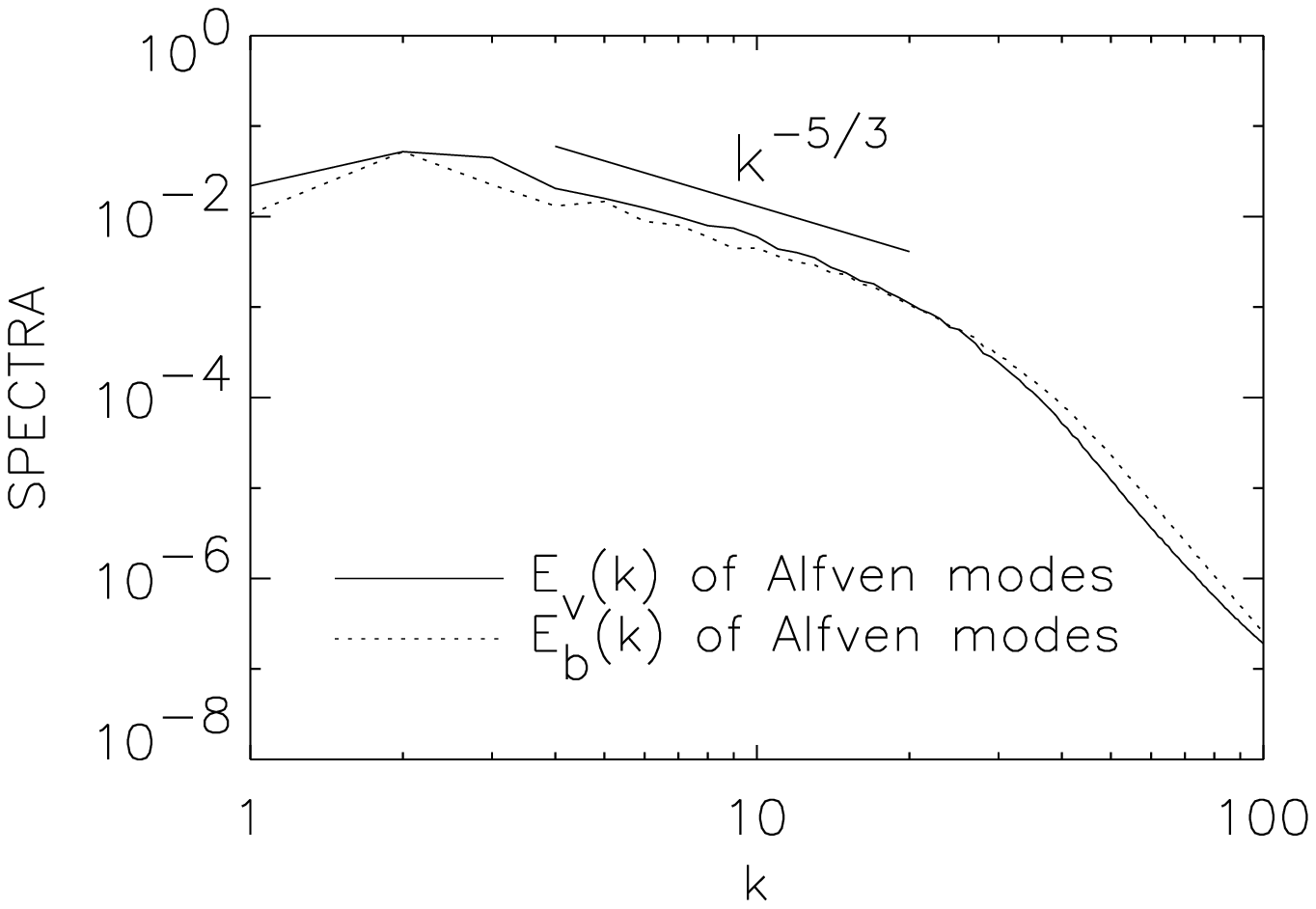}
\>
\includegraphics[width=.45\textwidth]{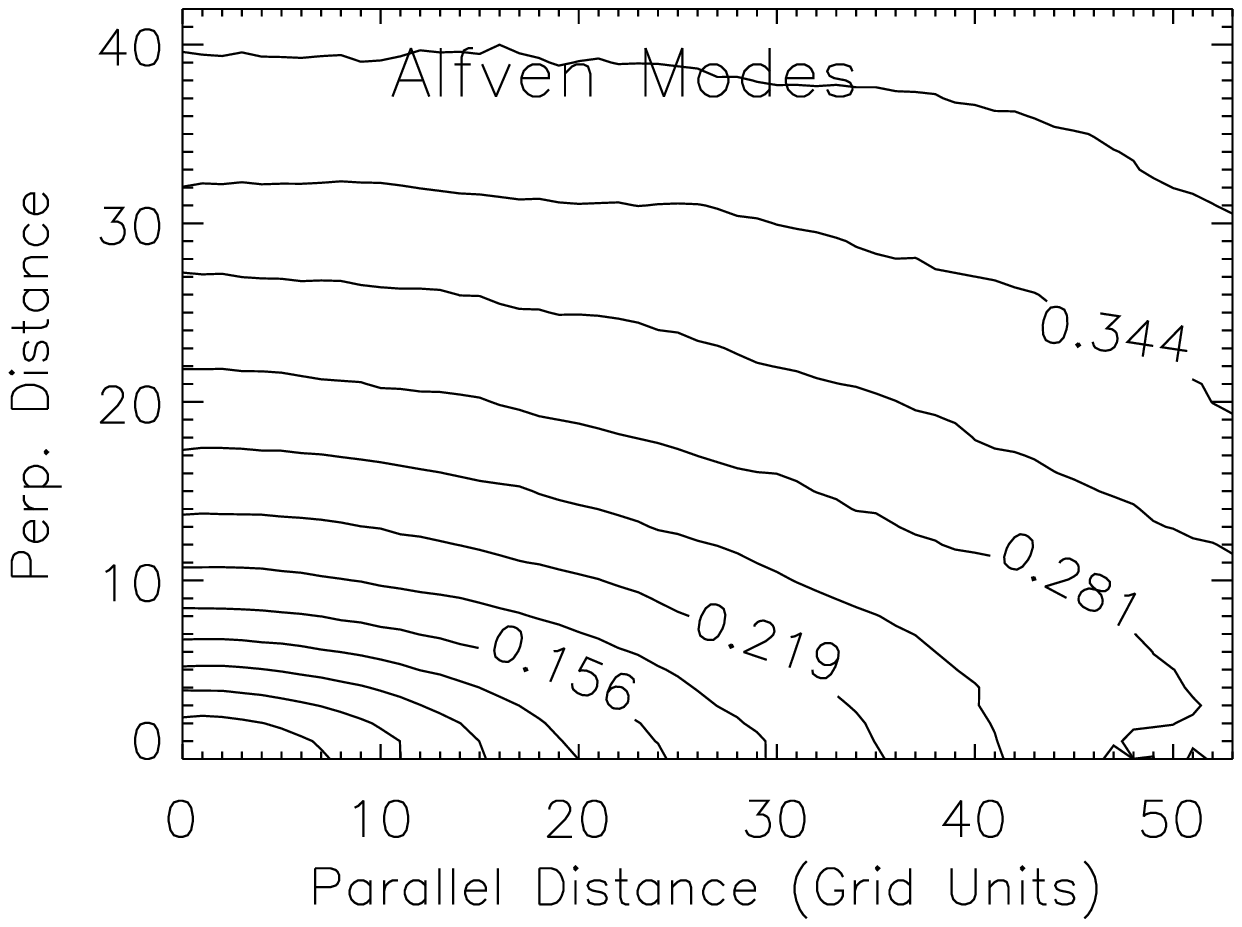}
\\
~~~~~~~~~~~~~~~~~~~~~~~~~~~~~~~(c)    \> ~~~~~~~~~~~~~~~~~~~~~~~~~(d) \\
\includegraphics[width=.50\textwidth]{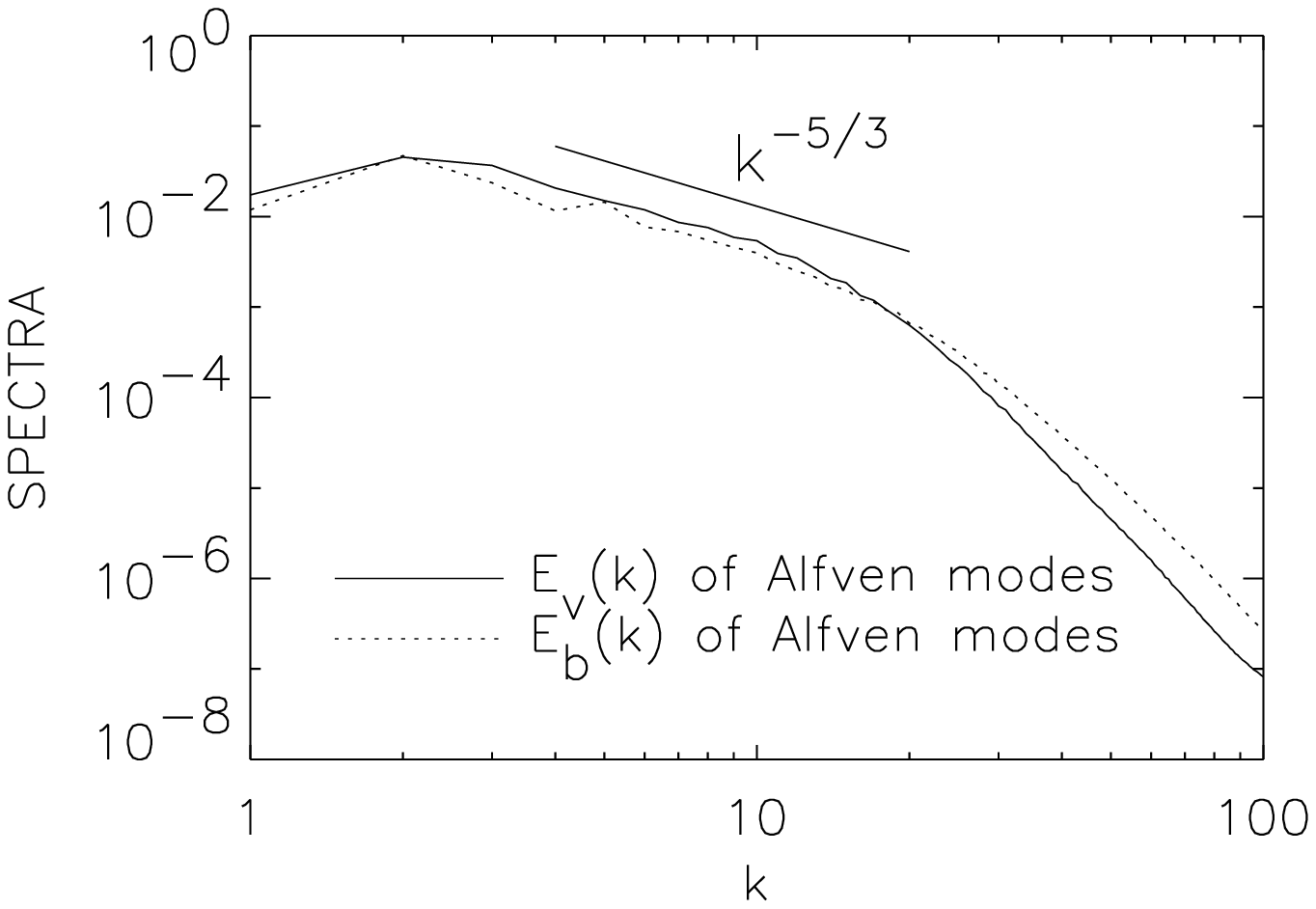}
\>
\includegraphics[width=.45\textwidth]{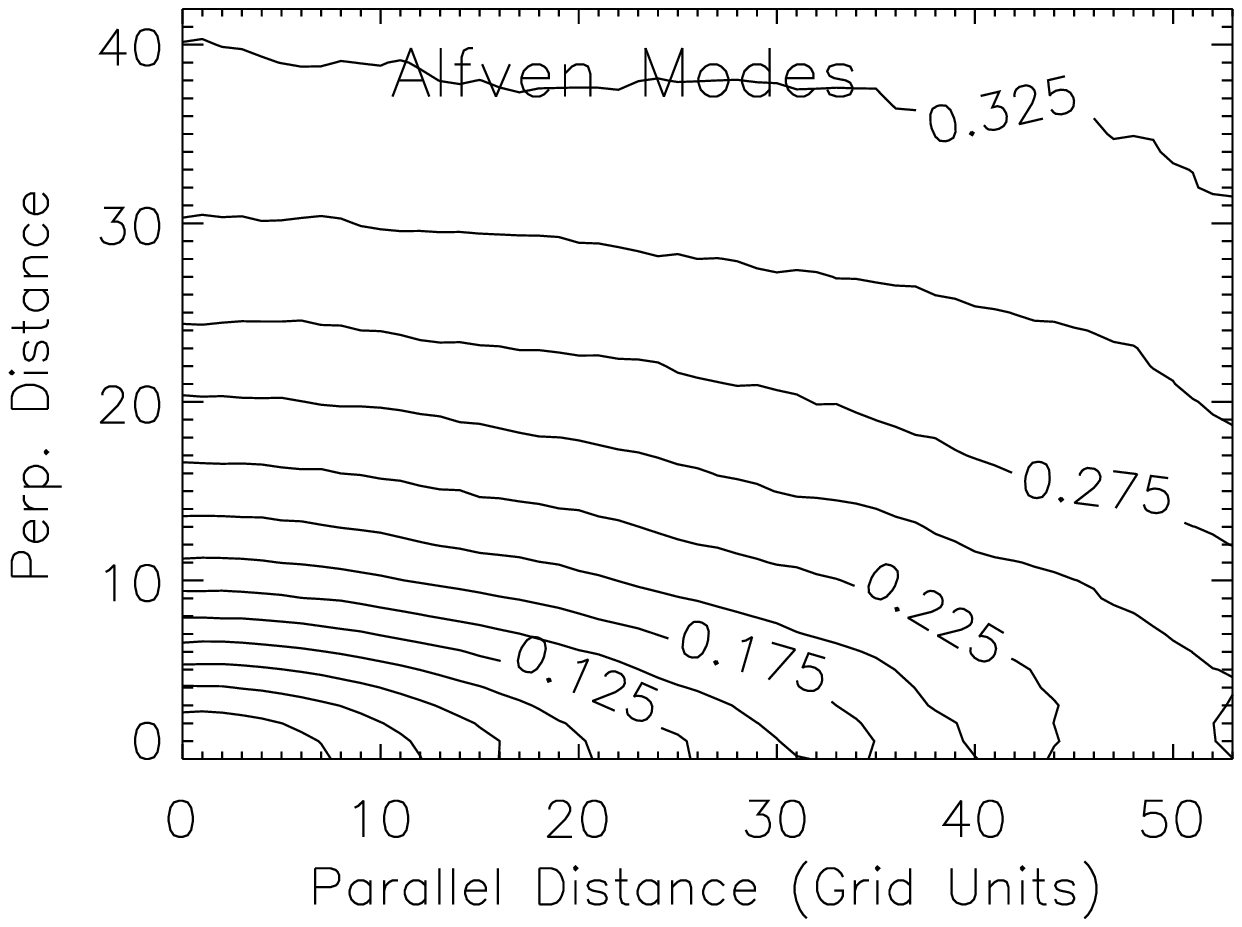}
\\
~~~~~~~~~~~~~~~~~~~~~~~~~~~~~~~(e)    \> ~~~~~~~~~~~~~~~~~~~~~~~~~(f) 
\end{tabbing}
\caption{
         Alfven modes. (a)\&(b): $M_s\sim 0.35$ ($\beta=4$).
         (c)\&(d): $M_s\sim 2.3$ ($\beta=0.2$).
         (e)\&(f): $M_s\sim 7$ ($\beta=0.02$).
         Spectra are compatible with Kolmogorov.
         Contours (or, eddy shapes) show scale-dependent anisotropy:
         smaller eddies are more elongated.
         Overall, the scalings are consistent with Goldreich \&
         Sridhar scalings.
}
\label{fig_result_alf}
\end{figure*}

{}Fig.~\ref{fig_result_alf}(a), (c), and (e) 
show that the spectra of Alfv\'{e}n waves follow
a Kolmogorov spectrum:
\begin{equation}
 \mbox{\it Spectrum of Alfv\'{e}n Waves:~~~}  
     E^{A}(k) \propto k_{\perp}^{-5/3},
\end{equation}
regardless of plasma $\beta$ or sonic Mach number $M_s$.

In Fig.~\ref{fig_result_alf}(b), (d), and (f),
we plot contours of equal second-order structure function for velocity
($SF_2({\bf r})=<|{\bf v}({\bf x}+{\bf r}) - 
                 {\bf v}({\bf x})|^2>_{avg.~over~{\bf x}}$)
obtained in local coordinate systems in which the parallel axis is aligned
with the local mean field 
(see \cite{ChoV00b};\cite{MarG01}; CLV02b).
The $SF_2$
 along the axis perpendicular to the local mean magnetic field
follows a scaling compatible with $r^{2/3}$.
The $SF_2$ along the axis parallel  to the local mean field follows
steeper $r^{1}$ scaling. 
The results are compatible with the GS95 model,
\begin{equation}
  \mbox{\it Anisotropy of Alfv\'{e}n Waves:~~~}  
  r_{\|}\propto r_{\perp}^{2/3}, \mbox{~or~} k_{\|}\propto k_{\perp}^{2/3},
\end{equation}
where
$r_{\|}$ and $r_{\perp}$ are the semi-major axis and semi-minor axis of eddies,
respectively \cite{ChoV00b}.   
When we interpret that contours represent eddy shapes, the above scaling 
means that smaller contours are more elongated.

\subsection{Slow modes in compressible MHD}
\begin{figure*}[!t]
\begin{tabbing}
\includegraphics[width=.50\textwidth]{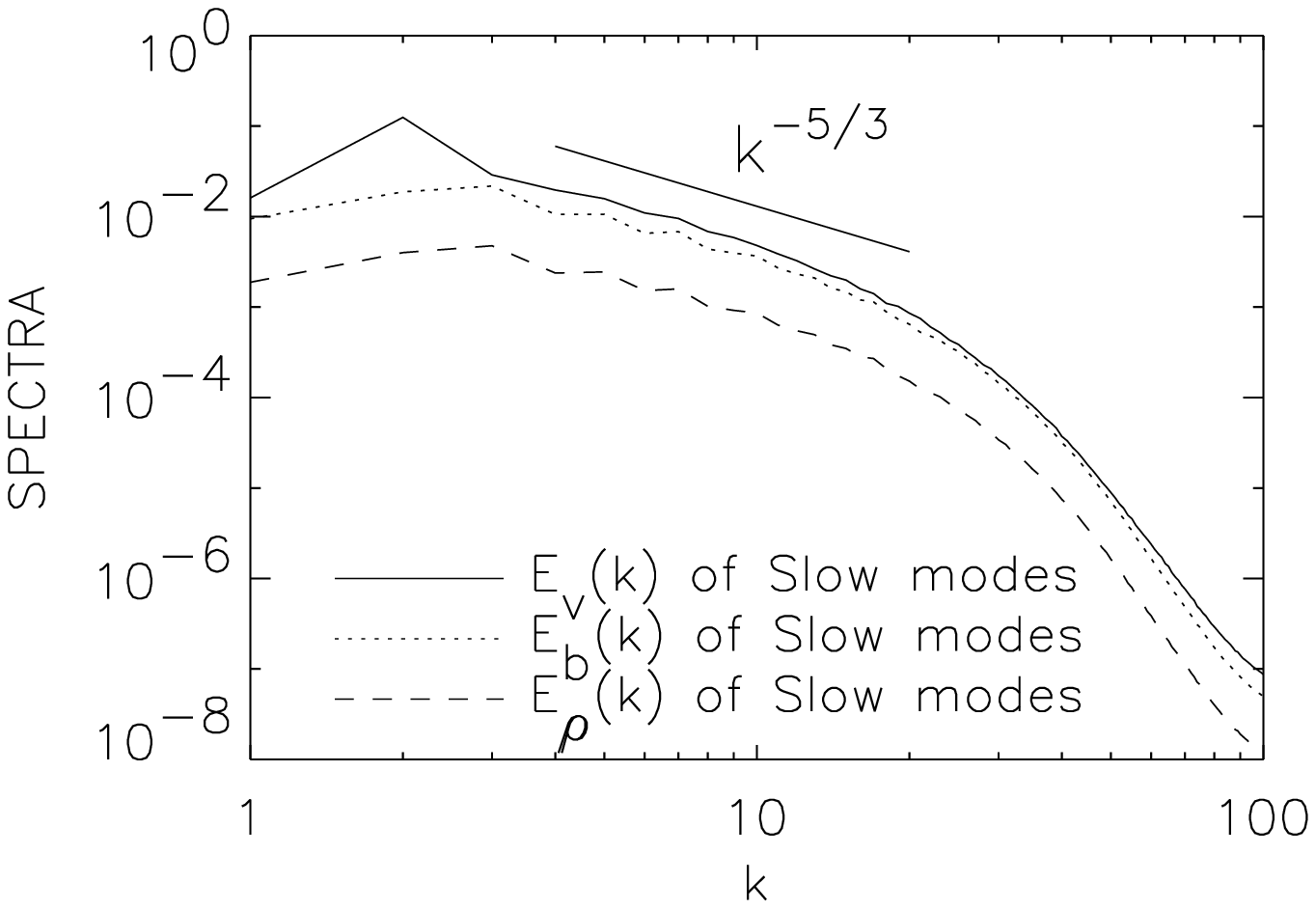}
\=
\includegraphics[width=.45\textwidth]{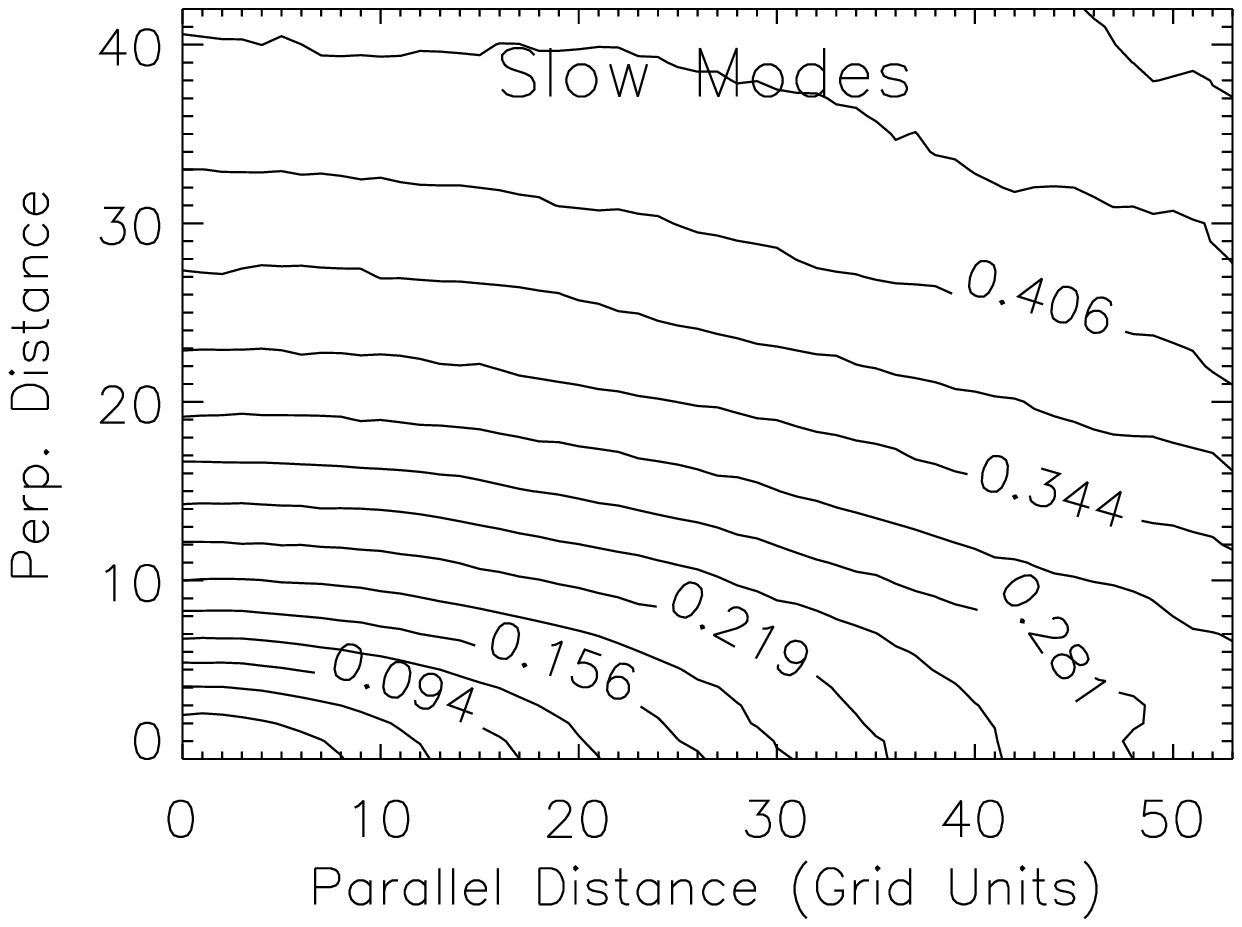}
\\
~~~~~~~~~~~~~~~~~~~~~~~~~~~~~~~(a)    \> ~~~~~~~~~~~~~~~~~~~~~~~~~(b) \\
\includegraphics[width=.50\textwidth]{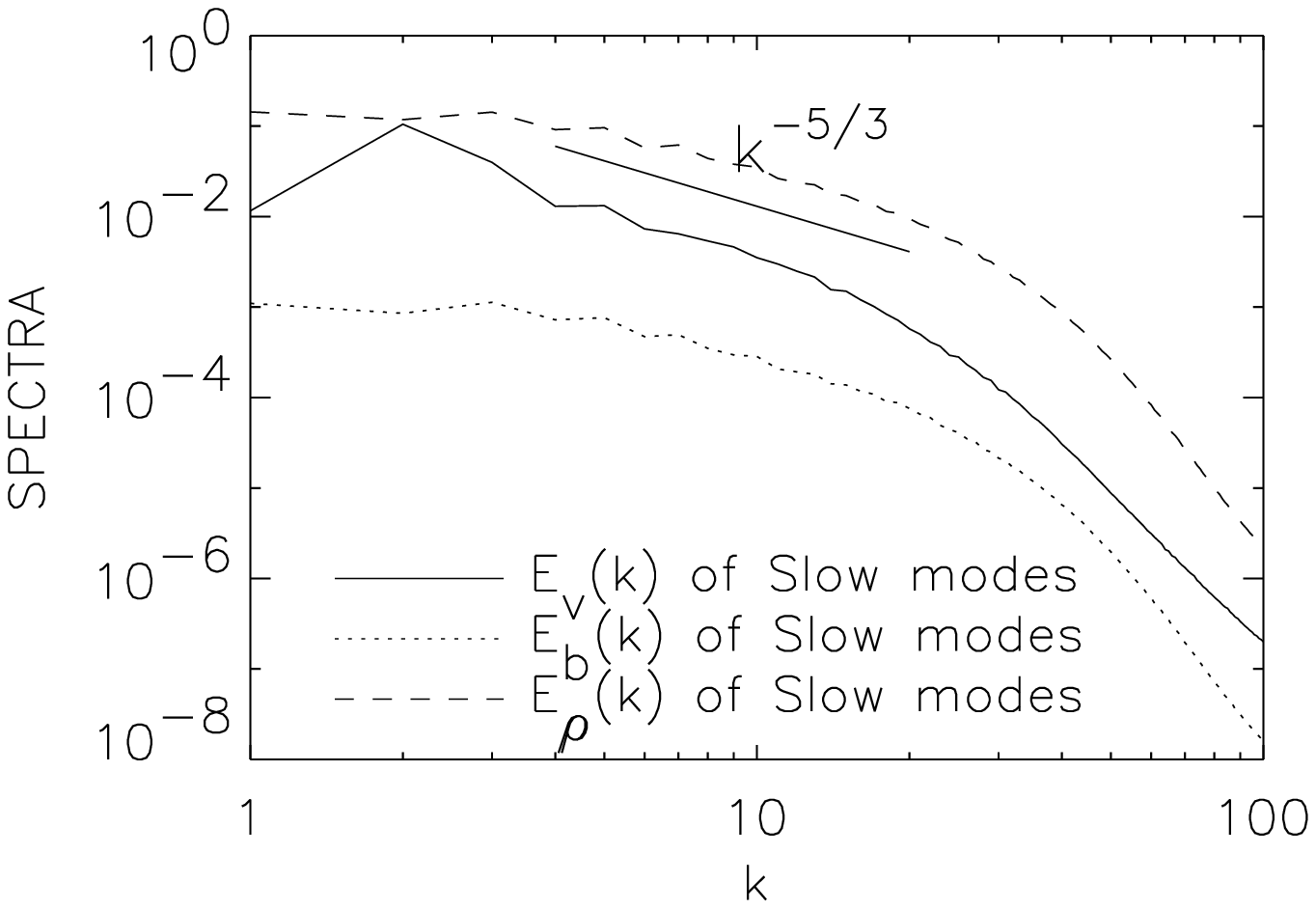}
\>
\includegraphics[width=.45\textwidth]{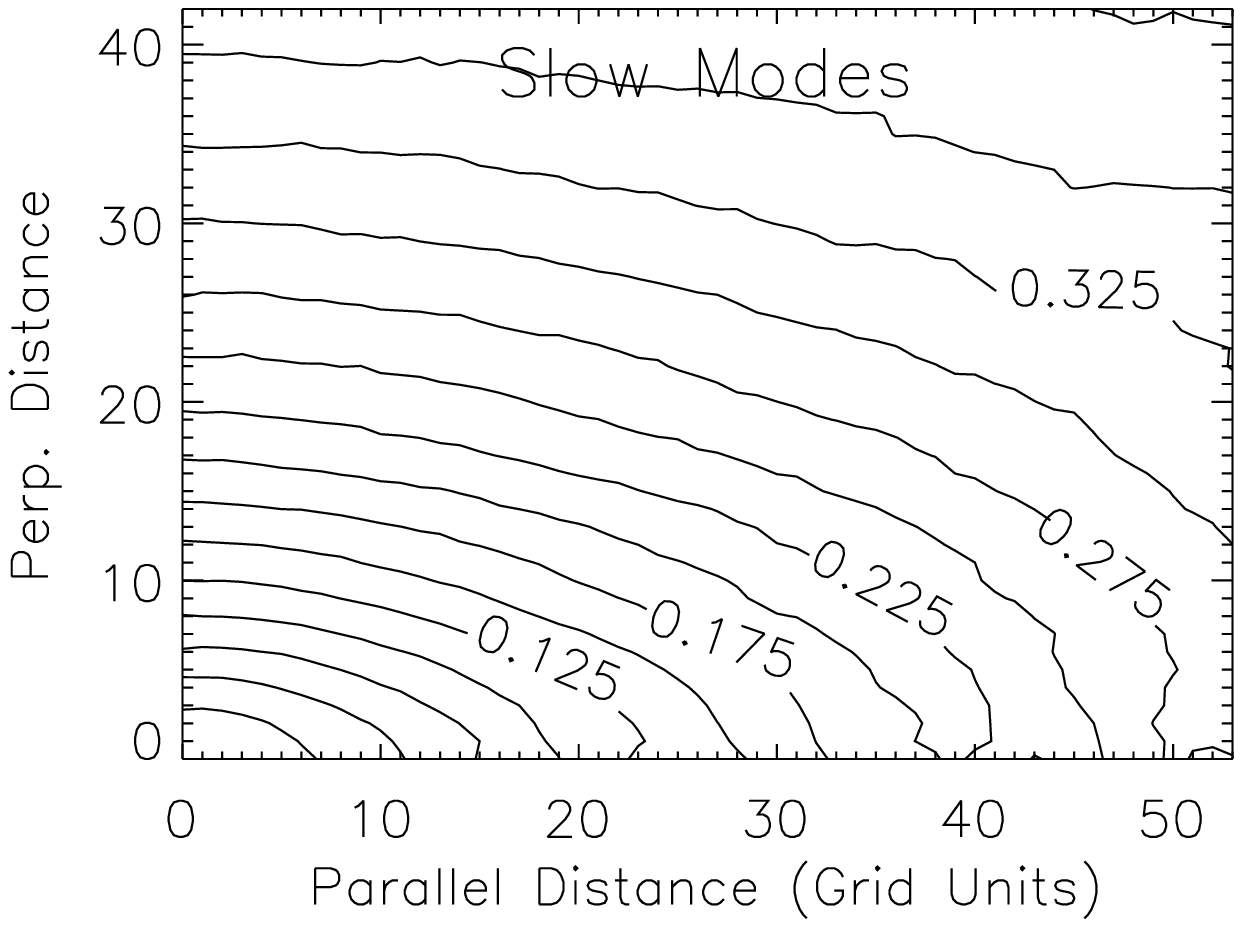}
\\
~~~~~~~~~~~~~~~~~~~~~~~~~~~~~~~(c)    \> ~~~~~~~~~~~~~~~~~~~~~~~~~(d) \\
\includegraphics[width=.50\textwidth]{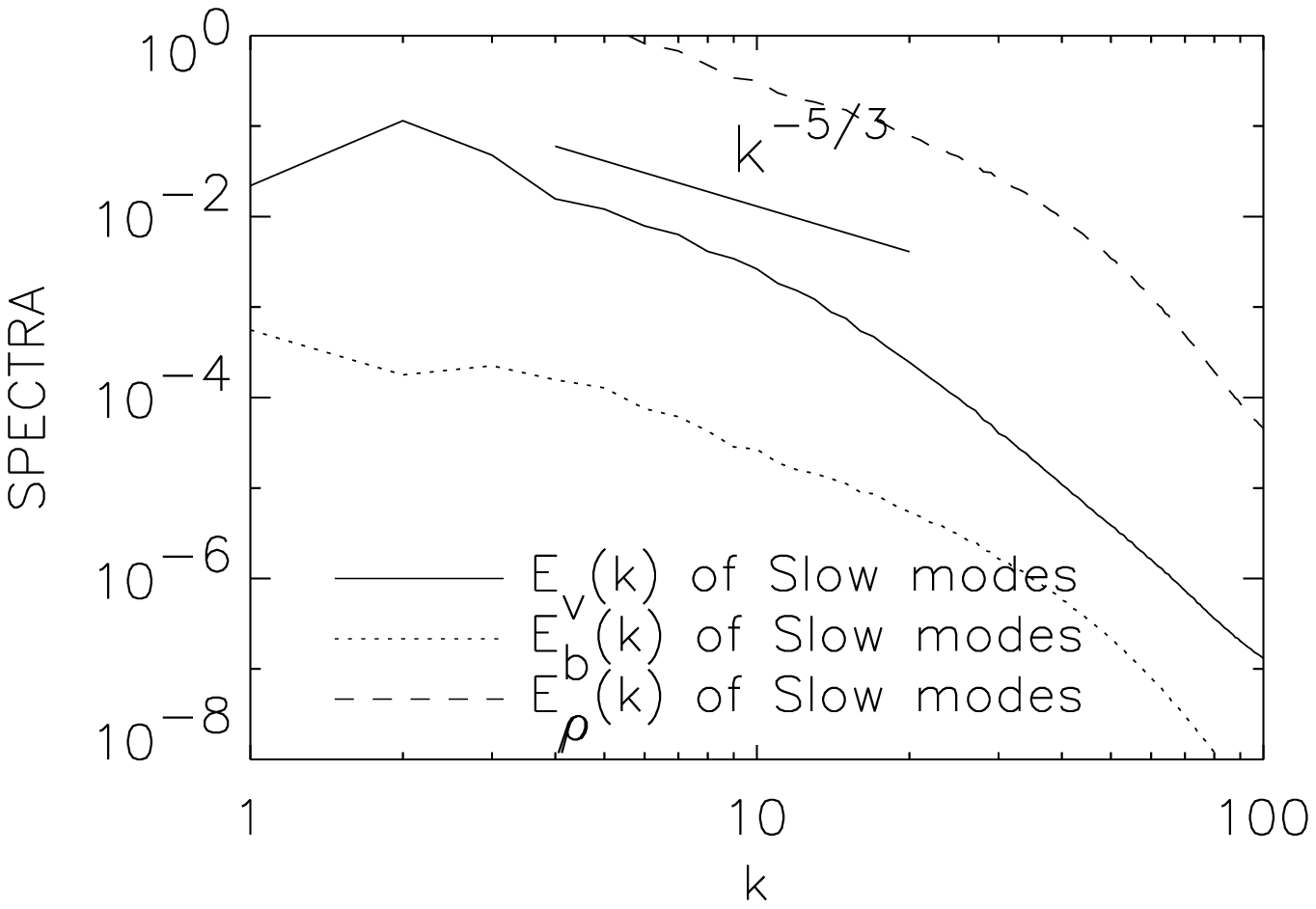}
\>
\includegraphics[width=.45\textwidth]{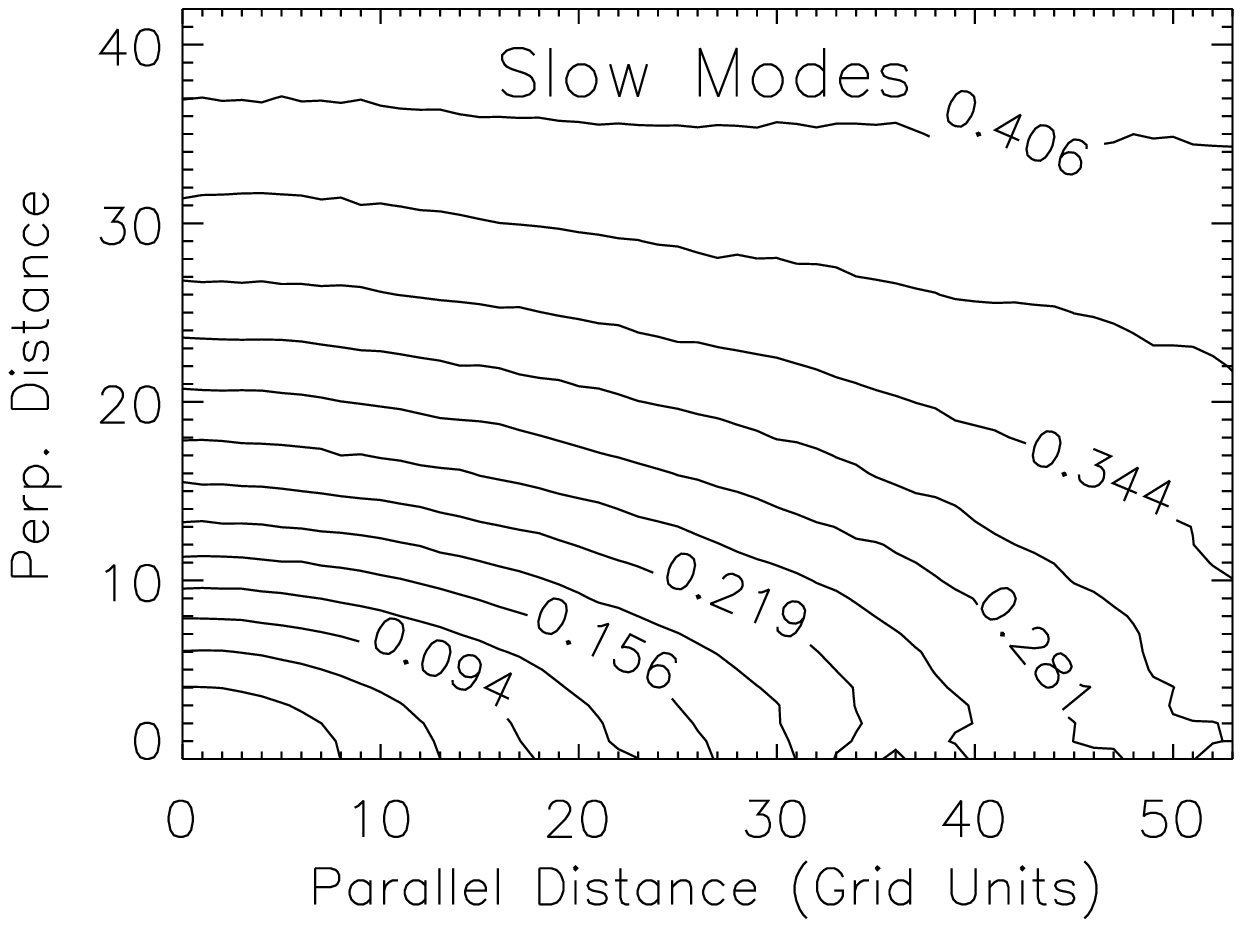}
\\
~~~~~~~~~~~~~~~~~~~~~~~~~~~~~~~(e)    \> ~~~~~~~~~~~~~~~~~~~~~~~~~(f) 
\end{tabbing}
\caption{
         Slow modes. (a)\&(b): $M_s\sim 0.35$ ($\beta=4$).
         (c)\&(d): $M_s\sim 2.3$ ($\beta=0.2$).
         (e)\&(f): $M_s\sim 7$ ($\beta=0.02$).
         In (a) and (c), spectra are compatible with Kolmogorov.
         In (e), spectra is uncertain.
         Contours (or, eddy shapes) show scale-dependent anisotropy:
         smaller eddies are more elongated.
         Overall, the scalings are consistent with Goldreich \&
         Sridhar scalings (except spectra in (e)).
}
\label{fig_result_slow}
\end{figure*}

The incompressible limit of slow waves is pseudo-Alfv\'{e}n waves.
Goldreich \& Sridhar \cite{GolS97} 
argued that
the pseudo-Alfv\'{e}n waves are slaved to the shear-Alfv\'{e}n 
(i.e.~ordinary Alfv\'{e}n)
waves, which
means that
pseudo-Alfv\'{e}n modes do not cascade energy for themselves. 
Lithwick \& Goldreich \cite{LitG01}   
made similar theoretical arguments 
for high $\beta$ plasmas and
conjectured similar behaviors of slow modes in low $\beta$ plasmas.
We confirmed that similar arguments are also applicable to slow waves
in low $\beta$ plasmas (CL02).
Indeed, energy spectra in 
Fig.~\ref{fig_result_slow}(a) and (c) are consistent with:
\begin{equation}
 \mbox{\it Spectrum of Slow Modes:~~~}  E^{s}(k) \propto k_{\perp}^{-5/3}.
\end{equation}
However, the kinetic energy spectrum for slow modes 
in Fig.~\ref{fig_result_slow}(e)
does not show
the Kolmogorov slope.
The slope is close to $-2$, which is suggestive of shock formation.
At this moment, it is not clear whether or not the $-2$ slope
is the true slope.
In other words, the observed $-2$ slope might be due to the limited
numerical resolution.
Runs with higher numerical resolution should
give the definite answer.

In Fig.~\ref{fig_result_slow}(b), (d), and (f),
contours of equal second-order 
velocity structure function ($SF_2$), 
representing eddy shapes,
show scale-dependent anisotropy: smaller eddies are more elongated.
The results are compatible with the GS95 model 
\begin{equation}
\mbox{\it Anisotropy of Slow Modes:~~~}  
  k_{\|}\propto k_{\perp}^{2/3}, \mbox{~or~} 
  r_{\|}\propto r_{\perp}^{2/3}, 
\end{equation}
where
$r_{\|}$ and $r_{\perp}$ are the semi-major axis and semi-minor axis of eddies,
respectively.

\subsection{Fast modes in compressible MHD}
\begin{figure*}[!t]
\begin{tabbing}
\includegraphics[width=.50\textwidth]{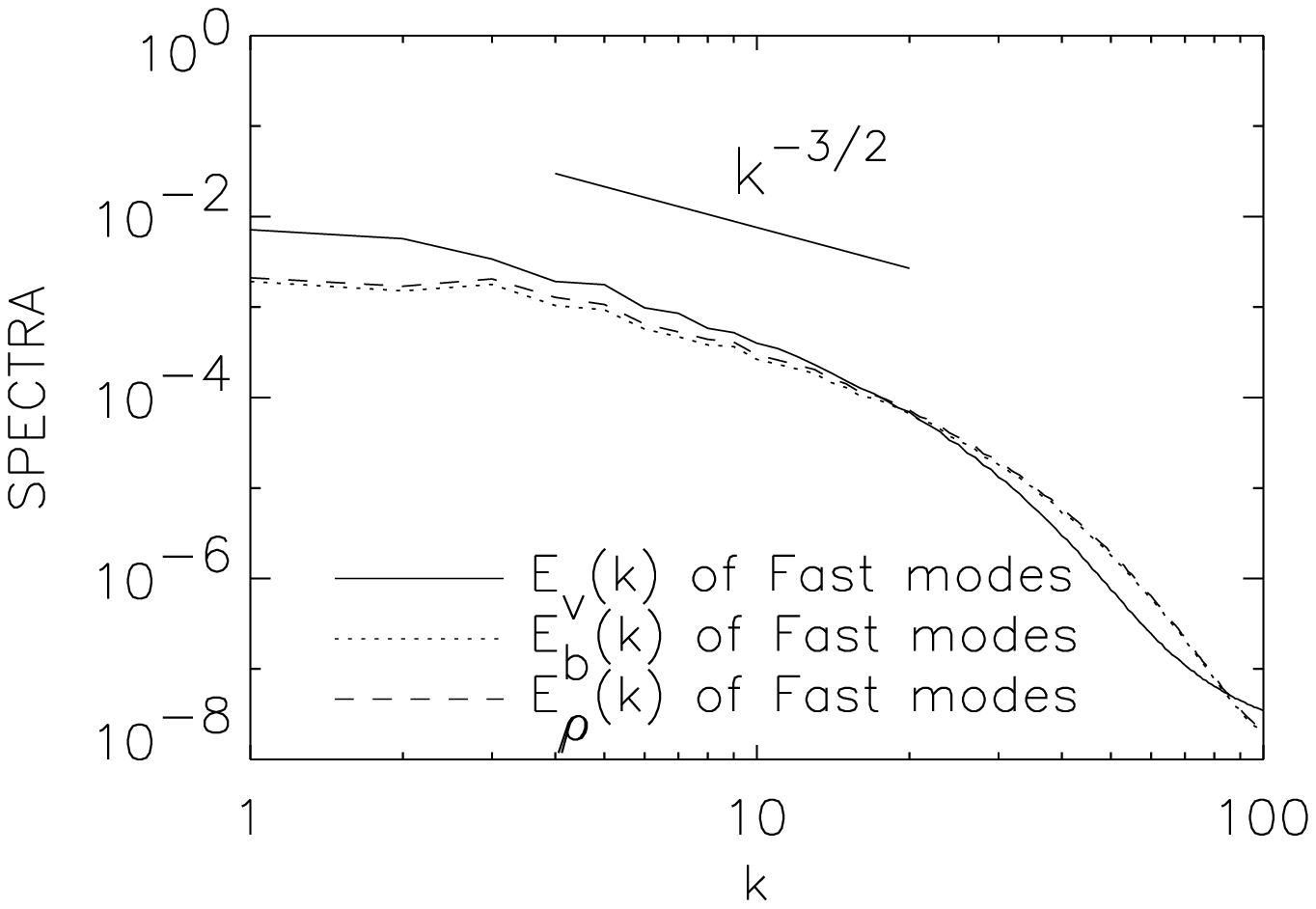}
\=
\includegraphics[width=.45\textwidth]{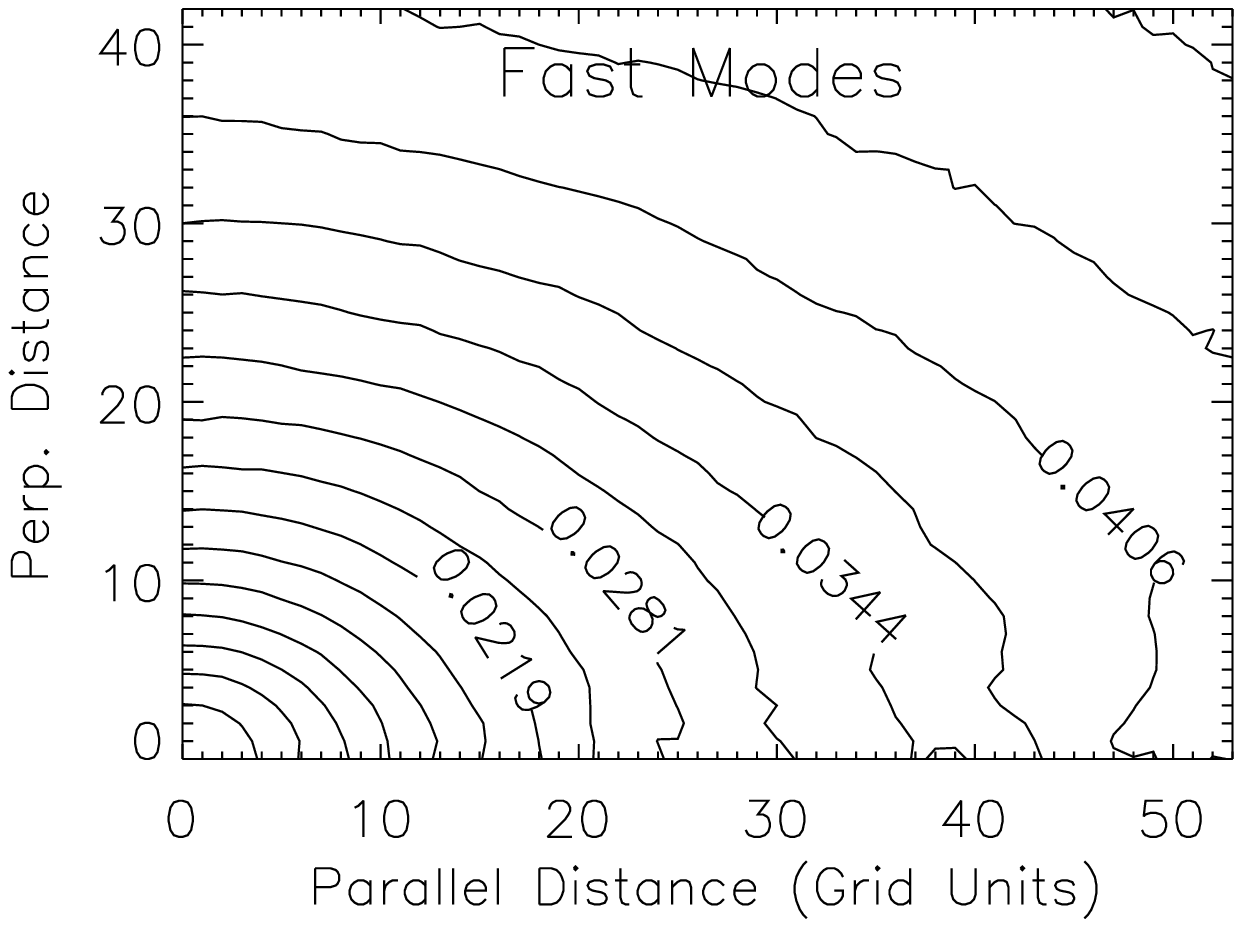}
\\
~~~~~~~~~~~~~~~~~~~~~~~~~~~~~~~(a)    \> ~~~~~~~~~~~~~~~~~~~~~~~~~(b) \\
\includegraphics[width=.50\textwidth]{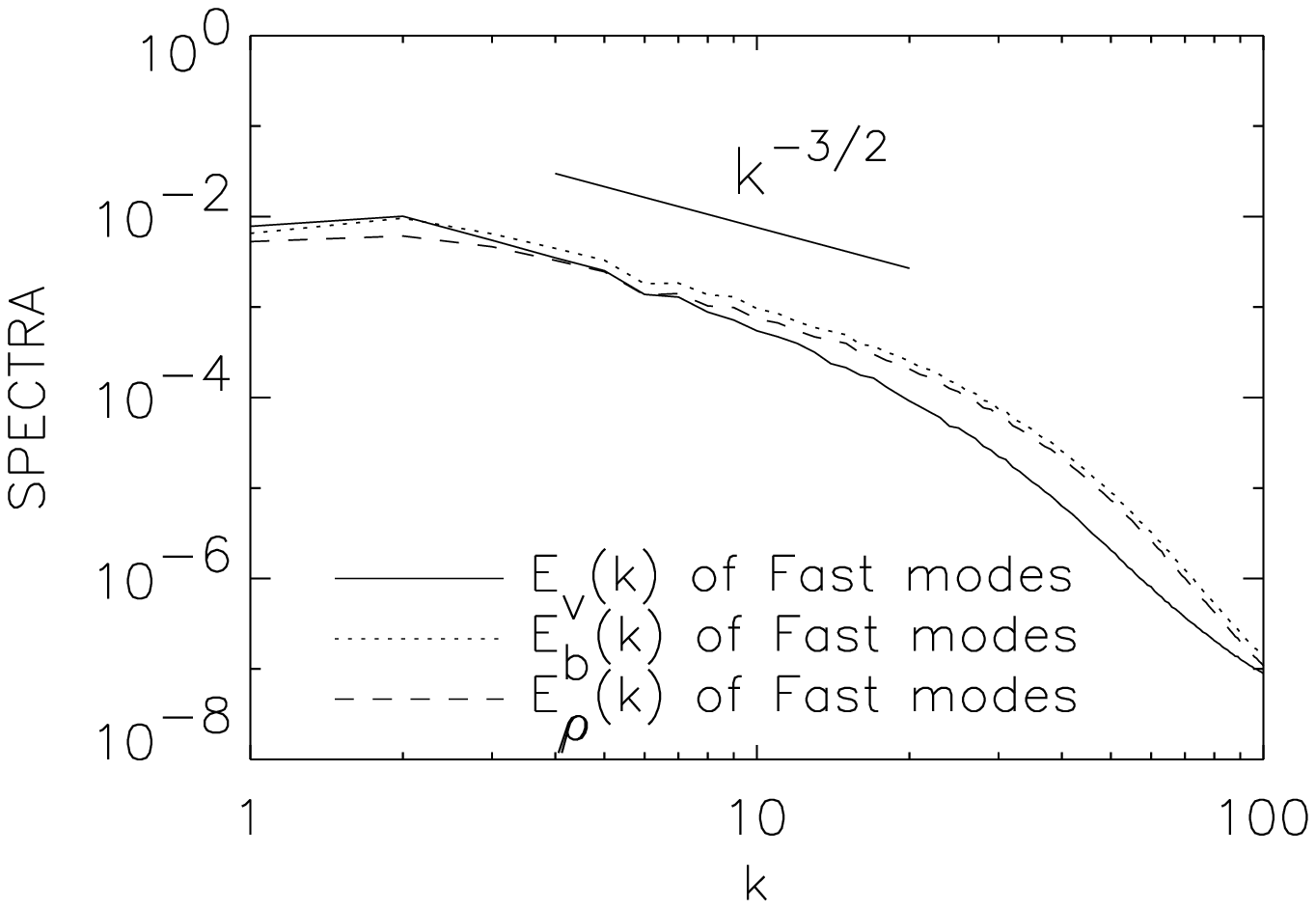}
\>
\includegraphics[width=.45\textwidth]{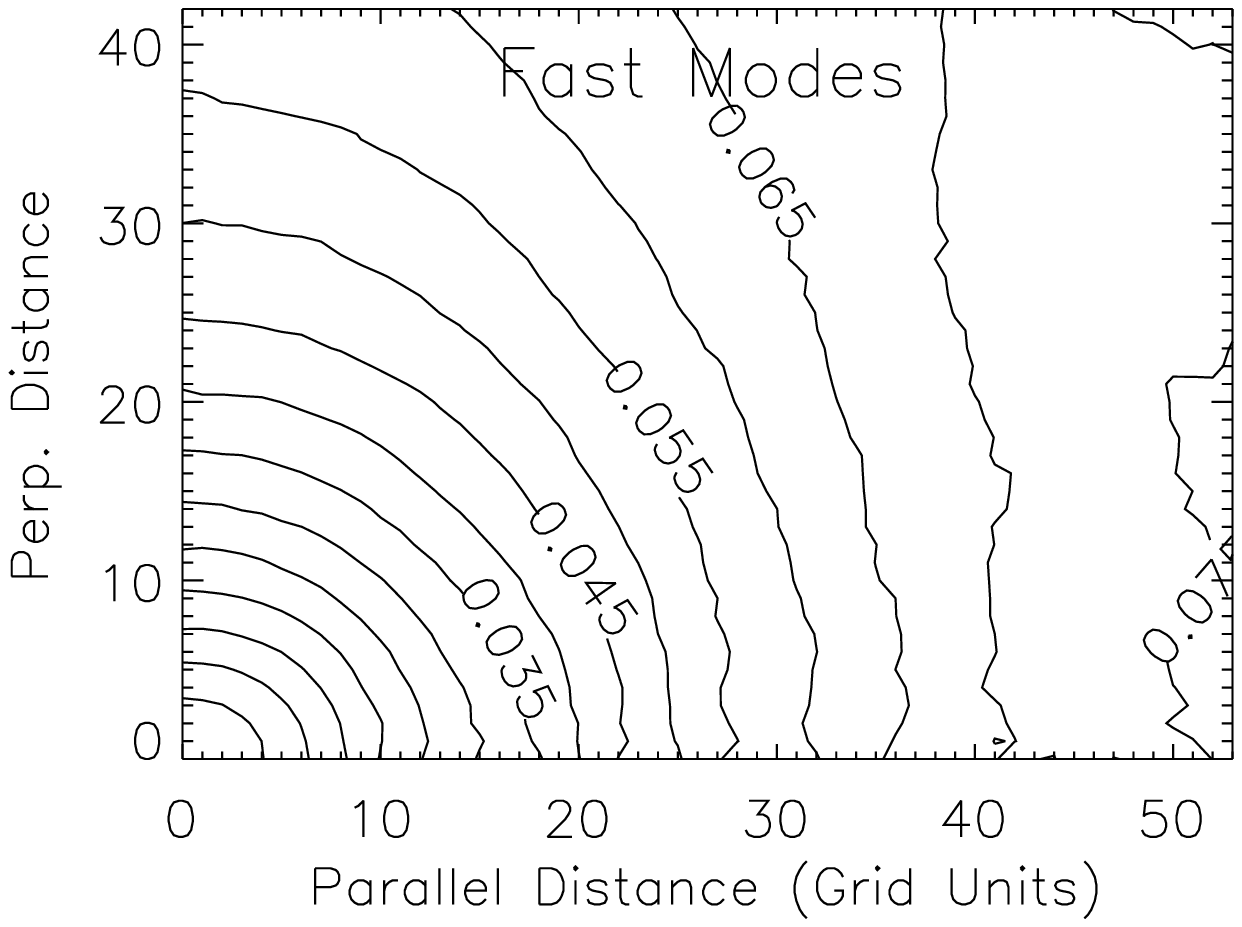}
\\
~~~~~~~~~~~~~~~~~~~~~~~~~~~~~~~(c)    \> ~~~~~~~~~~~~~~~~~~~~~~~~~(d) \\
\includegraphics[width=.50\textwidth]{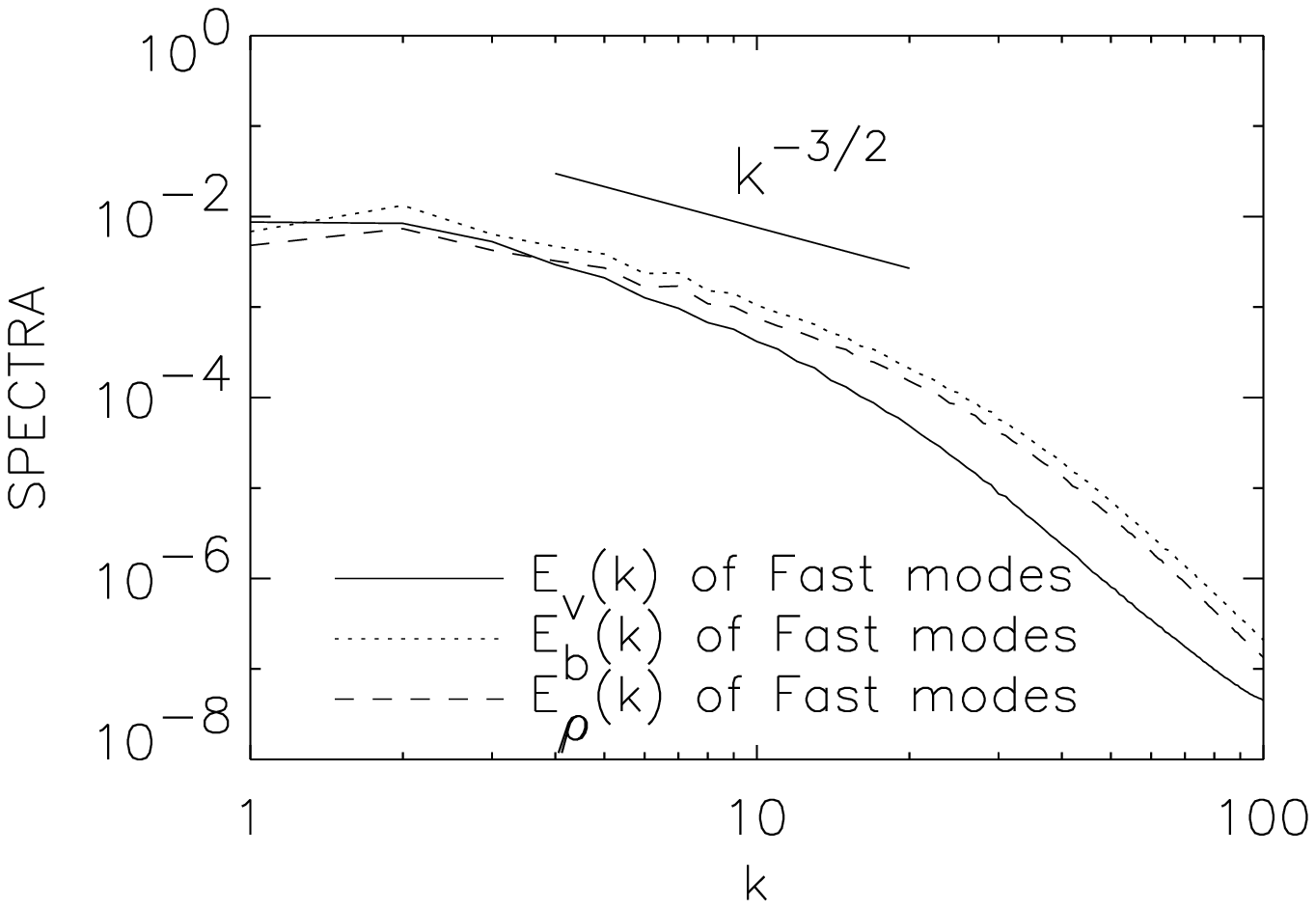}
\>
\includegraphics[width=.45\textwidth]{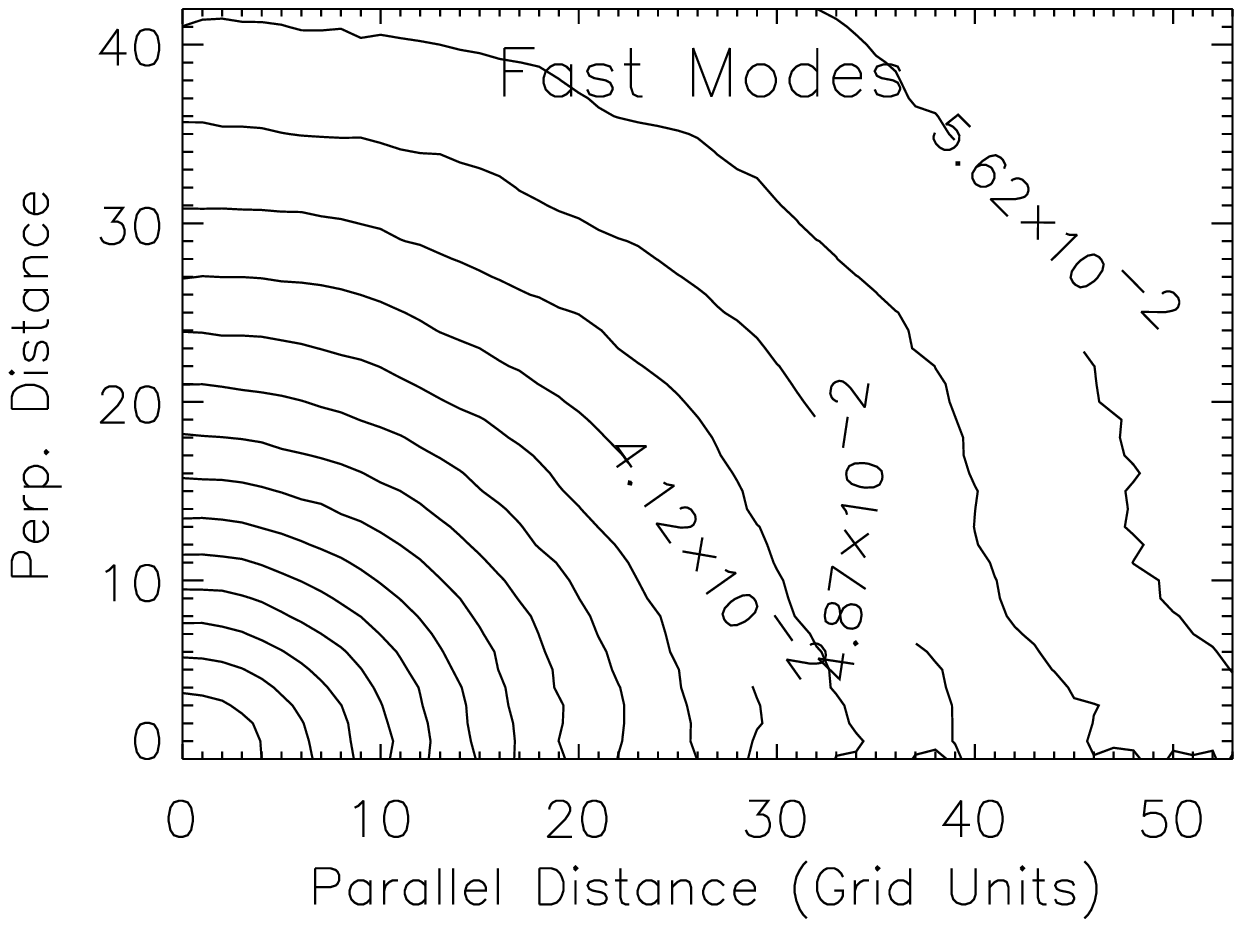}
\\
~~~~~~~~~~~~~~~~~~~~~~~~~~~~~~~(e)    \> ~~~~~~~~~~~~~~~~~~~~~~~~~(f) 
\end{tabbing}
\caption{
         Fast modes. (a)\&(b): $M_s\sim 0.35$ ($\beta=4$).
         (c)\&(d): $M_s\sim 2.3$ ($\beta=0.2$).
         (e)\&(f): $M_s\sim 7$ ($\beta=0.02$).
         Spectra are compatible with either $k^{-3/2}$ or Kolmogorov.
         Contours (or, eddy shapes) show isotropy.
         Overall, the scalings are consistent with those of
         acoustic turbulence.
}
\label{fig_result_fast}
\end{figure*}

{}Fig.~\ref{fig_result_fast}(b), (d), and (f)
show fast modes are isotropic.
{}The resonance conditions for the interacting fast waves are
$ 
\omega_1 + \omega_2 = \omega_3 \mbox{~~and~~}
  {\bf k}_1 + {\bf k}_2 = {\bf k}_3.
$ 
Since $ \omega \propto k$ for the fast modes,
the resonance conditions can be met only when
all three ${\bf k}$ vectors are collinear.
This means that the direction of energy cascade is 
{\it radial} in Fourier space.
This is very similar to acoustic turbulence, turbulence caused by interacting
sound waves 
\cite{Zak67,ZakS70,LvoLP00}.
Zakharov \& Sagdeev  \cite{ZakS70}   
found
$E(k)\propto k^{-3/2}$.
However, there is debate about
the exact scaling of acoustic turbulence.
Here we cautiously claim that our numerical results are compatible
with the Zakharov \& Sagdeev scaling:
\begin{equation}
\mbox{\it Spectrum of Fast Modes:~~~} E^f(k) \sim  k^{-3/2}.
\end{equation}
The eddies are isotropic (see also Fig.~\ref{fig_ani_fast}).

\section{Magnetic Field and Density Scalings}   \label{sect_c7}

We expect that isotropy/anisotropy of magnetic field is 
{\it similar to that of velocity} (see CL03).
However, anisotropy of density shows different behavior.
Density shows anisotropy for the high $\beta$ case.
But, for low $\beta$ cases, density shows more or less isotropic
structures. We suspect that
shock formation is responsible for isotropization of density.

To estimate the r.m.s. fluctuations, we use
the following linearized continuity and induction equations:
\begin{eqnarray}
 |\rho_k|  &=& (\rho_0 v_{k}/c) |\hat{\bf k}\cdot \hat{\bf \xi}|, \\
   |b_k|   &=& (B_0 v_{k}/c) |\hat{\bf B}_0\times \hat{\bf \xi}|,
\end{eqnarray}
where $c$ denotes velocity of slow or fast waves (equation (\ref{c_sf})).
{}From this, we obtain the r.m.s. fluctuations
\begin{eqnarray}
 (\delta \rho/\rho_0)_s  
       &=&  (\delta V)_s
           \langle |\hat{\bf k}\cdot \hat{\bf \xi}_s/c_s | \rangle, \\
 (\delta \rho/\rho_0)_f  
       &=& (\delta V)_f
           \langle |\hat{\bf k}\cdot \hat{\bf \xi}_f/c_f| \rangle, \\
 (\delta B/B_0)_s  
       &=& (\delta V)_s 
           \langle |\hat{\bf B}_0\times \hat{\bf \xi}_s/c_s| \rangle, \\
 (\delta B/B_0)_f  
       &=& (\delta V)_f
           \langle |\hat{\bf B}_0\times \hat{\bf \xi}_f/c_f| \rangle, 
\end{eqnarray}
where angled brackets denote a proper Fourier space average.
Generation of slow and fast modes velocity ($(\delta V)_s$ and $(\delta V)_f$)
depends on driving force.
Therefore, we may simply assume that 
\begin{equation}
  (\delta V)_{A} \sim (\delta V)_{s} \sim (\delta V)_{f},  \label{vavsvf}
\end{equation}
where we ignore constants of order unity.
However, when we consider mostly incompressible driving,
the generation fast modes may follow equation (\ref{eq_high2}).
In this case, the amplitude of fast mode velocity
is reduced by a factor of 
$ \left[ \frac{ V_A^2 + a^2 }{ (\delta V)^2_A } 
        \frac{ (\delta V)_A }{ V_A }   \right]^{-1/2}$:
\begin{equation}
  (\delta V)_{A} \sim (\delta V)_{s} 
   \sim  
   \left[ \frac{ V_A^2 + a^2 }{ (\delta V)^2_A } 
        \frac{ (\delta V)_A }{ V_A }   \right]^{1/2}
 (\delta V)_{f}.
   \label{vavsvf2}
\end{equation}
When we assume $(\delta V)_{A}\sim B_0$,
equation (\ref{vavsvf2}) reduces to equation (\ref{vavsvf})
in low $\beta$ plasmas.

\subsection{Low-$\beta$ case}
In this limit, $c_s \sim a \cos\theta$ and $c_f \sim V_A$.
Using equations (\ref{xis_lowbeta}) and (\ref{xif_lowbeta}),
we obtain
\begin{eqnarray}
 (\delta \rho/\rho_0)_s  
       &\sim& (\delta V)_s 
           \langle |\cos\theta/c_s| \rangle \sim (\delta V)_s/a, \\
 (\delta \rho/\rho_0)_f  
       &=& (\delta V)_f 
           \langle |\sin\theta/c_f| \rangle \sim (\delta V)_f/V_A, \\
 (\delta B/B_0)_s  
       &=& (\delta V)_s 
           \langle |\alpha \cos\theta \sin\theta/c_s| \rangle 
            \sim \alpha (\delta V)_s/a, \\
 (\delta B/B_0)_f  
       &=& (\delta V)_f 
           \langle | 1/c_f| \rangle \sim (\delta V)_f/V_A,   \label{bf_lowbeta}
\end{eqnarray}
where we ignore $\cos\theta$'s or $\sin\theta$'s.

When we assume 
$(\delta V)_{A}\sim (\delta V)_{s}\sim (\delta V)_{f}\sim B_0$, we get
\begin{eqnarray}
 (\delta \rho/\rho_0)_s  &\sim& M_s, \\
 (\delta \rho/\rho_0)_f  &=& \sqrt{\beta} M_s, \\
 (\delta B/B_0)_s  &=& \beta M_s, \\
 (\delta B/B_0)_f  &=& \sqrt{\beta} M_s
\end{eqnarray}
Therefore, in low $\beta$ plasmas, slow modes give rise to most of 
density fluctuations
(CL02).
On the other hand, magnetic fluctuation by slow modes is smaller than that
by fast modes by a factor of $\sqrt{\beta}$.

\subsection{High-$\beta$ case}
In this limit, $c_s \sim V_A \cos\theta$ and $c_f \sim a$.
Using equations (\ref{xis_highbeta}) and (\ref{xif_highbeta}),
we obtain
\begin{eqnarray}
 (\delta \rho/\rho_0)_s  
       &\sim& (\delta V)_s 
           \langle |\cos\theta \sin\theta/(\alpha c_s)| \rangle \nonumber \\
           &\sim& (V_A/a)(\delta V)_s/a, \\
 (\delta \rho/\rho_0)_f  
       &=& (\delta V)_f 
           \langle |1/c_f| \rangle \sim (\delta V)_f/a, \\
 (\delta B/B_0)_s  
       &=& (\delta V)_s 
           \langle |\cos\theta/c_s| \rangle 
            \sim  (\delta V)_s/V_A, \\
 (\delta B/B_0)_f  
       &=& (\delta V)_f 
           \langle | \sin\theta/c_f| \rangle \sim (\delta V)_f/a,  
\end{eqnarray}
where we ignore $\cos\theta$'s or $\sin\theta$'s.

Let us just assume that 
$(\delta V)_{A}\sim (\delta V)_{s}\sim B_0  \sim M_s^{-1} (\delta V)_{f}$
(cf. equation (\ref{vavsvf2})).
Then we have
\begin{eqnarray}
 (\delta \rho/\rho_0)_s  &\sim&  M_s/\sqrt{\beta} \sim M_s^2, \\
 (\delta \rho/\rho_0)_f  &\sim&  M_s^2, \\
 (\delta B/B_0)_s         &=& O(1), \\
 (\delta B/B_0)_f         &=& M_s^2.
\end{eqnarray}
The density fluctuation associated with slow modes
is $\sim M_s^2$, when
$(\delta V)_s\sim (\delta V)_{A} \sim V_A$.
This is consistent with Zank \& Matthaeus \cite{ZanM93}.    
The ratio of $(\delta \rho)_s$ to $(\delta \rho)_f$ is of order unity.
Therefore, both slow and fast modes give rise to similar amount of density
fluctuations.
Note that this argument is of order-of-magnitude in nature.
In fact, in our simulations for the high $\beta$ case, 
the r.m.s. density fluctuation by slow modes
is about twice as large as that by fast modes.
When we use equation (\ref{vavsvf}), 
we have a different result:
$ (\delta \rho)_s \sim (V_A/a)(\delta \rho)_f  < (\delta \rho)_f $.
It is obvious that
slow modes dominate magnetic fluctuations:
$(\delta B)_s > (\delta B)_f$ for both equations 
(\ref{vavsvf}) and (\ref{vavsvf2}).

\section{Slowly Evolving Fluctuations Below Viscous Cutoff} \label{sect_c8}

In hydrodynamic turbulence viscosity sets a minimal scale for
motion, with an exponential suppression of motion on smaller
scales.  Below the viscous cutoff the kinetic energy contained in a 
wavenumber band is 
dissipated at that scale, instead of being transferred to smaller scales.
This means the end of the hydrodynamic cascade, but in MHD turbulence
this is not the end of magnetic structure evolution.  For 
viscosity much larger than resistivity,
$\nu\gg\eta$, there will be a broad range of
scales where viscosity is important but resistivity is not.  
On these
scales magnetic field structures will be created 
by the shear from non-damped turbulent motions, which
amounts essentially to the shear from the smallest undamped scales.
The created magnetic structures would evolve through
generating small scale motions.
As a result, we expect
a power-law tail in the energy distribution, rather than an exponential
cutoff.  This  completely new regime
for MHD turbulence was reported in CLV02c.
Further research showed that there is a smooth connection between this
regime and small scale turbulent dynamo in high Prandtl number fluids
(see \cite{SchMC02}).

\begin{figure*}
  \includegraphics[width=0.32\textwidth]{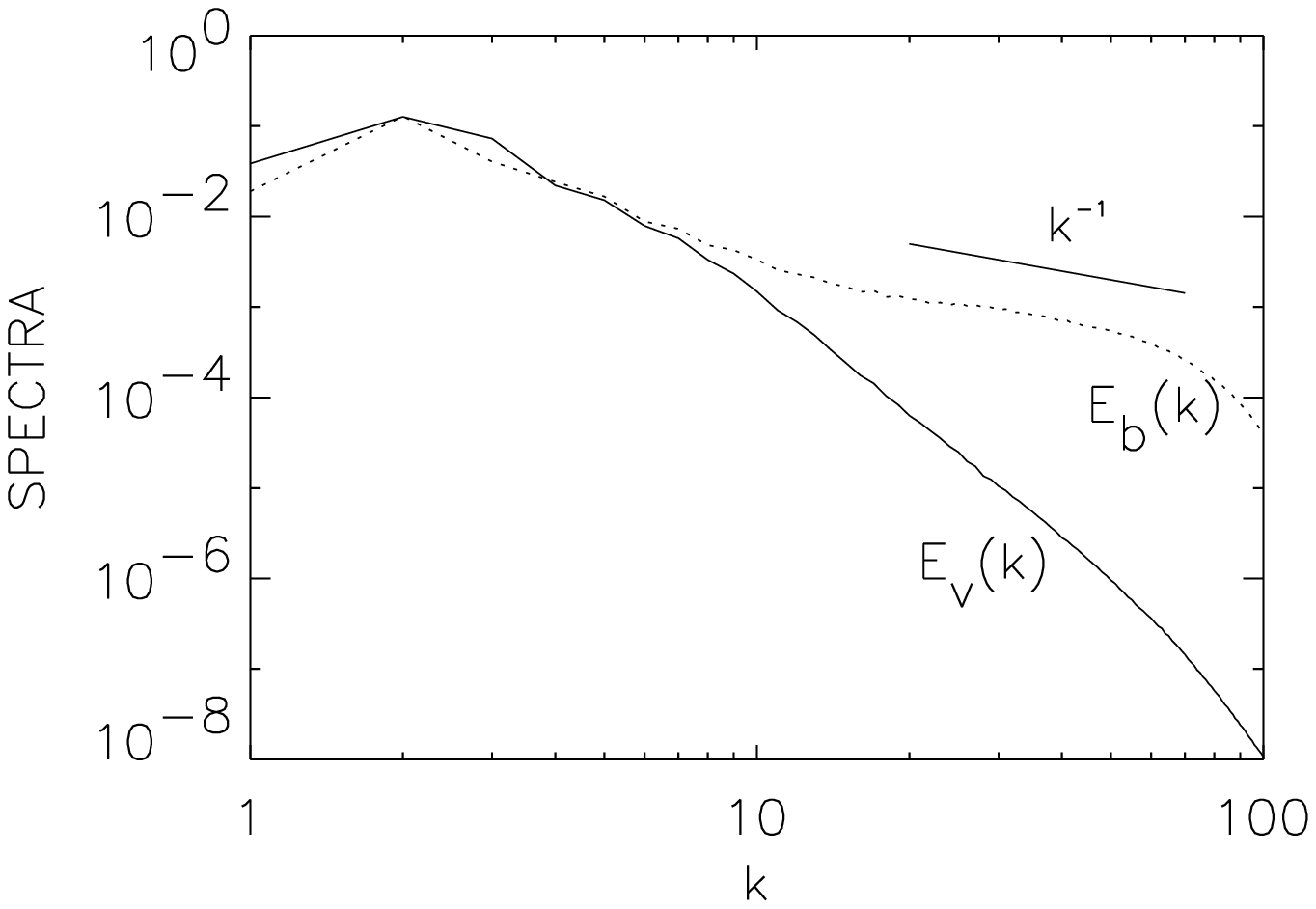}
  \includegraphics[width=0.32\textwidth]{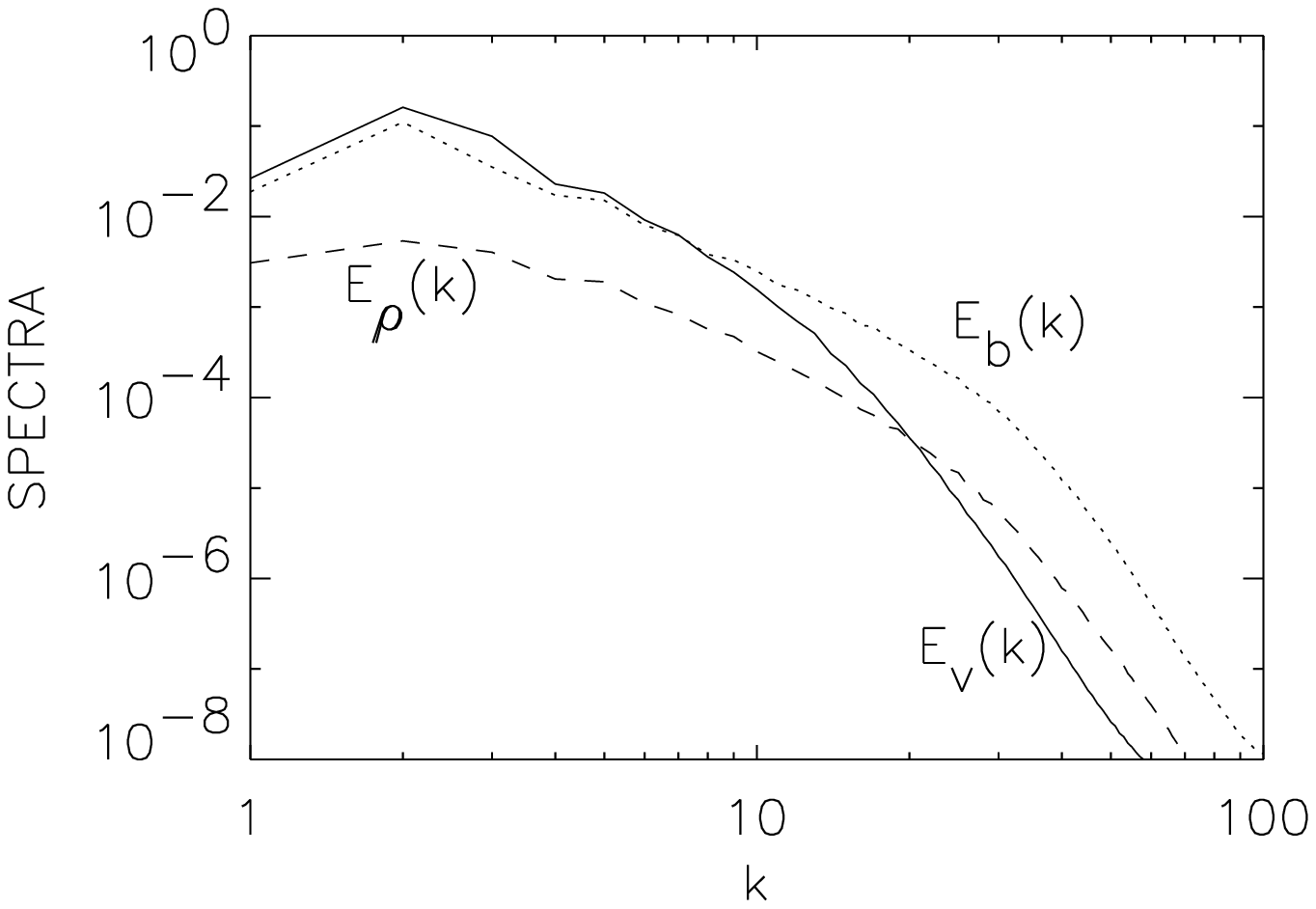}
  \includegraphics[width=0.32\textwidth]{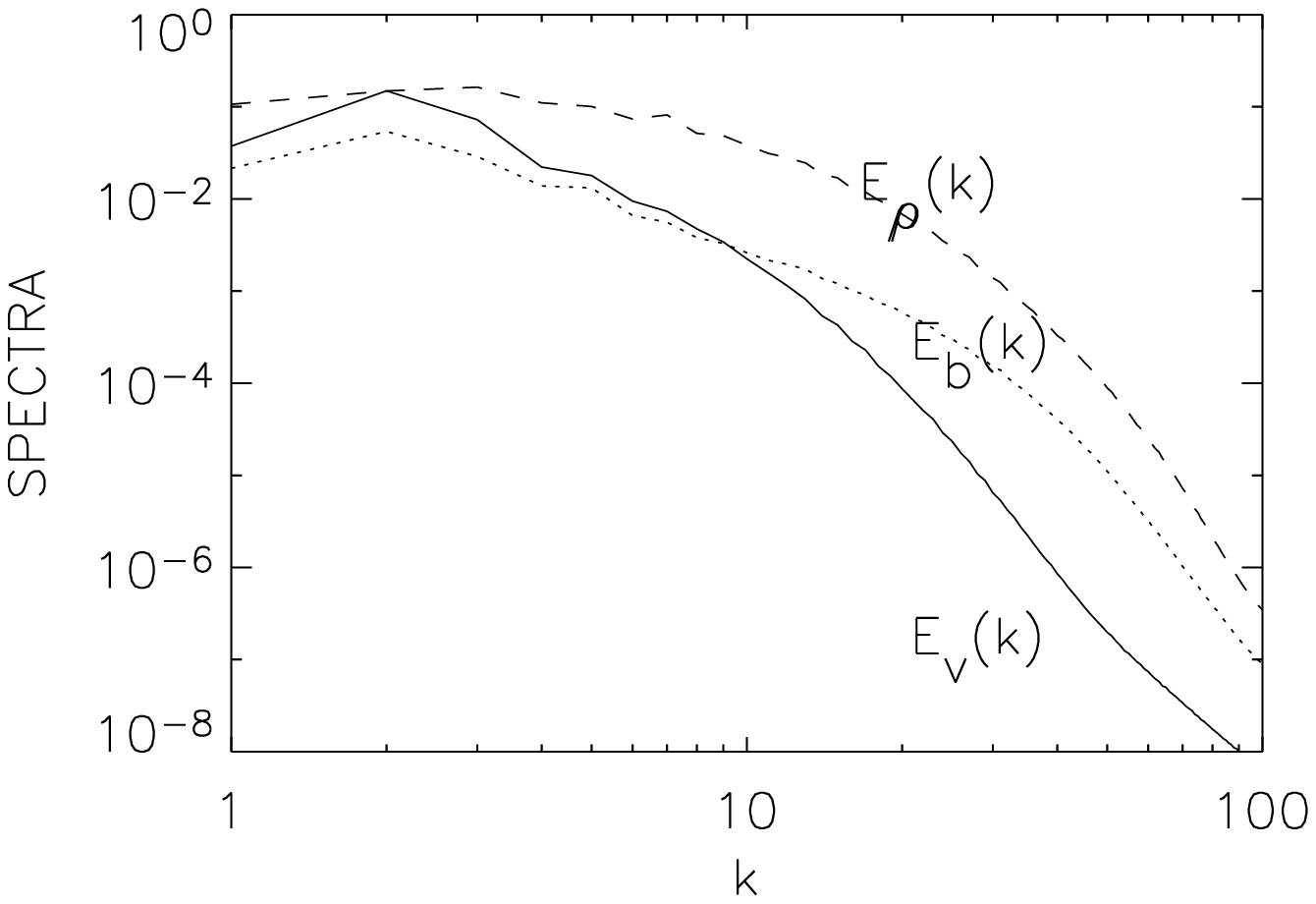} 
  \caption{
      Viscously damped regime (viscosity $>$ magnetic diffusivity).
      Due to large viscosity, velocity damps after $k\sim10$.
    {\it (a) Left:} Incompressible case with $384^3$ grid points. 
        Magnetic spectra show a shallower slope ($E_b(k)\propto k^{-1}$)
        below the velocity damping scale.
       {}From  CLV02c.     
    {\it (b) Middle:} Compressible case with $216^3$ grid points (high $\beta$).
        Magnetic and density spectra show structures below the
        velocity damping scale at $k\sim10$.
        The spectral slope is poorly defined
        because it was relatively hard to
        achieve very small magnetic diffusivity in the compressible run.
        {}From  CL03.
    {\it (c) Right:} Compressible case with low $\beta$.
        Density structures are vividly enhanced.
}
\label{fig_viscous}
\end{figure*}

CLV02c explored this regime numerically with a grid of $384^3$ and a 
physical viscosity for velocity damping. The kinetic Reynolds number was
around 100. 
We achieved a very small magnetic diffusivity by the use of
hyper-diffusion.
The result is presented in  Fig.~\ref{fig_viscous}a.
A theoretical model for this new regime and its
consequences for stochastic reconnection  \cite{LazV99}
can be found in Lazarian, Vishniac, \& Cho \cite{LazVC03}.  
It explains the spectrum $E(k)\sim k^{-1}$ as a cascade of magnetic
energy to small scales under the influence of shear at the 
marginally damped scales. The mechanism is based on the solenoidal
motions and therefore the compressibility should not alter the physics
of this regime of turbulence.

CL03 showed that the new regime of turbulence is also valid for
compressible MHD. 
We use the same physical viscosity as in incompressible case (see
CLV02c).
We rely on numerical diffusion, which is much smaller than physical
viscosity, for magnetic field.
The inertial 
range is much smaller due to numerical reasons, but it is clear that
the new regime of MHD turbulence persists. The magnetic fluctuations,
however, compress the gas and thus cause fluctuations in density.
The amplitude of density perturbations is higher in low $\beta$ plasma
(see Fig.~\ref{fig_viscous}c).
This is a new (although expected) phenomenon compared to our earlier
incompressible calculations. These density fluctuations may have important
consequences for the small scale structure of the ISM. 

\section{Discussion} \label{sect_c9}

\subsection{Range of applicability}

One may argue that for the first time ever we have universal scaling
relations that describe turbulence over a wide range of plasma
$\beta$ and Mach numbers. This remarkable fact entails a lot of
astrophysical implications.

\vspace{0.3cm}
{\it Is this consistent with the observational data discussed above?}\\
Yes, we would claim that the observed spectra that are
consistent with the Kolmogorov scaling arise naturally. 
Indeed, Alfven and slow modes exhibit
GS95 type scaling. Most of the energy in those modes correspond
to perturbations perpendicular to magnetic field. The scaling
for the perpendicular motions is Kolmogorov-like, i.e. $E(k)\sim
k_{\bot}^{-5/3}$. However, this fact does not imply that
it is OK to use Kolmogorov scalings to solve astrophysical 
problems (see \S\ref{sect_SF}). 
It can be a serious mistake to disregard scale-dependent
anisotropy of MHD turbulence.
Fast modes are isotropic and have a bit different scaling. 
A more
careful analysis of the observational data is necessary to detect
the signature of fast modes. 

\vspace{0.3cm}
{\it Is super-Alfvenic turbulence different?}\\
In the paper above 
we considered the cases in which the Alfven speed 
associated with
the mean magnetic field is slightly faster than the r.m.s. fluid velocity.
This regime is called ``sub-Alfvenic'' regime.
If initially the turbulent energy is larger 
than magnetic energy, we are in the regime of 
so-called ``super-Alfvenic'' turbulence. 
In this regime the growth of the magnetic 
field is expected through so called ``turbulent dynamo'' 
(see \cite{ChoV00a}; \cite{MulB00}; CLV02a).
The magnetic energy 
at scale $l$ in this regime grows exponentially 
with the characteristic rate of the eddy turnover time. Thus we expect
to reach equipartition between magnetic and kinetic energies
at the energy injection scale. At smaller scales
the turbulence becomes sub-Alfvenic and our earlier considerations
should be applicable. Although the decomposition of MHD turbulence
described above does not work for irregular
magnetic field characterizing super-Alfvenic turbulence, CL03
results are suggestive that our considerations 
about Alfven, slow and fast modes
are applicable to 
this regime. 

\subsection{Compressible MHD turbulence and star formation}  \label{sect_SF}

{\it How fast does MHD turbulence decay?}\\ 
This question has fundamental
implications for {\it star formation} (see \cite{McK99}). 
Indeed, it was thought originally that
magnetic fields would prevent turbulence from fast decay. Later 
(see \cite{MacKB98,StoOG98}; and review \cite{VazOP00})
this was reported not to be true. 
However, fast decay was erroneously
associated with the coupling between compressible and incompressible 
modes. The idea was that incompressible motions 
quickly transfer their energy to the compressible modes, which get
damped fast by direct dissipation (presumably through shock formation).

Our calculations support the relation in eq.~(\ref{eq_high2}).
According to it the coupling of Alfven and compressible motions is important
only at the energy injection scales where $\delta V_l\sim V_A$. 
As the turbulence evolves the perturbations become smaller and the
coupling less efficient. Typically for numerical simulations
the inertial range is rather small and this could explain why
marginal coupling of modes was not noticed.

Our results show that MHD turbulence damping does not depend on whether
the fluid is compressible or not. The incompressible motions damp also within
one eddy turnover time. This is the consequence of the fact that within
the strong turbulence\footnote{For a formal definition of
strong, weak and intermediate turbulence see Goldreich \& Sridhar 
\cite{GolS97}
and CLV02a, but
here we just mention in passing that in most astrophysically important
cases the MHD turbulence is ``strong''.} mixing motions perpendicular
to magnetic field are hydrodynamic to high order (CLV02b) and the 
cascade of energy induced by those motions is similar to the hydrodynamic 
one, i.e. energy cascade happens within an eddy turnover time. 

\vspace{0.3cm}
{\it Is the decay always fast for compressible MHD turbulence?}\\
This issue does need further investigation. However, our preliminary
answer to this question is ``no''. Indeed, incompressible MHD computations
(see \cite{MarG01}; CLV02b)
show that the
rate of turbulence decay depends on the degree of turbulence 
imbalance\footnote{This quantity is also called cross helicity
(see \cite{MatGM83}).},    
i.e.
the difference in the energy fluxes moving in opposite directions.  
The strongly imbalanced incompressible turbulence was shown to persist
longer than its balanced counterpart. This enabled CLV02b to speculate that 
this may enable energy transfer between clouds and may
explain the observed turbulent linewidths of GMCs without evident 
star formation. 
Our results above show a marginal coupling of compressible and 
incompressible
modes. This is suggestive that the results obtained in incompressible 
simulations are applicable to compressible environments
if amplitudes of perturbations are not large.
The complication arises from the existence of the parametric instability
\cite{DelVL01}
that happens as the density 
perturbations reflect Alfven waves
and grow in amplitude. This instability eventually controls the degree
of imbalance that is achievable. However, the growth rate of the instability
is substantially slower than the Alfven wave oscillation rate. Therefore,
if we take into account that interstellar sources are intermittent not
only in space, but also in time, the transport of turbulent 
energy described in CLV02b may be feasible.

\vspace{0.3cm}
{\it What is the density structure that we expect to see?}\\
First of all, we do not expect to see tight correlation between density
and magnetic field. Such sort of correlation is expected in the traditional
static picture of the ISM. Introduction of turbulence in the picture
of ISM complicates the analysis (see discussion in 
\cite{VazOP00}; CLV02a).
Our results confirm earlier claims
(e.g. CLV02a; \cite{PasV02})
that magnetic field
- density correlations may be weak. First of all, some magnetic field
fluctuations are related to Alfvenic turbulence which does not compress
the medium. Second, slow modes in low $\beta$ plasmas are essentially
density perturbations that propagate along magnetic field and which
marginally perturb magnetic fields.

On the small scales we expect to see structures that anti-correlate
with magnetic field and caused by the new regime of turbulence below
the ambipolar damping scale. We mentioned above that the simulations
in Cho, Lazarian \& Vishniac (CLV02c) 
and theoretical calculations
in Lazarian, Vishniac \& Cho  \cite{LazVC03}   
show that the magnetic field
in a newly discovered regime of MHD turbulence can
produce a shallow spectrum $E(k)\sim k^{-1}$ spectrum 
of magnetic fluctuations. Calculations in CL03 
suggest that this will translate in the corresponding
shallow spectrum of density. 
For cold neutral medium (CNM; see Draine \& Lazarian \cite{DraL98}
for a list of idealized phases) the spectrum
of density fluctuations can protrude from a fraction of parsec to
a scale of $\sim 100$ AU (see CL03).

\subsection{Astrophysical significance of fast modes}

Results of CL02 and CL03 show that the Alfven and slow modes
exhibit Goldreich-Sridhar scale-dependent anisotropy. 
However, it would be very wrong to forget that fast modes are isotropic.
It is possible to show that in many instances that difference makes
fast modes very important. Consider two examples.

\begin{figure*}
  \includegraphics[width=0.49\textwidth]{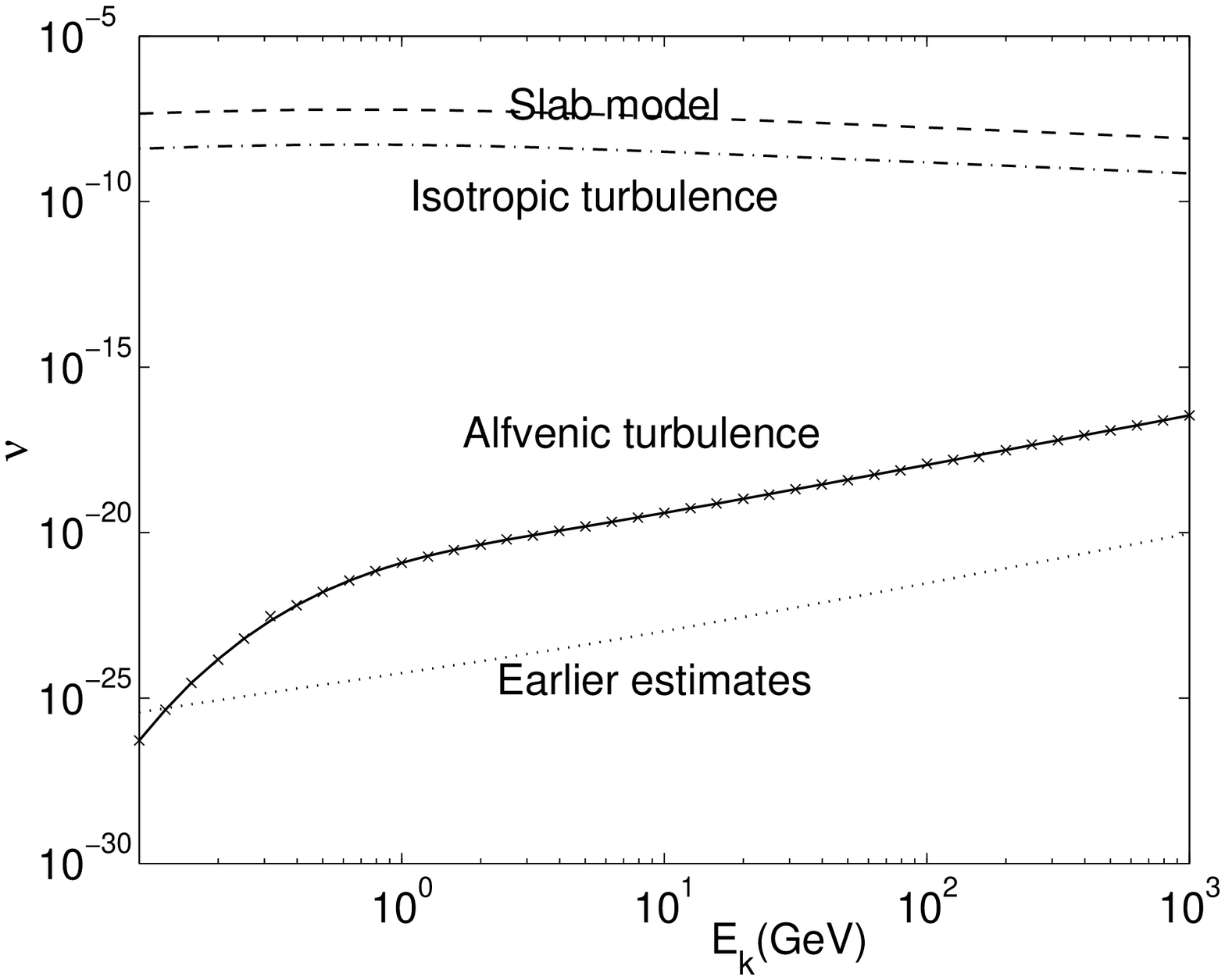}
\hfill
  \includegraphics[width=0.49\textwidth]{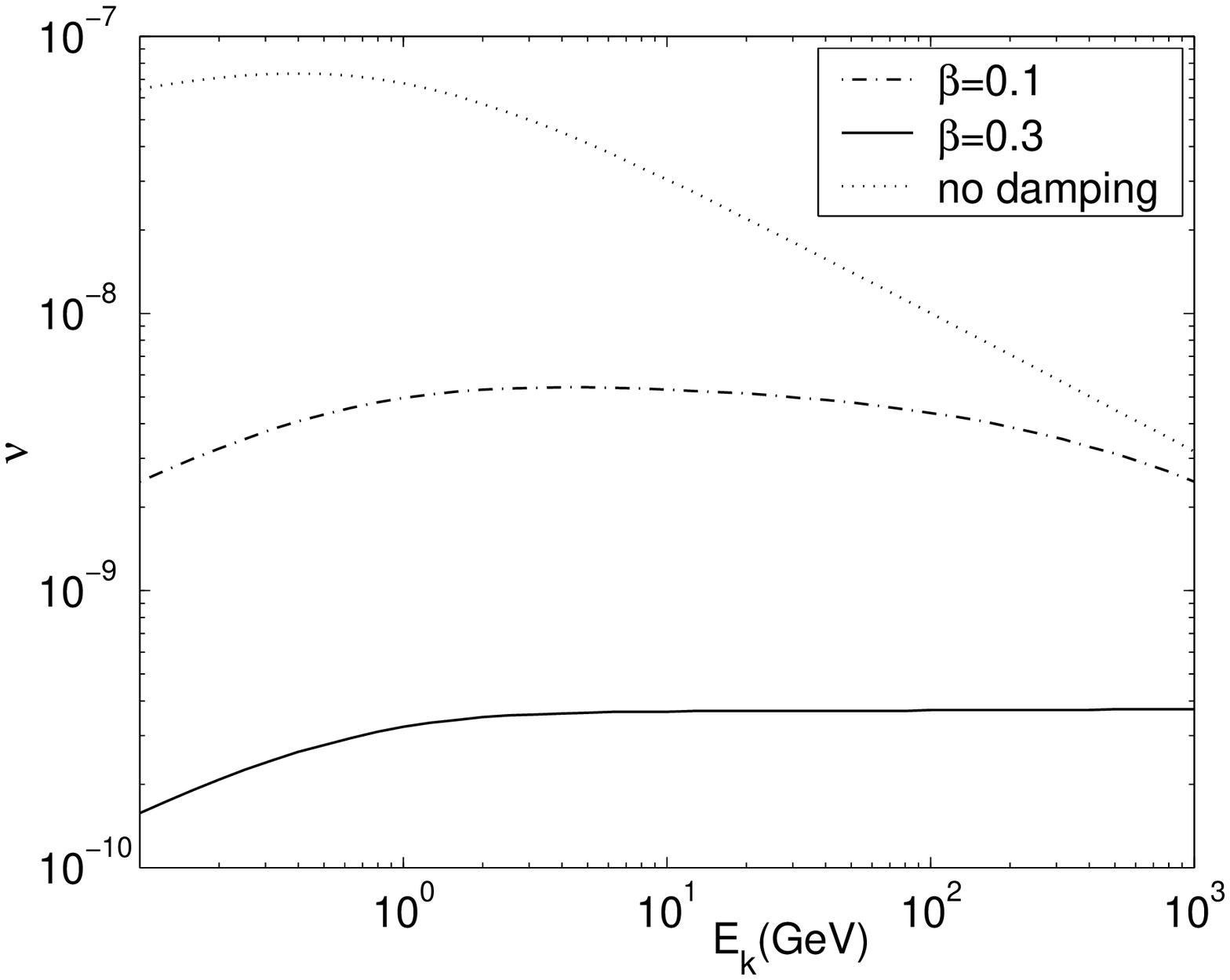} \\
  \caption{Rate of cosmic ray scattering 
        by Alfvenic ({\it left panel}) and fast ({\it right Panel}) modes
of MHD turbulence. The scattering by Alfvenic modes is negligible, although
it is substantially larger than an earlier estimate in Chandran 
\cite{Cha01}.
The default for many researchers is Alfvenic turbulence with Kolmogorov
spectrum (upper curve on the left panel). Fast modes
scatter cosmic rays
much more efficiently in spite of being partially damped in the ISM 
the dependence of scattering on plasma $\beta$
is a new prediction. From Yan \& Lazarian \cite{YanL02}.
}
\label{fig_cosray}
\end{figure*}

\vspace{0.3cm}
{\it Cosmic Ray Propagation}.\\
The propagation of cosmic rays is mainly determined by their interactions
with electromagnetic fluctuations in the interstellar medium. 
The resonant interaction of cosmic ray particles
with MHD turbulence has been repeatedly suggested as the main
mechanism for scattering and isotropizing cosmic rays. 
In these
studies, it is usually assumed that the turbulence is
{\it isotropic} with a Kolmogorov spectrum (e.g. \cite{SchM98}). 
Yan \& Lazarian \cite{YanL02}   
identified fast modes as being responsible
for cosmic ray scattering. The scattering by Alfvenic turbulence,
which is the default for most of the theoretical
constructions, is from 15 to 5 orders smaller than it is usually
obtained using Kolmogorov model of Alfvenic turbulence 
(see Fig.~\ref{fig_cosray}). 

\vspace{0.3cm}
{\it Dust Grain Dynamics}.\\
Turbulence induces relative dust grain motions and leads to 
grain-grain collisions. These collisions
determine grain size distribution, which affects most dust properties,
including starlight absorption and  H$_2$ formation. 
Unfortunately, as in the
case of cosmic rays, earlier work appealed
to hydrodynamic turbulence to predict grain relative velocities.
Lazarian \& Yan   \cite{LazY02}       
and Yan \& Lazarian  \cite{YanL03}     
considered motions of charged grains in MHD turbulence and identified
the direct interaction of the charged grains with fast modes as
the principal mechanism for acceleration of grains with radius
larger than $\sim 10^{-5}$~cm. Those modes
can acceleration provide grains with supersonic velocities 
(see fig.~{\ref{fig_grain}).

\begin{figure*}
  \includegraphics[width=0.49\textwidth]{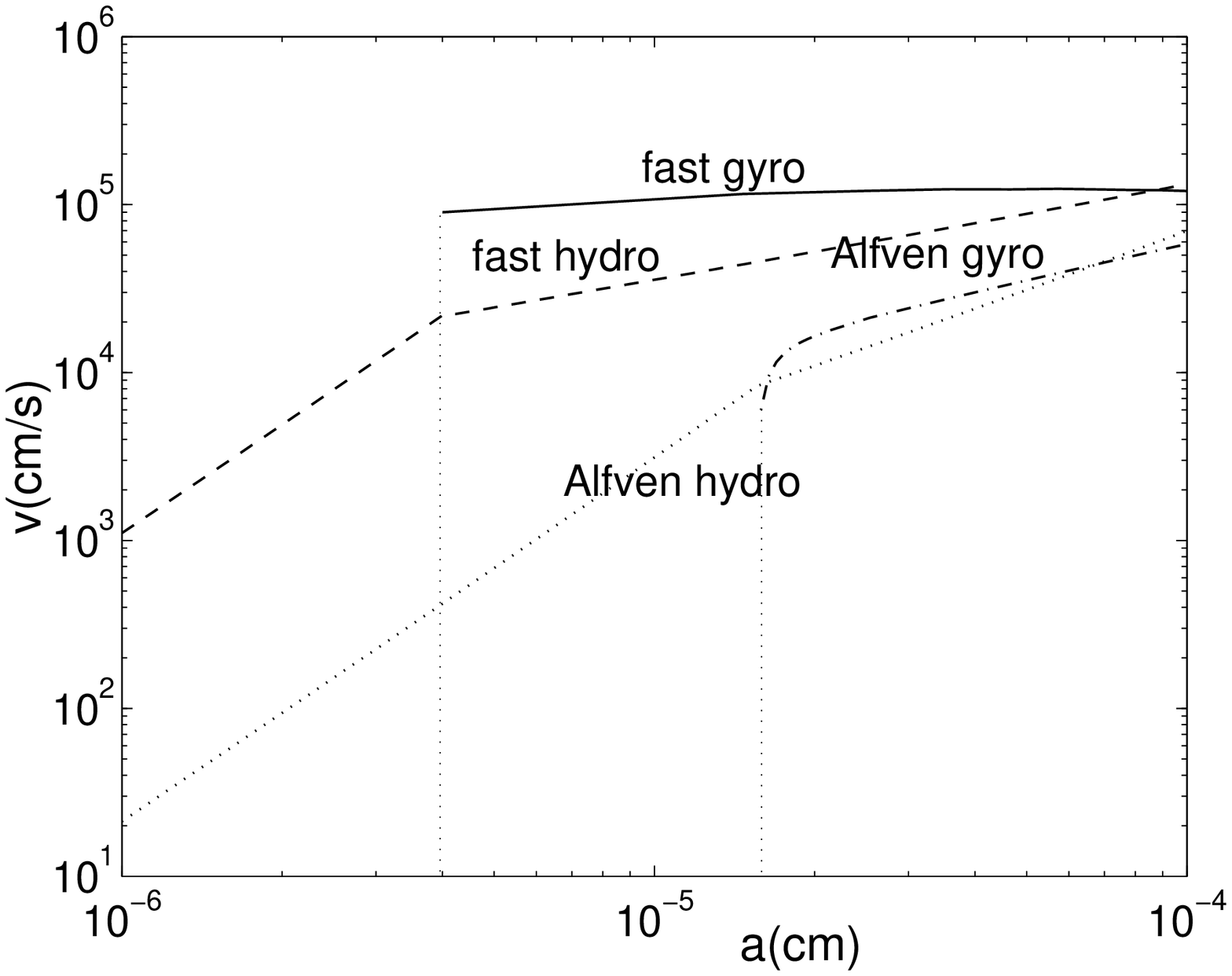}
\hfill
  \includegraphics[width=0.49\textwidth]{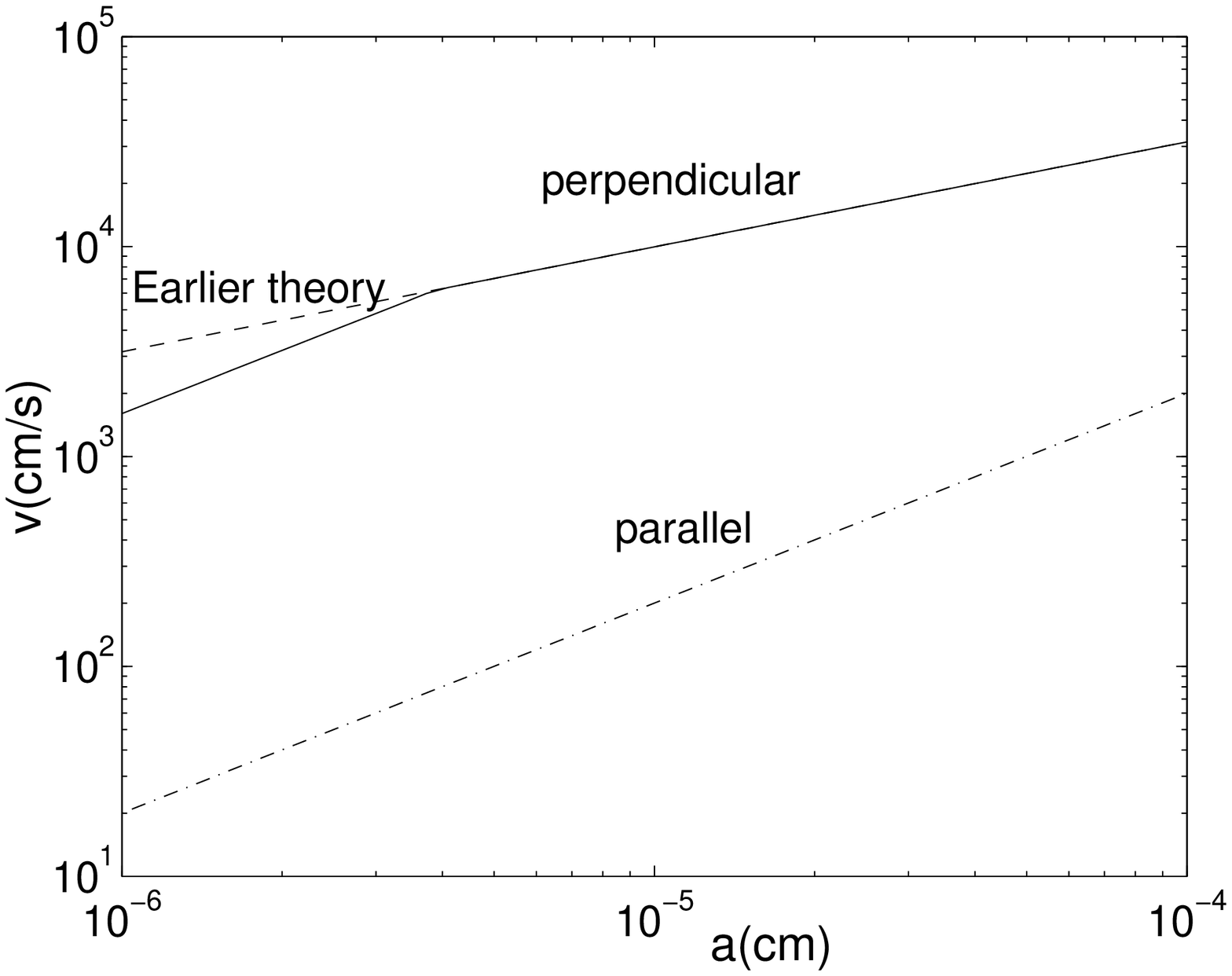}
  \caption{A new mechanism of grain acceleration
is based on the direct interaction of charged grains
with MHD turbulence. 
Gyroresonance of charged grains with fast modes
dominates the acceleration of charged grains in Cold Neutral Medium (CNM).
Within dark clouds the interaction of MHD turbulence and charged
grains is important only for small grains.
 When grains decouple from magnetic field,
Kolmogorov theory provides an OK estimate for grain velocities.
{}From Yan \& Lazarian \cite{YanL03} and
Lazarian \& Yan \cite{LazY02}.
}
\label{fig_grain}

\end{figure*}

\subsection{Other examples: from HII regions to gamma ray busts}

Lithwick \& Goldreich \cite{LitG01}   
addressed the issue of the origin of density
fluctuations within HII regions. There the gas
pressure is larger than the magnetic pressure (the `high $\beta$' regime)
and they conjectured that
fast waves, which are essentially sound waves, would be decoupled from
the rest of the cascade.
They found that density fluctuations are due to the slow mode 
and the entropy mode, which are passively mixed by shear Alfv\'en waves and 
follow a Kolmogorov spectrum.
Our results in CL03
suggest that fast modes may be also an important source of
density fluctuations. In addition, results on the new regime of turbulence
(CLV02c, CL03) indicate that the new
regime of turbulence can fluctuations on very small scales and this entails
resumption of the turbulent cascade \cite{LazVC03}
 that was not considered 
in \cite{LitG01}.      

The whole machinery of MHD turbulence scalings
is required to deal with turbulence in gamma ray bursts (see \cite{LazPY03}).
There both fast and Alfven waves can transfer
their energy to emitting electrons. However, the ways that they transfer
their energy are different and this may result in important observational
consequences.

Heating of ISM and of Diffuse Ionized Gas (DIG), in particular,
is another issue where imbalanced MHD turbulence is important
(see CLV02a). Compression of molecular clouds by MHD turbulence (see
\cite{MyeL98}),      
stochastic magnetic reconnection \cite{LazV99,LazVC03}
are other examples
when it is essential to know the fundamental properties of compressible
MHD.

\section{Summary}   \label{sect_sum}
In the paper, we have 
studied generation of compressible MHD modes.
We have presented the 
statistics of compressible MHD turbulence
for high, intermediate, and low $\beta$ plasmas 
and for different sonic and Alfven Mach numbers. 
For subAlfvenic turbulence we provided the decomposition
of turbulence into Alfven, slow and fast modes.
We have found that the generation of compressible modes 
by Alfvenic modes
is suppressed and, contrary to the common belief, the drain of
energy from Alfven to compressible modes is marginal along the
cascade. As the result the Alfvenic modes form a separate
cascade with the properties similar to those of Goldreich-Sridhar
cascade in incompressible media. As Alfven modes shear slow modes
they impose their scaling on them. 
On the contrary, fast modes
show isotropy for both magnetic-
and gas-pressure dominated plasmas. 
The new insight into compressible MHD entails
important astrophysical consequences
that range from the dynamics of star
formation to the dynamics of gamma ray busts.

{\bf Acknowledgments:}
We thank Ethan Vishniac, Peter Goldreich, 
Bill Matthaeus, Chris McKee, 
and Annick Pouquet
for stimulating discussions.
We acknowledge the support of NSF Grant AST-0125544.
This work was partially supported by NCSA
under AST010011N and
utilized the NCSA Origin2000.

\appendix
\section*{Appendix}

Let us consider a small perturbation
in the presence of a strong mean magnetic field.
We write density, velocity, pressure, and magnetic field as the sum of 
constant and fluctuating parts:
$\rho \rightarrow \rho_0+\rho$, ${\bf v} \rightarrow {\bf v}_0+{\bf v}$, 
$P \rightarrow P_0+p$,
and ${\bf B} \rightarrow {\bf B}_0+{\bf b}$, respectively.
We assume that ${\bf v}_0=0$ and that
perturbation is small : $\rho << \rho_0$, etc.
Ignoring the second and higher order contributions, we
can rewrite the MHD equations as follows:
\begin{eqnarray}
  \frac{\partial \rho}{\partial t} + \rho_0 \nabla \cdot {\bf v} = 0, 
\\
  \rho_0 \frac{\partial {\bf v}}{\partial t} + \nabla (a^2 \rho) 
                  - \frac{1}{4\pi} (\nabla \times {\bf b})\times {\bf B}_0=0,
    \\
 \frac{\partial {\bf b}}{\partial t}
      +\nabla \times \left[ {\bf v} \times {\bf B}_0 \right]=0,
\end{eqnarray}
where we assume a polytropic equation of state: 
$p=a^2 \rho$ with $a^2=\gamma p_0/\rho_0$.
We follow arguments in Thompson (1962) to derive magnetosonic waves.
Let ${\bf \xi}({\bf r},t)$ be the displacement vector, so that
$\partial {\bf \xi}/\partial t = {\bf v}$.
Assuming that the displacements vanish at $t=0$, we can integrate
the equations as follows
\begin{eqnarray}
  \rho + \rho_0 \nabla \cdot {\bf \xi} = 0, \\
  \ddot{\bf \xi} = a^2 \nabla (\nabla \cdot {\bf \xi})
   + (\nabla \times {\bf b})\times {\bf B}_0/4\pi\rho_0, 
                                               \label{momen_1stO_intm}\\
  {\bf b} = \nabla \times (\xi \times {\bf B}_0).
\end{eqnarray}
The momentum equation (eq. \ref{momen_1stO_intm}) becomes
\begin{eqnarray}
 \ddot{\bf \xi}=a^2 \nabla (\nabla \cdot {\bf \xi}) +
             \left[ \nabla \times \left( \nabla \times (\xi \times {\bf B}_0) 
                                \right)
             \right] \times {\bf B}_0 /4\pi\rho_0  \nonumber \\
      ~~ =a^2  \nabla (\nabla \cdot {\bf \xi}) +
          \nabla ( B_0^2 \nabla \cdot {\bf \xi} -
         {\bf B}_0 \cdot \nabla {\bf B}_0 \cdot{\bf \xi})/4\pi\rho_0 
                                                               \nonumber \\
     ~~~  - ({\bf B}_0 \cdot \nabla)^2 {\bf \xi}/4\pi\rho_0 
   + \left[{\bf B}_0 ( {\bf B}_0 \cdot \nabla) \nabla \cdot {\bf \xi} 
     \right]/4\pi\rho_0 
\end{eqnarray}
Using $\alpha=a^2/V_A^2=\beta(\gamma/2)$, $V_A=B_0/4\pi\rho_0$, we have
\begin{equation}
\ddot{\bf \xi}/V_A^2-\nabla [ (\alpha+1) \nabla \cdot {\bf \xi}-  
                     (\hat{\bf B}_0 \cdot \nabla)(\hat{\bf B}_0 \cdot {\bf \xi})
                   ]                                    \nonumber \\
  - (\hat{\bf B}_0 \cdot \nabla)^2 {\bf \xi}
  + (\hat{\bf B}_0 \cdot \nabla)(\nabla \cdot {\bf \xi}) \hat{\bf B}_0
  = 0
\end{equation}
In Fourier space
 the equation becomes
\begin{equation}
  \ddot{\bf \xi}/V_A^2 + k \hat{\bf k}[(\alpha +1)k \xi_k - k_{\|} \xi_{\|}]
      + k_{\|}^2 {\bf \xi} - k_{\|} k \xi_{k} \hat{k}_{\|} =0,
  \label{eq_in_FSP}
\end{equation}
where $\xi_k={\bf \xi}\cdot \hat{\bf k}$, 
$\xi_{\|}={\bf \xi}\cdot \hat{\bf k}_{\|}$, $\hat{\bf k}={\bf k}/k$,
and $\hat{\bf k}_{\|}$ is unit vector parallel to ${\bf B}_0$ 
(i.e. $\hat{\bf k}_{\|} = \hat{\bf B}_0$).
Assuming $\ddot{\bf \xi}=-\omega^2 {\bf \xi}=-c^2k^2 {\bf \xi}$,
we can rewrite (\ref{eq_in_FSP}) as
\begin{equation}
   (c^2/V_A^2 -\cos^2 \theta){\bf \xi}
    - [(\alpha+1) \xi_k-\cos \theta \xi_{\|}]\hat{\bf k}
    + \cos \theta \xi_{k} \hat{k}_{\|} =0,
   \label{xi_in_k}
\end{equation}
where $\cos \theta = k_{\|}/k$ and $\theta$ is the angle between
${\bf k}$ and ${\bf B}_0$.

Using $\hat{\bf k}=\sin\theta \hat{\bf k}_{\perp}+\cos\theta \hat{\bf k}_{\|}$,
we get
\begin{eqnarray}
(c^2/V_A^2 -\cos^2 \theta){\bf \xi}
  - [ (\alpha+1)\xi_k-\cos \theta \xi_{\|} ]\sin\theta \hat{\bf k}_{\perp} 
   \nonumber \\ 
  - \{ [ (\alpha+1)\xi_k-\cos \theta \xi_{\|}]\cos\theta 
  - \cos\theta \xi_{k} \} \hat{\bf k}_{\|}=0.
\end{eqnarray}
Writing ${\bf \xi}=\xi_{\perp} \hat{\bf k}_{\perp}
       + \xi_{\|} \hat{\bf k}_{\|}
       + \xi_{\varphi} \hat{\bf \varphi}$,
we get
\begin{eqnarray}
   (c^2/V_A^2 -\cos^2 \theta)\xi_{\perp}-[(\alpha+1)\xi_k-\cos \theta \xi_{\|}]
       \sin\theta =0,      \label{eq_perp}
    \\
    (c^2/V_A^2 -\cos^2 \theta)\xi_{\|}-
    [ \alpha\xi_k-\cos \theta \xi_{\|}]\cos\theta=0,  \label{eq_par}
    \\
     (c^2/V_A^2 -\cos^2 \theta)\xi_{\varphi} = 0.     \label{eq_psi}
\end{eqnarray}

The non-trivial solution of equation (\ref{eq_psi}) is the Alfven wave, whose
dispersion relation is
$\omega/k = V_A \cos\theta$.
The direction of the displacement vector for Alfven wave is parallel to
the azimuthal basis $\hat{\bf \varphi}$:
\begin{equation}
   \hat{\bf \xi}_A = -\hat{\bf \varphi} 
         = \hat{\bf k}_{\perp} \times \hat{\bf k}_{\|}.
\end{equation}

Let us consider solutions of equations (\ref{eq_perp}) and (\ref{eq_par}).
Using $\xi_k=\xi_{\perp}\sin\theta+\xi_{\|}\cos\theta$, we get
\begin{eqnarray}
   (c^2/V_A^2 -\cos^2 \theta)\xi_{\perp}-(\alpha+1)\sin^2\theta \xi_{\perp}-
                              \alpha \cos \theta \sin\theta \xi_{\|}=0,
    \\
    (c^2/V_A^2 -\cos^2 \theta)\xi_{\|}-
   \alpha\sin\theta \cos\theta \xi_{\perp}-
    (\alpha-1)\cos^2 \theta \xi_{\|}=0.
\end{eqnarray}
Rearranging these, we get
\begin{eqnarray}
   (c^2/V_A^2 -\alpha \sin^2\theta-1)\xi_{\perp}-
                              \alpha \cos\theta \sin\theta \xi_{\|}=0,
                   \label{kpar_and_kperp1}
    \\
    (c^2/V_A^2 -\alpha \cos^2 \theta)\xi_{\|}-
   \alpha \sin\theta \cos\theta \xi_{\perp}=0.    \label{kpar_and_kperp2}
\end{eqnarray}
Combining these two, we get
\begin{eqnarray}
  (c^2/V_A^2 -\alpha \sin^2\theta-1)
  (c^2/V_A^2 -\alpha \cos^2 \theta)
  \nonumber \\
   = \alpha^2 \sin^2\theta \cos^2\theta .
\end{eqnarray}
Therefore, the dispersion relation is
\begin{equation}
   c^4/V_A^4 - (1+\alpha) c^2/V_A^2 + \alpha \cos^2\theta = 0.
\end{equation}
The roots of the equation are
\begin{equation}
   c_{f,s}^2= \frac{1}{2} V_A^2 \left[ (1+\alpha) \pm 
               \sqrt{ (1+\alpha)^2 - 4\alpha \cos^2\theta } \right], 
        \label{c_sf}
\end{equation}
where subscripts `f' and 's' stand for `fast' and 'slow' waves, respectively.

We can write
\begin{eqnarray}
   {\bf \xi}=\xi_{\|} \hat{\bf k}_{\|} + \xi_{\perp} \hat{\bf k}_{\perp} 
   \propto \left[\frac{\xi_{\|}k_{\perp}}{\xi_{\perp}k_{\|}}\right]
            k_{\|} \hat{\bf k}_{\|} 
     + 
      k_{\perp} \hat{\bf k}_{\perp}.
\end{eqnarray}
Plugging eq. (\ref{c_sf}) into eq.~(\ref{kpar_and_kperp1}) and 
(\ref{kpar_and_kperp2}), we get
\begin{eqnarray}
   \left[ \frac{1+\alpha}{2} \pm \frac{\sqrt{D}}{2} 
                -\alpha \sin^2\theta -1               \right] \xi_{\perp}
     = \alpha \cos\theta \sin\theta \xi_{\|}, \\
   \left[ \frac{1+\alpha}{2}\pm \frac{\sqrt{D}}{2} 
                -\alpha \cos^2\theta \right] \xi_{\|}
     = \alpha \cos\theta \sin\theta \xi_{\perp},
\end{eqnarray}
where $D=(1+\alpha)^2-4{\alpha} \cos^2{\theta}$.
Using $k_{\|}=k\cos\theta$ and 
$k_{\perp}=k\cos\theta$, we get
\begin{eqnarray}
   \left[ \frac{-1+\alpha}{2}\pm \frac{\sqrt{D}}{2} 
   \right] \xi_{\perp}k_{\|}
                -\alpha \sin^2\theta   \xi_{\perp}k_{\|}
     = \alpha \cos^2\theta \xi_{\|}k_{\perp}, \\
   \left[ \frac{1+\alpha}{2}\pm \frac{\sqrt{D}}{2} 
   \right]\xi_{\|}k_{\perp}
                -\alpha \cos^2\theta \xi_{\|}k_{\perp}
     = \alpha  \sin^2\theta \xi_{\perp}k_{\|}.
\end{eqnarray}
Arranging these, we get
\begin{equation}
    \frac{ \xi_{\|}k_{\perp} }{ \xi_{\perp}k_{\|} }
    = \frac{ -1 + \alpha \pm \sqrt{D} }{ 1+\alpha \pm \sqrt{D} },
\end{equation}
where the upper signs are for fast mode and the lower signs for slow mode.
Therefore, we get
\begin{eqnarray}
   \hat{\bf \xi}_s \propto 
     ( -1 + \alpha - \sqrt{D} )
            k_{\|} \hat{\bf k}_{\|} 
     + 
     ( 1+\alpha - \sqrt{D} ) k_{\perp} \hat{\bf k}_{\perp},
  \label{eq_xis_new}
\\
   \hat{\bf \xi}_f \propto 
     ( -1 + \alpha + \sqrt{D} )
            k_{\|} \hat{\bf k}_{\|} 
     + 
     ( 1+\alpha + \sqrt{D} ) k_{\perp} \hat{\bf k}_{\perp}. 
   \label{eq_xif_new}
\end{eqnarray}
The slow basis $\hat{\bf \xi}_s$ lies between $\hat{\bf k}_{\|}$ and
$-\hat{\bf \theta}$.
The slow basis $\hat{\bf \xi}_f$ lies between $\hat{\bf k}_{\perp}$ and
$\hat{\bf k}$ (Fig.~\ref{fig_separation}).
Here overall sign of $\hat{\bf \xi}_s$ and $\hat{\bf \xi}_f$ is not important.

When $\alpha \rightarrow 0$,
equations (\ref{eq_xif_new}) and (\ref{eq_xis_new}) becomes
\begin{eqnarray}
   \hat{\bf \xi}_s \approx  \hat{\bf k}_{\|}
     -(\alpha \sin\theta \cos\theta)\hat{\bf k}_{\perp}, \label{xis_lowbeta}
\\
   \hat{\bf \xi}_f \approx  (\alpha \sin\theta \cos\theta) \hat{\bf k}_{\|}
                         +\hat{\bf k}_{\perp}.    \label{xif_lowbeta}     
\end{eqnarray}
In this limit, $\hat{\bf \xi}_s$ is mostly proportional to $\hat{\bf k}_{\|}$
and $\hat{\bf \xi}_f$ to $\hat{\bf k}_{\perp}$.
When $\alpha \rightarrow \infty$,
equations (\ref{eq_xif_new}) and (\ref{eq_xis_new}) becomes
\begin{eqnarray}
   \hat{\bf \xi}_s \approx  -\hat{\bf \theta}
             +(\sin\theta \cos\theta/\alpha)\hat{\bf k}, \label{xis_highbeta}
\\
   \hat{\bf \xi}_f \approx
                 (\sin\theta \cos\theta/\alpha) \hat{\bf \theta}
                                   +\hat{\bf k}.    \label{xif_highbeta}
\end{eqnarray}
When $\alpha = \infty$, slow modes are called {\it pseudo}-Alfvenic modes.

We can obtain slow and fast velocity component 
by projecting Fourier velocity component 
${\bf v}_{\bf k}$ onto $\hat{\bf \xi}_s$ and $\hat{\bf \xi}_f$, respectively.

{}To separate slow and fast magnetic modes, 
we assume the linearized continuity equation
($\omega \rho_k = \rho_0 {\bf k} \cdot  {\bf v}_k$) and
the induction equation
($\omega {\bf b}_k = {\bf k} \times ({\bf B}_0 \times {\bf v}_k)$)
are {\it statistically} true.
{}From these, we get Fourier components of density
and {\it non-Alfv\'{e}nic} magnetic field:
\begin{eqnarray}
\rho_k 
       &=&(\rho_0 \Delta v_{k,s}/c_s) \hat{\bf k}\cdot \hat{\bf \xi}_s
      +(\rho_0 \Delta v_{k,f}/c_f) \hat{\bf k}\cdot \hat{\bf \xi}_f
        \nonumber \\
       &\equiv&\rho_{k,s}+\rho_{k,f},   \label{eq_rho}   \\
b_k    &=&
       (B_0 \Delta v_{k,s}/c_s) |\hat{\bf B}_0\times \hat{\bf \xi}_s|
      +(B_0 \Delta v_{k,f}/c_f) |\hat{\bf B}_0\times \hat{\bf \xi}_f|
     \nonumber \\
    &\equiv&b_{k,s}+b_{k,f},   \label{eq_b1}  \\
    &=& \rho_{k,s} (B_0/\rho_0)
      (|\hat{\bf B}_0\times \hat{\bf \xi}_s|/\hat{\bf k}\cdot \hat{\bf \xi}_s)
                                     \nonumber   \\
      &+& \rho_{k,f} (B_0/\rho_0)
      (|\hat{\bf B}_0\times \hat{\bf \xi}_f|/\hat{\bf k}\cdot \hat{\bf \xi}_f),
             \label{eq_b2}
\end{eqnarray}
where $\Delta v_k \propto v_k^+-v_k^-$ (superscripts `+' and `-'
represent opposite directions of wave propagation) 
and subscripts `s' and `f' stand for `slow' and
`fast' modes, respectively. 
{}From equations (\ref{eq_rho}), (\ref{eq_b1}), and (\ref{eq_b2}),
we can obtain $\rho_{k,s}$, $\rho_{k,f}$, $b_{k,s}$, and $b_{k,f}$
 in Fourier space.

\end{document}